\documentclass{aa}  
\usepackage{graphicx}
\usepackage{txfonts}
\usepackage{amssymb}
\usepackage{natbib}
\usepackage[FIGTOPCAP]{subfigure}
\usepackage{longtable}
\usepackage{pdflscape} % for 'landscape' environment
\usepackage{cancel}
\usepackage{color}
\usepackage[dvipsnames]{xcolor}

\usepackage{scrextend}
\usepackage{hyperref}

\newcommand{\specialcell}[2][c]{%
\renewcommand{\arraystretch}{1.2}
  \begin{tabular}[#1]{@{}l@{}}#2\end{tabular}}

\def\Hii{H\,{\sc ii}}

\def\kms{km\,s$^{-1}$}

\def\msun{M$_{\odot}$}

\def\Msol{M$_\odot$}

\def\ngc6357{NGC\,6357}
\def\g333{G333.6-0.2}
\def\mys{MYStIX}

\def\Moba{21}
\def\Mob{16}
\def\Ma{5}
\def\Myso{1}
\def\Mlt{55}
\def\Muncl{4}
\def\Mtot{81}

\def\Goba{8}
\def\Gob{4}
\def\Ga{4}
\def\Gyso{1}
\def\Glt{27}
\def\Guncl{3}
\def\Gtot{39}

\def\NGCoba{31}
\def\NGCob{22}
\def\NGCa{9}
\def\NGCyso{4}
\def\NGClt{81}
\def\NGCuncl{7}
\def\NGCtot{123}

\def\TotalSpec{243}

\def\minagem{1}
\def\maxagem{3}
\def\minagengc{0.5}
\def\maxagengc{3}

\def\maxageg{3}
\begin{document} 

\title{The young stellar content of the giant \Hii\ regions M8, G333.6-0.2, and NGC\,6357 with VLT/KMOS\thanks{Based on observations collected at the European Southern Observatory at Paranal, Chile (ESO program 095.C-0048).}}
\titlerunning{The young stellar content of M8, G333.6-0.2, and NGC\,6357.} 

  \author{\mbox{M.C. Ram\'irez-Tannus}\inst{1,2} 
         \and
         \mbox{J. Poorta}\inst{2}
         \and
         \mbox{A. Bik}\inst{3}
         \and
         \mbox{L. Kaper}\inst{2}
         \and 
         \mbox{A. de Koter}\inst{2,4}
         \and
         \mbox{J. De Ridder}\inst{4}
         \and
         \mbox{H. Beuther}\inst{1}
         \and
         \mbox{W. Brandner}\inst{1}
         \and
         \mbox{B. Davies}\inst{5}
         \and
         \mbox{M. Gennaro}\inst{6}
         \and 
         \mbox{D. Guo}\inst{1}
         \and
         \mbox{T. Henning}\inst{1}
         \and
         \mbox{H. Linz}\inst{1}
         \and
         \mbox{T. Naylor}\inst{7}
         \and
         \mbox{A. Pasquali}\inst{8}
         \and
         \mbox{O.H. Ram\'irez-Agudelo}\inst{9,10}
         \and
         \mbox{H. Sana}\inst{4}
}

  \institute{
        Max Planck Institute for Astronomy, Königstuhl 17, 
            D-69117 Heidelberg, Germany\\
     \email{ramirez@mpia.de, j.poorta@uva.nl}
        \and
        %2
            %1
            Astronomical Institute `Anton Pannekoek', University of Amsterdam,
             Science Park 904, 1098 XH Amsterdam, The Netherlands;\\
        \and
        %3
    	    Department of Astronomy, Stockholm University, 
    	    Oskar Klein Center, SE-106 91 Stockholm, Sweden
		\and
		%4
		    Institute of Astronomy, 
		    KU Leuven, Celestijnenlaan 200 D, 3001 Leuven, Belgium
        \and
        %5
            Astrophysics Research Institute, Liverpool John Moores University, Egerton Wharf, Birkenhead CH41 1LD, UK
        \and
        %6
            Space Telescope Science Institute, 3700 San Martin Drive, 
            Baltimore, MD 21218, USA
        \and
        %7
            School of Physics, University of Exeter, Stocker Road, Exeter EX4 4QL, United Kingdom
        \and
        %8
            Astronomisches Rechen-Institut, Zentrum für Astronomie der Universität Heidelberg, 
            Mönchhofstrasse 12-14 69120 Heidelberg, Germany
        \and
        %9
            UK Astronomy Technology Centre, 
            Royal Observatory Edinburgh, Blackford Hill, Edinburgh, EH9 3HJ, UK
        \and
        %10
            German Aerospace Center (DLR), Institute for the Protection of Maritime Infrastructures, Fischkai 1, D-27572 Bremerhaven, Germany
            }
            
   \date{\today}

\abstract
{The identification and characterisation of populations of young massive stars in (giant) \Hii\ regions provides important constraints on \textit{i)} the formation process of massive stars and their early feedback on the environment, and \textit{ii)}  the initial conditions for population synthesis models predicting the evolution of ensembles of stars.}
{We identify and characterise the stellar populations of the following young giant \ion{H}{ii} regions: M8, \g333, and \ngc6357.}
{We have acquired $H$- and $K$-band spectra of around 200  stars using The K-band Multi Object Spectrograph (KMOS) on the ESO {\it Very Large Telescope}. The targets for M8 and \ngc6357\ were selected from the Massive Young Star-Forming Complex Study in Infrared and X-ray (\mys), which combines X-ray observations with near-infrared and mid-infrared data. For \g333, the sample selection is based on the near-infrared colours combined with X-ray data. We introduce an automatic spectral classification method in order to obtain temperatures and luminosities for the observed stars. We analysed the stellar populations using their photometric, astrometric, and spectroscopic properties and compared the position of the stars in the Hertzprung-Russell diagram with stellar evolution models to constrain their ages and mass ranges.}
{We confirm the presence of candidate ionising sources in the three regions and report new ones, including the first spectroscopically identified O stars in \g333.
In M8 and \ngc6357, two populations are identified: (i) OB main-sequence stars ($M > 5$~\msun) and (ii) pre-main sequence stars ($M\approx0.5-5~\rm{M_{\odot}}$). The ages of the clusters are $\sim$\minagem-\maxagem~Myr, $<\maxageg$~Myr, and $\sim$\minagengc-\maxagengc~Myr for M8, \g333, and \ngc6357, respectively. 
We show that \mys\ selected targets have $>$ 90\% probability of being members of the H\,{\sc ii} region, whereas a selection based on near infrared (NIR) colours leads to a membership probability of only $\sim$70\%.
}
{}


  \maketitle
%
%________________________________________________________________

\section{Introduction}

The assembly of massive stars is a key problem in modern astronomy. From an observational point of view, two lines of research are being pursued  to better understand their formation process \citep[]{2007ARA&A..45..481Z,2017arXiv170600118M}. One approach is aimed at the very early stages of star formation, observing the cold cores in which massive stars are being formed. The central objects are surrounded by extended disks and produce molecular outflows that can be studied at sub-mm and radio wavelengths \citep{2016A&ARv..24....6B, 2017A&A...602A..59C,2017A&A...603A..10B, 2016MNRAS.462.4386I}. A second strategy is to study the final phase and outcome of the formation process, characterising the recently formed stars and looking for remnant signatures of their assembly. 

Here, we adopt the second strategy. Finding and characterising young massive stars is important for two reasons. First, the properties of the newly formed massive stars, such as mass, rotational velocity, magnetic field, and multiplicity provide us with information about the formation process. Second, the statistical distributions of these properties – once a large enough sample is scrutinised – define the initial conditions for population synthesis models in which the evolution of entire populations of massive stars is traced.

As formation takes place in opaque natal clouds dispersed throughout the Galactic disk, severe extinction constitutes an observational challenge when studying these young massive stars at optical or shorter wavelengths. Therefore, the youngest stars are studied at near-infrared (NIR) wavelengths, where the extinction is much lower. 
The technical development of instrumentation and detectors in this wavelength regime is also an advantage when studying the youngest populations of stars. For example, by studying stellar populations via NIR spectroscopy, \citet{2012ApJ...744...87B, 2014A&A...561A..12B} present evidence of an age spread (2-3~Myr) in the giant \Hii\ region W3~Main, where the most massive star is 35~\msun\ (O5.6-O7.5V). They conclude that the high UV flux from this and other massive stars affects the disk fraction of the lowest mass stars.  \citet{2016A&A...589A..16W} analysed 14 O-type stars with masses between 20 and 120~\msun\ in the giant \Hii\ region W49, and concluded that the cluster age is around 1.5~Myr. 
\citet{paper1} performed a study of eleven massive young stellar object (mYSO) candidates in the giant \Hii\ region M17, with masses ranging from 6 to 25~\Msol\ and a likely age $<1$~Myr, and confirmed the pre-main-sequence (PMS) nature of six objects. This constitutes a small but unique sample of massive stars observed just after their formation in a very young cluster. 
Interestingly, they observed a lack of double-lined spectroscopic binaries and measured a narrow range of measured radial velocities ($-10 < v_{\rm rad} < 20$~\kms) which is in stark contrast to the overall properties of small and large samples of fully formed, main-sequence massive stars \citep[e.g.][]{2009AJ....137.3358M,2014ApJS..215...15S, 2012ApJ...751....4K, 2012A&A...538A..74P, 2014ApJS..213...34K, 2015A&A...580A..93D}. The samples studied in these works are characterised by large radial-velocity amplitudes ($\Delta v_{\rm rad} > 100$~\kms), from which one can conclude that 40\% to 50\% of OB-type binaries have periods of one month or less.

Spectroscopic studies aiming to classify (massive) stars in star-forming regions have the problem that cluster membership determination using NIR photometry suffer from pollution by 
fore- and background stars \citep[typically $\sim$ 50 \%][]{2012ApJ...744...87B,2016A&A...589A..16W}. The Massive Young Star-Forming Complex Study in Infrared and X-ray \citep[\mys,][]{2013ApJS..209...26F} applied a novel approach in which X-ray, NIR, and mid-IR (MIR) observations are combined. 
Since X-rays are tracers of pre-main sequence low- and intermediate mass stars (through emission produced by coronal activity), and of massive stars (through emission produced by shocks in their stellar winds or colliding winds), this approach 
allows for a more efficient removal of fore- and background stars and hence more robust constraints on cluster membership. 
NIR multiplexed instruments like KMOS on the ESO {\it Very Large Telescope} allow for the simultaneous acquisition of tens of spectra in complex environments. 
The combination of this observing strategy with input catalogues containing \mys\ Probable Complex Members \citep[MPCM;][]{2013ApJS..209...32B} provides an efficient way of obtaining spectra of young massive stars in regions with high extinction. Gaia DR2  \citep{2016A&A...595A...1G, 2018A&A...616A...1G} information on distances and proper motions, as well as spectroscopic followup, provides a reliable way to assess cluster membership. We follow this approach here as well, and cross correlate the \mys\ and NIR photometry methods to assess their reliability. The latter methodologies are the main ones to use at large distances, where Gaia becomes unreliable or unfeasible.

In this paper we present $H$ and $K$ band spectroscopic observations performed with VLT/KMOS surveying three giant \ion{H}{ii} regions (M8, \g333, and \ngc6357) in order to characterise their stellar populations. The three star forming regions observed, presented in Section~\ref{P4:sec:sampleRegions}, were selected to sample a range in cluster morphologies. 
In Section~\ref{P4:sec:obs_red} we provide an overview of the target selection procedure. In Section~\ref{P4:sec:kmos} we describe the KMOS observations and in Section~\ref{P4:sec:datred} the data reduction procedure.
In Section~\ref{P4:sec:phot} we present the stellar populations of the three giant \Hii\ regions by placing them in near-infrared (NIR) and mid-infrared (MIR) colour-magnitude (CMD) and colour-colour diagrams (CCD). 
In Section~\ref{P4:sec:GaiaDR2} we match our sources with the Gaia DR2 catalogues in order to use their astrometric properties to further assess cluster membership. In Section~\ref{P4:sec:mYSOs} we describe the massive young stellar objects detected in our sample. 
We present a detailed NIR spectral classification of the massive stars and the PMS population in Section~\ref{P4:sec:specClass}. With these results we confirm or reject the massive star nature of the candidates presented by \citet{2017ApJ...838...61P}. 
We combine all the results to place the stars in Hertzsprung-Russell diagrams (HRDs) and obtain an estimate of the age of the regions and the mass range of the observed stars in Section~\ref{P4:sec:HRD}. 
We discuss our results in Section~\ref{P4:sec:Discussion} and summarise our findings in Section~\ref{P4:sec:Conclusion}.

\section{Sample of giant \Hii\ regions}
\label{P4:sec:sampleRegions}

 We observed three \Hii\ regions with different mass ranges and physical sizes: \g333, M8, and \ngc6357\ (Table~\ref{P4:tab:regs}). The location of the selected targets is shown in Figures~\ref{P4:fig:M8_targets}, \ref{P4:fig:G333_targets}, and \ref{P4:fig:NGC6357_targets}. We have colour-coded the targets according to their temperature derived from our spectral classification (see Sections~\ref{P4:sec:specClass} and \ref{P4:sec:HRD}). The candidate mYSOs (Section~\ref{P4:sec:mYSOs}) are labelled as magenta diamonds. Below, we describe each region in some detail.

\subsection{M8}

\begin{figure*}[ht]
\centering
{
\includegraphics[width=0.85 \hsize]{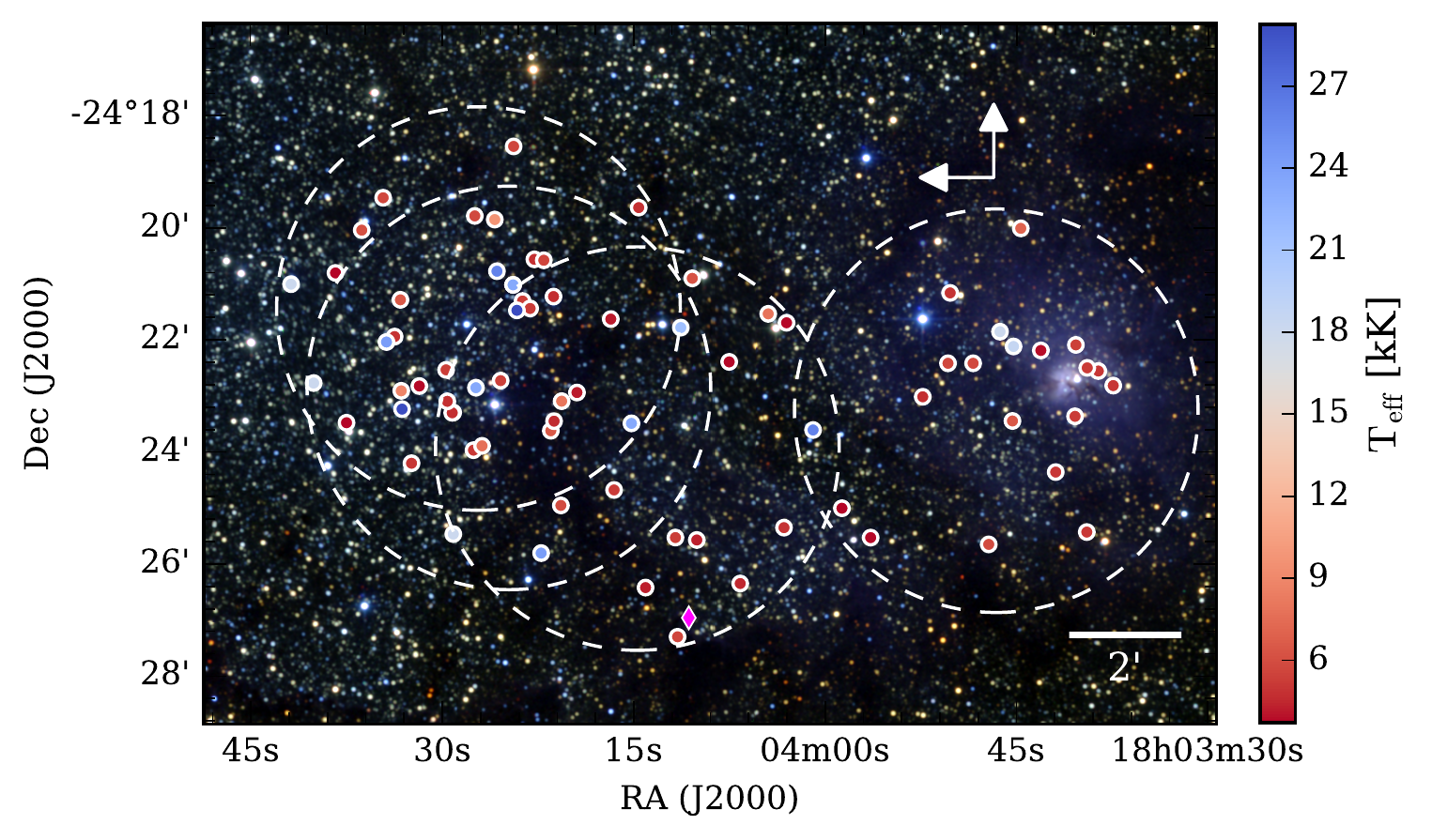}
}
\caption{2MASS near-infrared colour image of the Lagoon Nebula (M8) with the 81 KMOS targets over-plotted. North is up, east is to the left. 
The white dashed circles show the 7.2' field-of-view of KMOS for each of the pointings. The colour of the targets represents the effective temperature according to their spectral type (see Sect.~\ref{P4:sec:HRD}). The objects for which no spectral classification has been obtained are coloured grey. The magenta diamond corresponds to a massive YSO. The number of KMOS targets in this region is 81.
}
\label{P4:fig:M8_targets}
\end{figure*}

Also known as the Lagoon Nebula (Figure~\ref{P4:fig:M8_targets}), the M8 \Hii\ region spans a $\sim$50'$\times$40' region in the plane of the sky and is located at a distance of about 1.3~kpc in the Sagittarius-Carina arm in front of the Galactic centre region. It is associated with the open cluster NGC~6530 containing several O-type stars and about 60 B stars \citep{2013ApJS..209...26F}. The principal source of ionising flux in the \Hii\ region is the O4 star 9~Sgr  \citep{2008hsf2.book..533T}. The western part of the nebula is concentrated into a compact \Hii\ region with much denser ionised gas than the main body of the Lagoon Nebula. This region is called the Hourglass Nebula and it is powered by the O7 star Herschel~36. Since M8 is close to the Galactic plane and projected on top of the Galactic bulge, it is difficult to distinguish between the cluster members and foreground and background stars. Many molecular clumps are scattered through the nebula, suggesting that star formation is triggered by ionisation fronts produced by 9~Sgr and Herschel~36 \citep{2002ApJ...580..285T}.
The MPCM catalogue for M8 contains 2056 sources  \citep{2013ApJS..209...26F}.
Figure~\ref{P4:fig:M8_targets} shows a 2MASS colour image ($J$, $H$, and $K$) of the region. The western compact \Hii\ region around Herschel~36 is visible as diffuse emission. Overplotted as coloured dots are the cluster members for which we obtained KMOS spectra; their colour represents the effective temperature according to their spectral classification.

\citet{2007MNRAS.374.1253A} performed intermediate resolution spectroscopy of PMS stars in M8. They identified 27 classical T~Tauri stars, seven weak-lined T~Tauri stars and three PMS emission objects with spectral type G. They identified the stars to be younger than 3~Myr and between 0.8 and 2.5~$\rm M_{\odot}$. \citet{2007A&A...462..123P, 2019A&A...623A.159P} identified 237 PMS stars as members including 53 binaries. \citet{2010MNRAS.407.1170K} identified 64 class 0/I and 168 class II objects using \emph{Spitzer} near-infrared photometry. 

\subsection{\g333\ }

\begin{figure*}
\centering
{
\includegraphics[width=0.75\hsize]{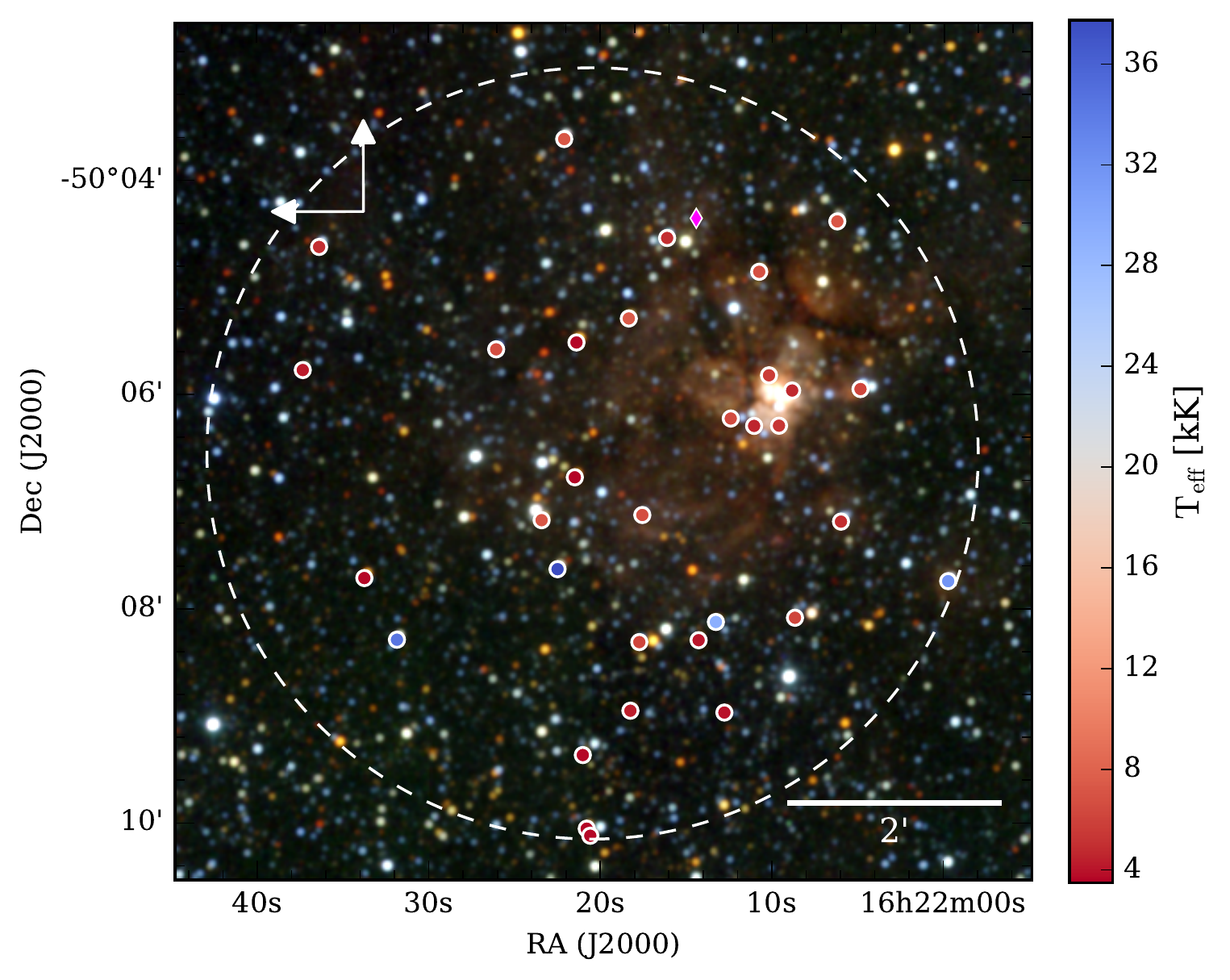}
}
\caption{Same as Figure~\ref{P4:fig:M8_targets} for \g333. The number of KMOS targets in this region is 39.}
\label{P4:fig:G333_targets}
\end{figure*}

\g333\ (Figure~\ref{P4:fig:G333_targets}) is an ultra-compact \Hii\ (UCHII) region embedded in the giant \Hii\ region RCW106. It is the most prominent massive star forming region in the G333 complex. Located at a distance of $\sim2.6$~kpc \citep{2005AJ....129.1523F}, it is compact (11'' in diameter) and dusty, and hosts powerful ionising stars \citep{2013A&A...558A.119K}. \g333\ is embedded in a massive molecular clump of 15000~$\rm M_{\odot}$ \citep{2004A&A...426..119M}. \citet{2014ApJS..213....1T, 2018ApJS..235...43T} detected more than 100 X-ray sources clustered in the central region, and over 500 sources in the field of view. The central sources would be responsible for powering the \ion{H}{ii} region. \citet{2018ApJS..235...43T} also detected diffuse X-ray emission that they associated with plasma flowing out of the embedded cluster. To our knowledge, there are no age estimates for \g333\ reported in the literature.

\subsection{\ngc6357}

\begin{figure*}
\centering
{
\includegraphics[width=0.85\hsize]{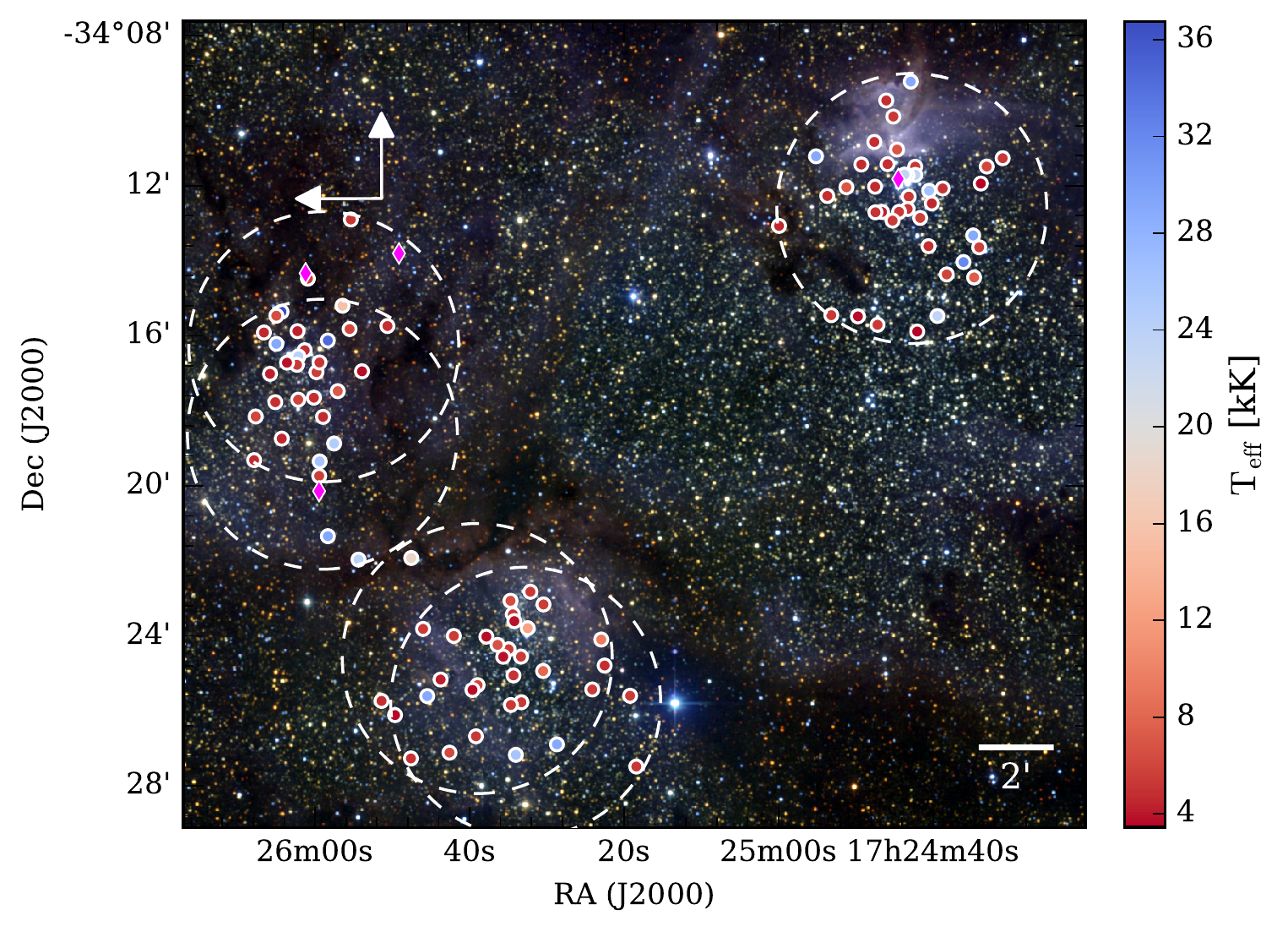}
}
\caption{Same as Figure~\ref{P4:fig:M8_targets} for \ngc6357. The number of KMOS targets in this region is 123.}
\label{P4:fig:NGC6357_targets}
\end{figure*}

\ngc6357\ is a major star-forming region in the Sagittarius-Carina spiral arm (Figure~\ref{P4:fig:NGC6357_targets}). Its brightest \Hii\ region is G353.1+0.9 located near the northern rim of the cloud and ionised by the massive stellar cluster Pis~24 \citep{2013ApJS..209...26F}. \citet{2007ApJ...660.1480M} determined that Pis~24 contains two very massive stars of about 100~$\rm M_{\odot}$ each. The H$\alpha$ image of \ngc6357\ shows a shell containing a Wolf-Rayet star; this could be a super-bubble produced by an earlier generation of massive stars. This hypothesis is supported by the discovery of an older X-ray population around Pis~24 \citep{2007ApJS..168..100W}. \citet{2015A&A...573A..95M} found that Pis~24 is very young ($\sim$1-3~Myr) and that it probably consists of a few sub-clusters. They found evidence that the most massive stars triggered the current star formation episode. The MPCM catalogue for this region contains 2235 sources  \citep{2013ApJS..209...26F}.

%__________________________________________________________________

\section{Observations and data reduction}
\label{P4:sec:obs_red}

The observations presented here were obtained between April and August 2015 using the {\it K-band Multi Object Spectrograph} \citep[KMOS,][]{2013Msngr.151...21S} mounted at the ESO {\it Very Large Telescope} (VLT). For M8 and \ngc6357\ the target selection was based on the results of the \mys\ team \citep[][]{2013ApJS..209...26F, 2013ApJS..209...32B,2017ApJ...838...61P}. \g333\ is not covered by \mys, therefore we adopted a different method in order to select the targets.

\begin{table}
\centering
\caption{Properties of the observed \Hii\ regions.}             % title of Table
\begin{minipage}{\hsize}
\centering
\renewcommand{\arraystretch}{1.4}
\setlength{\tabcolsep}{3pt}
\begin{tabular}{lccccc}
\hline
\hline
\Hii\ & $\alpha$ & $\delta$  & Diameter & Distance \\
region & (J2000) & (J2000)  & \arcmin & pc \\
\hline                        
   M8 & $18^h$ $03^m$41\fs6 & -24\degr22\arcmin43\arcsec   & 10 & $1336^{+76}_{-68}$\footnote{\label{P4:foot:kuhn}\citet{2019ApJ...870...32K}} \\
   \g333\ & $16^h$ $22^m$12\fs3 & -50\degr05\arcmin56\arcsec & 7 & $2540\pm0.71$\footnote{This work} \\
   \ngc6357\ & $17^h$ $24^m$14\fs8 & -34\degr20\arcmin06\arcsec & 27 & $1770^{+140}_{-120}$\footref{P4:foot:kuhn} \\
\hline
\vspace{-15pt}																	
\end{tabular}
\renewcommand{\footnoterule}{}
\end{minipage}
\label{P4:tab:regs}																	
\end{table}	

\subsection{Target selection in M8 and \ngc6357}

M8 and \ngc6357\ are included in the \mys\ project \citep{2013ApJS..209...26F, 2013ApJS..209...32B} which provides a selection of MPCM based on the cross-matching of three samples: (i) X-ray sources with IR counterparts, variability, and/or spatial distribution consistent with being PMS stars; (ii) MIR excess sources consistent with having circumstellar material (i.e. possible YSOs with or without X-ray counterparts); and (iii) known OB stars from \citet{2013ApJS..209...32B}, which do not necessarily have X-ray counterparts.

As the MPCM catalogue contains many more sources than we can possibly observe with KMOS, we applied several selection criteria to obtain a prioritised list of targets. We limited our sample to objects with $K_{\mathrm{s}} < 13.0$~mag. Due to possible saturation of the KMOS detector we could not observe sources brighter than $K_{\mathrm{s}} = 9.0$~mag. The main aim of the project is to classify the stellar spectra in order to derive the overall characteristics of the host regions. Therefore, our selection focuses on stars without infrared excess, as these offer the best chance of a direct NIR detection of the stellar photosphere. We identify three priority classes as follows.

\smallskip
\noindent

As \emph{Priority one} targets we selected all stars listed as known OB stars in the MPCM catalogue. We added a sample of \mys\ candidate OB stars from \citet{2017ApJ...838...61P}. These candidate X-ray emitting OB stars are identified via two methods: 1) By fitting \citet{2004astro.ph..5087C} models to the 1--8~$\mu$m spectral energy distribution and then selecting all objects with $\mathrm{L > 10^4 \,L_{\odot}}$, which approximately corresponds to main sequence B1 stars or earlier; 2) By inspecting the NIR CCD and CMD for the regions and identifying sources with X-ray detection that are reddened main-sequence stars according to their position in the CCD and that have dereddened $J$-band magnitudes $M_J\leq-2.4$~mag. By performing SED fitting, \citet{2017ApJ...838...61P} reported five new candidate OB stars (of which we observed two) in addition to the 28 previously published ones (of which we observed 12) in M8 (see Table~\ref{P4:tab:M8physicalProp}). In \ngc6357 there were 16 previously published OB stars (of which we observed 4). \citet{2017ApJ...838...61P} increased this sample with 26 new candidate OB stars (of which we observed 17): 25 of those selected via SED fitting and one more selected via its NIR colours (see Table~\ref{P4:tab:NGC6357physicalProp}).

\smallskip
\noindent

For the \emph{Priority two} targets, we constructed colour-colour diagrams based on the MPCM $JHK_{\mathrm{s}}$ photometry, and selected the sources without infrared excess. This is similar to method two described above, but without a $J$-band magnitude limit. The targets already included under priority one and those with $9 < K_{\rm s} < 13$~mag are removed from this subsample. In M8 we observed 55 priority two sources and in \ngc6357 82. 

\smallskip
\noindent

As low priority sources (\emph{Priority three}) we added the infrared excess sources rejected in the selection of the priority two sources. We observed 12 sources in M8 and 20 in \ngc6357.

\subsection{Target selection in \g333}

In the case of G333.6-0.2 we selected 23 stars by matching the X-ray observations of this region with the VISTA DR2 catalogue.
We carried out the target selection using the {\it Chandra} X-ray catalogue of \cite{2014ApJS..213....1T}, but in Table~\ref{P4:tab:catalogues} we report matches to the more recent catalogue of \cite{2018ApJS..235...43T}.
This does not change any of the pairings but it does update the X-ray catalogue names.

Many of our objects are saturated in VISTA DR2, and so we replaced any $J$, $H$, or $K$ magnitudes brighter than 13~mag with those of the nearest 2MASS star within 3\arcsec.
Both the VISTA and 2MASS magnitudes should be viewed with some caution as they are often for objects which are crowded or superimposed on nebulosity.
We did not replace the positions of saturated stars since VISTA is less susceptible to crowding than 2MASS, and VISTA positions remain relatively precise to quite bright magnitudes (we found that in the range $9<K<10$~mag 66\% of 2MASS-VISTA matches are within 0.15\arcsec\ of each other in this region). 

The matches were carried out using the Bayesian method described in \cite{2013ApJS..209...30N}, which assesses the likelihood of an X-ray/IR paring using the uncertainty in position assigned to each X-ray source by the source extraction.
It also allows for the fact that brighter sources in the IR are more likely to be the counterparts of a given X-ray source, and reduces the probability of a match if there are other possible counterparts.
The mean probability of an infrared star being the counterpart is 97 percent, with the lowest being 82 percent.

We complemented this sample by adding 18 stars from the 2MASS catalogue with $H-K>$1 mag in order to select some of the sources with high extinction. 
Also, these targets were split in three priority classes, where priority one contains 19 X-ray matched sources without near-infrared excess. 
Four X-ray matched sources show an infrared excess and are added as priority two sources. 
Finally, 18 2MASS sources without an X-ray match are the priority three  sources.

\subsection{VLT/KMOS observations}
\label{P4:sec:kmos}

KMOS consists of 24 deployable integral field units (IFUs) and three spectrographs \citep{2013Msngr.151...21S}. The IFUs are divided into three groups and the light is redirected into one of three spectrographs, each spectrograph disperses the light of eight IFUs. We divided the \Hii\ regions in several fields and for each field we took a long (300~s) and a short (30~s) exposure in order to cover the faint ($13 < K < 11$~mag) and bright ($K < 11$~mag) targets, respectively. For each exposure we applied a 9-point dither pattern both for the $H$- and the $K$-band grating. The fields observed per \Hii\ region are shown in Figures~\ref{P4:fig:M8_targets}, \ref{P4:fig:G333_targets}, and \ref{P4:fig:NGC6357_targets}. In the case of \g333\ and the upper right field in \ngc6357\ we observed the same field twice, once for the bright and once for the faint stars. We used one IFU per spectrograph (three IFUs in total) to perform sky observations.

A set of calibration images was taken in addition to each of the observations. This set consists of dark frames, flat fields and an arc-lamp exposure; the dark and arc-lamp frames are taken at six different rotator angles. A telluric standard star was observed in one IFU per spectrograph.

\subsection{Data reduction}
\label{P4:sec:datred}

The KMOS observations were reduced using the KMOS pipeline version 3.12.3 \citep{2013A&A...558A..56D} in combination with the ESO Recipe Execution Tool (esorex). These routines were used to calibrate and reconstruct the cubes of the science and standard-star observations as described by \citet{2013A&A...558A..56D}. The sky subtraction was performed using the standard KMOS routines, and the telluric correction was done using the default template for telluric correction: one telluric star is observed in one IFU for each of the three spectrographs. This telluric spectrum is then used to perform the telluric correction for all stars in that spectrograph. The data reduction performed by the pipeline can be divided in three basic steps: {\it i)} flat field correction, {\it ii)} extraction of the spectra, and {\it iii)} telluric correction, instrumental effects correction, and flux calibration. 

The first and third steps use procedures that are standard for NIR spectroscopy. Given that we are doing integral field spectroscopy, the second step is different from the standard procedure for other instruments. In the case of KMOS, the reconstruction of the 3D data cubes is performed by interpolating from look-up tables which provide the ($x$,$y$,$\lambda$) location in the final cube for each pixel on the detector. The interpolation is performed based on weighted averages of neighbouring points; therefore the weighted average cannot have a higher value than the neighbour pixel with highest flux. This procedure is problematic for observations of spatially compact, but spectrally extended objects, like our targets. The spatial axes are in that case under-sampled, resulting in (at times quasi-periodic) ripples which appear in the spectra after interpolation and reconstruction. This problem is described in detail in section~4.2 of \citet{2013A&A...558A..56D}. The ripple pattern is prominent and different for each star and is severe in some cases, as can be seen in the spectra displayed in Figure~\ref{P4:fig:M8_SpecExamples} and in the online figures of this paper. For these reasons, we were not able to remove the ripples from the spectra. In the cases where the ripples are severe, they hamper our ability to extract physical information from the spectral lines.

The spectra have a resolving power $R$ between 3570 to 4880 over the $1.46-2.40$~$\mu$m range, corresponding to $\sim80 - 60~{\rm km\,s^{-1}}$ respectively, per resolution element. \citet{2015ApJ...798...23L} performed a series of tests of the reliability of the KMOS pipeline wavelength calibration. They computed the radial velocity shifts from spaxel to spaxel in one IFU as well as the shifts for different IFUs. They conclude that, both  within each IFU and among different IFUs, the wavelength calibration obtained from the KMOS pipeline is accurate at a level of a few {${\rm km\,s^{-1}}$}. They establish that the wavelength calibration by the KMOS pipeline is well suited for bulk kinematic studies, but if one would like to measure the radial velocity of an individual star, the wavelength calibration should be refined.

The normalisation procedure for the OBA stars was performed in two steps. We first fitted a second degree polynomial to all the spectra. Then we visually inspected all the spectra and selected the ones of sufficient quality (not too many ripples and detectable lines, which corresponds to signal-to-noise ratio, S/N > 10) to be classified spectroscopically. From this first step we could also distinguish the late-type objects and the YSOs from the OBA stars. We then re-normalised the OBA stars using the interactive spectrum normaliser specnorm.py\footnote{https://github.com/ishivvers/astro/blob/master/specnorm.py} which allows to select continuum points of the spectrum interactively and fits a cubic spline through the selected points.

\begin{figure*}[ht!]
{
\includegraphics[width=0.98\hsize]{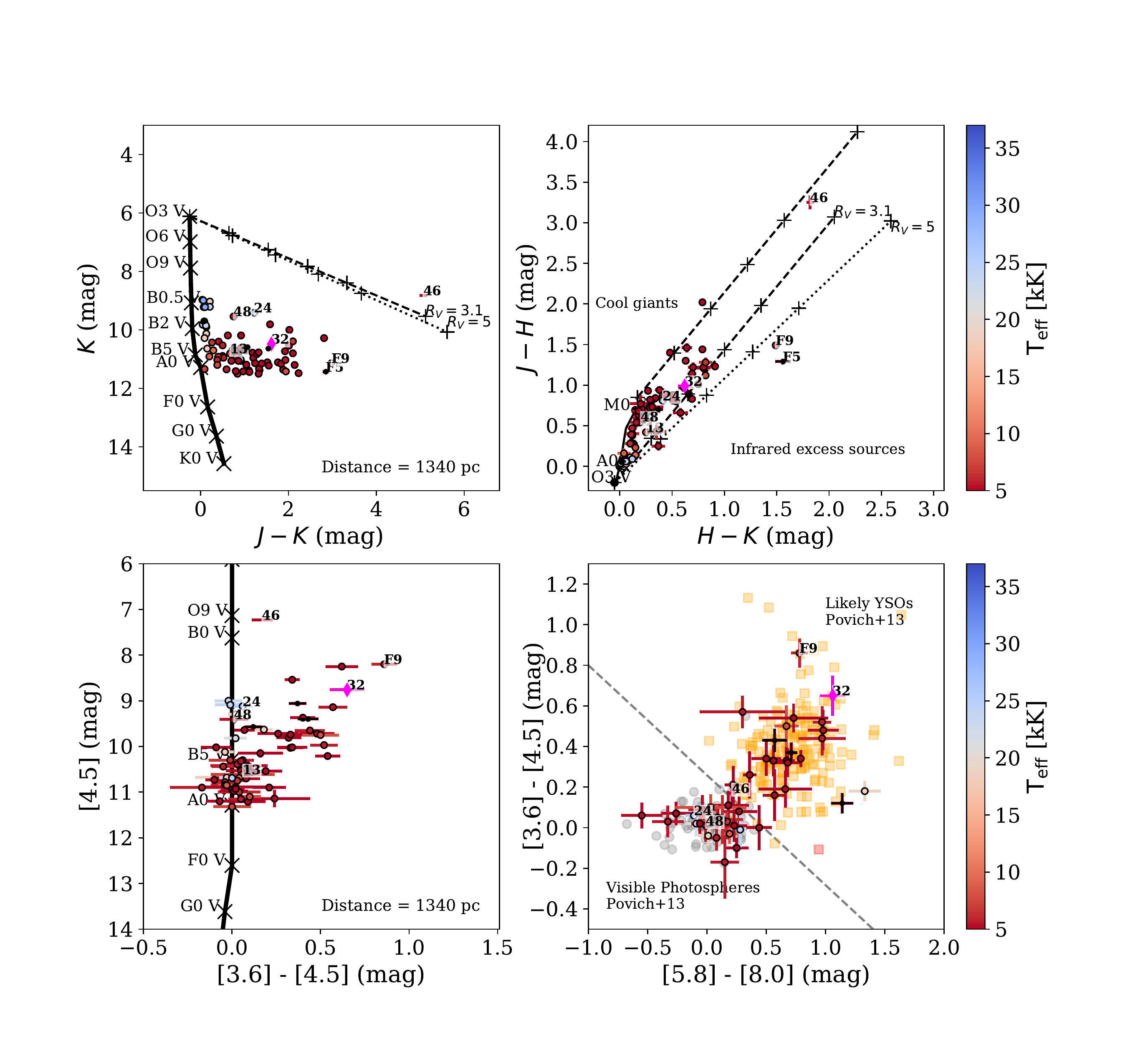}
}
\caption{NIR (top) and MIR (bottom) colour-magnitude (left) and colour-colour (right) diagrams for M8. The objects have been colour-coded based on their effective temperature as determined through spectral classification (Section~\ref{P4:sec:specClass}). 
The magenta diamonds are the objects with CO bandheads in emission and the black dots the unclassified objects. 
The specific objects mentioned in the text are labelled with their respective names. 
The location of the zero-age main sequence at the distance of M8 is shown with a solid black line \citep{1997ApJ...489..698H, 2000asqu.book.....C}. 
The black-dashed line in the top-left panel represents the reddening line for an O3~V star for $R_V=3.1$. In the top-right panel we show the reddening lines of an O3~V (left) and a M0~V (right) star. In both cases the crosses on top of the reddening lines correspond to $A_V$ values of 5, 10, 15, 20, and 30~mag. The grey-dotted lines show the reddening line of an O3~V star for $R_V=5$. 
The dashed line in the bottom-right panel divides the CCD into the areas dominated by YSOs (grey dots) and visible photospheres (orange squares) in the MIRES catalogue \citep{2013ApJS..209...31P}.
}
\label{P4:fig:CMD_CCD_M8}
\end{figure*}

\begin{figure*}
% \centering
{
\includegraphics[width=0.98\hsize]{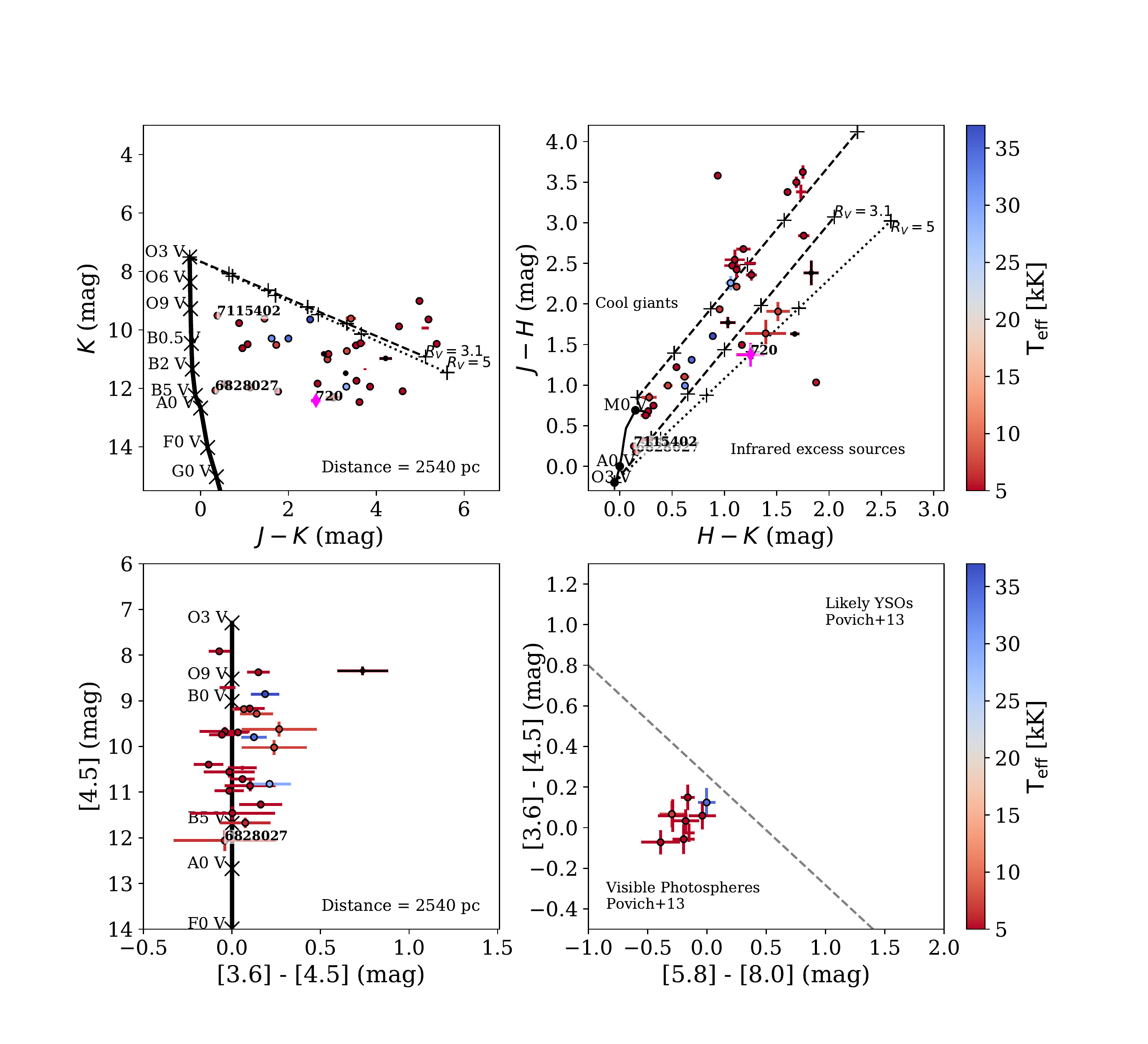}
}
\caption[]{Same as Figure~\ref{P4:fig:CMD_CCD_M8} for stars in \g333.}
\label{P4:fig:CMD_CCD_G333}
\end{figure*}   

\begin{figure*}
% \centering
{
\includegraphics[width=0.98\hsize]{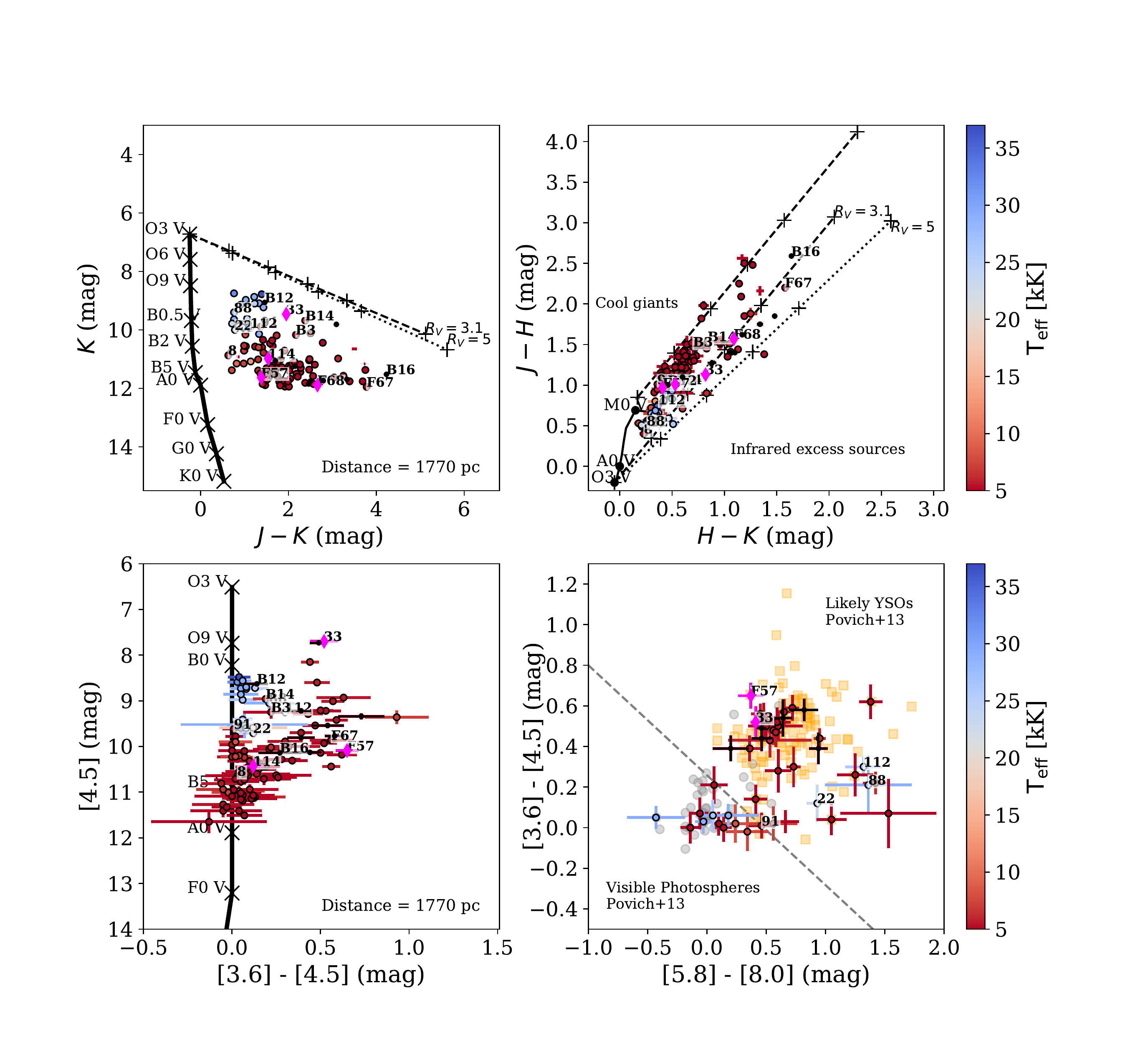}
}
\caption[]{Same as Figure~\ref{P4:fig:CMD_CCD_M8} for our stars in \ngc6357.}
\label{P4:fig:CMD_CCD_NGC6357}
\end{figure*}

\section{Photometric data}
\label{P4:sec:phot}

The photometric data were taken from the \mys\ catalogue \citep{2013ApJS..209...26F, 2013ApJS..209...28K, 2013ApJS..209...29K} which includes the $JHK$-band magnitudes from the {\it United Kingdom Infra-red Telescope} (UKIRT) wide field camera \citep{2007MNRAS.379.1599L, 2008MNRAS.391..136L} or from the 2MASS point source catalogue \citep{2006AJ....131.1163S}, plus the four {\it Spitzer} IRAC bands \citep[GLIMPSE;][]{2003PASP..115..953B}. In the online tables described in Table~\ref{P4:tab:catalogues} we list the photometry for each of the observed objects.

The near- and mid-infrared colour-magnitude diagrams (CMDs) and colour-colour diagrams (CCDs) for each of the regions are shown in Figures~\ref{P4:fig:CMD_CCD_M8} to \ref{P4:fig:CMD_CCD_NGC6357}. In these diagrams too, we have colour coded the sources according to the spectral classification described in Section~\ref{P4:sec:specClass}. The diagrams show the zero-age main sequence (solid line) and reddening lines (dashed and dotted lines). We adopted the extinction law by \citet{1989ApJ...345..245C} with the total-to-selective extinction value $\rm R_V$ set to 3.1 (dashed lines) and 5 (dotted lines). In the NIR diagram one can distinguish a reddened main sequence ($\rm A_V \sim 2$~mag in M8 and $\rm A_V \sim 5$~mag in \ngc6357) and a population of lower mass pre-main-sequence stars. 

For \g333\ we could not rely on the \mys\ catalogue. We used the $JHK$ CMD and CCDs of this region in order to check for membership of our selected targets. The spread of the objects on the CMD and the CCD is sizeable and shows that in \g333\ the extinction varies strongly from sight line to sight line.

In the in the \mys\ IR Excess Source catalogue (MIRES) catalogue compiled by \citet{2013ApJS..209...31P} the authors identified probable YSO members of the \mys\ regions by combining IR SED fitting, IR colour cuts, and spatial clustering analysis. 
In the MIR CCDs of M8 and \ngc6357\ we plot the likely YSOs and IR excess sources from the MIRES catalogue (orange squares) as well as the sources that \citet{2013ApJS..209...31P} identify as having visible photospheres (grey dots) in order to identify the region in the CCDs that is dominated by YSOs. \g333\ is not covered by the MIRES catalogue, so we used the cut found in M8 and \ngc6357\ to show the YSO dominated region in the MIR CCD. In the bottom-right panels of Figures~\ref{P4:fig:CMD_CCD_M8} to \ref{P4:fig:CMD_CCD_NGC6357} the cut used to separate YSOs and visible photospheres is shown with a dashed-grey line.

The CCD and CMD diagrams presented here are further discussed in Section \ref{P4:sec:PhotVsSpt}, where they are also used to confirm or reject cluster membership.

\section{Astrometric properties of our sample}
\label{P4:sec:GaiaDR2}

\begin{figure*}
    {\includegraphics[width=0.98\hsize]{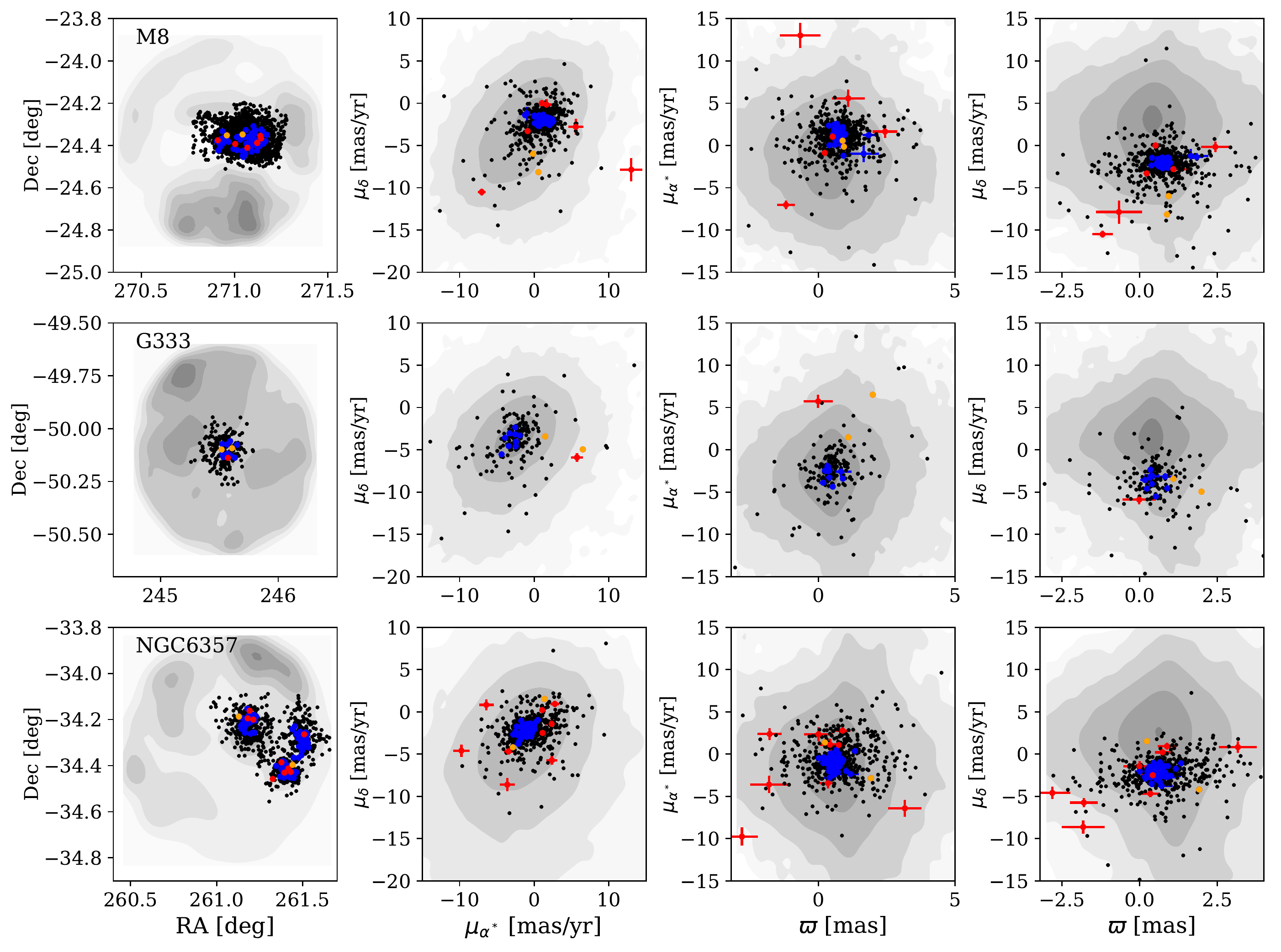}
    }
\caption{Astrometric properties of the here studied cluster stars for M8 \textit{(up)}, \g333\ \textit{(centre)} and \ngc6357\ \textit{(bottom)}. The blue, red and orange correspond to stars in our catalogues. The blue, red and orange dots correspond to the stars in our catalogue for which a reliable cross-match was found. The blue dots are confirmed cluster members, and the red and orange dots are candidate non-cluster members for which we verified the quality of the astrometric fit. The sources in orange have a reliable Gaia DR2 astrometric solution and therefore we confirm that they are not cluster members; for the sources in red we do not confirm nor reject cluster membership. The grey contours represent the density of Gaia DR2 sources brighter than G=19 within a radius of 0.5 degrees from the central coordinates listed in Table~\ref{P4:tab:regs} and the black dots show all the \mys\ sources with Gaia counterparts within $1\arcsec$. The first column shows the equatorial Gaia sky positions. Columns two to four show the parallax $\varpi$ vs the proper motion $(\mu_{\alpha^*}, \mu_{\delta})$.
}
 \label{P4:fig:All_GAIA}
\end{figure*}

We cross-matched our catalogue with the Gaia DR2 archive \citep{2016A&A...595A...1G, 2018A&A...616A...1G,2018A&A...616A...2L}. 
Because of the strong and highly variable extinction in the \Hii\ regions, 
we did not use the magnitude as a cross-match parameter, but relied on sky coordinates only, demanding a separation no larger than 1\arcsec. In addition we limited ourselves to $G\le 19$
to reduce the risk of false matches with very faint stars. 
If more than one match was found within 1\arcsec, the brighter one was chosen. In addition, we extracted all Gaia sources with $G\le 19$ within a radius
of 0.5 degree from the central coordinates given in Table~\ref{P4:tab:regs}. The astrometric properties for the stars in the three clusters are shown in Figure~\ref{P4:fig:All_GAIA}. The sky positions
of the Gaia sources are shown with blue, orange, and red dots in the first column of this figure. The grey contours represent the density of Gaia sources in the local neighbourhood and the black dots show all \mys\ sources (in M8 and \ngc6357) and X-ray sources (in \g333) that have Gaia counterparts within 1\arcsec. We note that for \g333\ only 12 sources of our catalogue have Gaia counterparts that are brighter than $G=19$.

The parallaxes $(\varpi)$ and the proper motions $(\mu_{\alpha^*}, \mu_{\delta})$
of the Gaia sources are shown in columns two to four of Figure~\ref{P4:fig:All_GAIA}. In blue we show our catalogue stars, and the red and orange dots correspond to the stars among our catalogue stars that are probable outliers based on their astrometric properties $(\varpi, \mu_{\alpha^*}, \mu_{\delta})$. We checked whether these properties are sufficiently reliable to reject their membership.  
We computed the renormalised unit weight error (RUWE) which quantifies the quality and the 
reliability of the astrometric fit \citep{LL:LL-124}. 
Following the suggestion of the latter authors, we retain only those sources
for which RUWE~$< 1.4$ as astrometrically reliable.

For each of the clusters we find two outliers (plotted in orange in the Figure~\ref{P4:fig:All_GAIA}) in $(\varpi, \mu_{\alpha^*}, \mu_{\delta})$ space among the cluster stars in our catalogues, which have nevertheless a reliable Gaia DR2 astrometric solution.
For M8 these are sources F13 (2MASS~J18035006-2421079) and F24 (2MASS~J18041025-2420524). 
For G333.6-0.2 the sources are 515756828027 (2MASS~J16222585-5005334) and 515757115402 (2MASS~J16220464-5005556). 
For \ngc6357\ the astrometric solution of sources 91 and 8 (2MASS~J17243100-3411143) indicate that the stars may not be cluster members.

The Gaia distances computed using the confirmed members (blue dots) are $1.34\pm0.07$~kpc for M8 and $1.77\pm0.12$~kpc for \ngc6357, which are in agreement with the distances derived by \citet{2018ApJ...864..136B} and \citet{2019ApJ...870...32K} using Gaia parallaxes of more complete samples. We derive a distance of $2.54\pm0.71$~kpc for \g333 which is in agreement with the reported distance to this region by \citet{2005AJ....129.1523F}.

%__________________________________________________________________

\section{Massive young stellar objects}
\label{P4:sec:mYSOs}

In the sample selection we gave priority to objects with no IR excess. Nevertheless, our KMOS sample includes six objects that we classified as candidate massive young stellar objects (mYSOs): object 32 in M8, objects 33, F57, 114, and F68 in \ngc6357, and object 720 in \g333\ . Their position in the CMD and CCD (Section~\ref{P4:sec:phot}) is consistent with a mYSO and/or class~II nature. We classify these objects based on the presence of double-peaked emission lines and/or CO overtone emission, indicative of a rotating circumstellar disk. CO overtone emission is produced in high-density (10$^{10}$ -- 10$^{11}$~cm$^{-3}$) and high-temperature (2500 -- 5000~K) environments \citep{1995ApJS..101..309C}. CO is easily dissociated, and therefore must be shielded from the strong UV radiation coming from the star. These conditions occur in the inner regions of accretion disks, which makes the CO bandheads an important tool to trace the inner disk structure around mYSOs \citep[e.g. ][]{2006A&A...455..561B, 2013MNRAS.429.2960I,paper1}. The shape of the CO lines probes the Keplerian rotation profile of the circumstellar disk \citep{2004A&A...427L..13B, 2004ApJ...617.1167B, 2010MNRAS.402.1504D, 2010MNRAS.408.1840W, 2013MNRAS.429.2960I}. 
The blue shoulder in the bandheads is a measure of the inclination of the disk: an extended blue shoulder indicates a high inclination angle (i.e. near `edge-on' view). 

In Figure~\ref{P4:fig:CO_Brlines} we plot the Brackett series profiles and the CO first overtone bandheads for the six candidate YSOs in our KMOS sample. Here we provide a description of the individual mYSO candidates:

\begin{figure*}[ht]
    {\includegraphics[width=0.9\hsize]{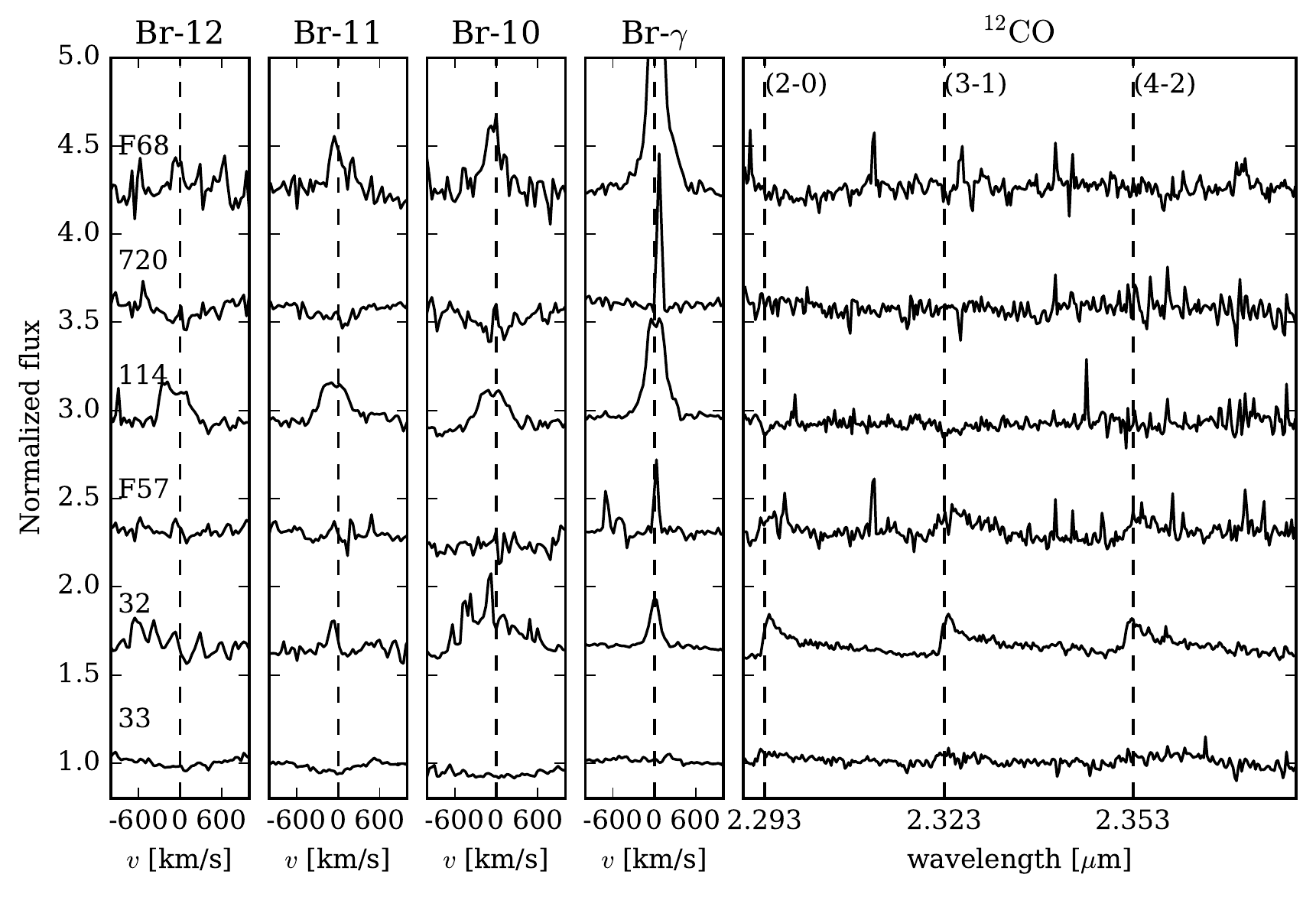}
    }
\caption{{\em Left}: Brackett series line profiles for the six YSOs in our sample: object 32 in M8, objects 33, F57, 114, and F68 in \ngc6357, and object 720 in \g333. {\em Right}: CO first overtone bandheads CO(2-0), CO(3-1), CO(4-2).}
 \label{P4:fig:CO_Brlines}
\end{figure*}

\emph{NGC\,6357-F68:} We do not detect any CO bandheads in this object, but the Brackett (Br) lines are in emission. Br$\gamma$ is very strong in emission on top of a photospheric absorption line. Br-12, Br-11, and Br-10 appear to be contaminated by residuals of the data reduction. This source is not part of the MIRES catalogue \citep{2013ApJS..209...31P}.

\emph{\g333\ -720:} CO bandhead emission is not present. The Brackett lines are in emission superposed on an absorption line; the emission is very narrow and likely has a nebular origin. In order to confirm the presence of a circumstellar disk, more and better signal-to-noise observations are needed.

\emph{NGC\,6357-114:} This object does not show CO bandheads in emission, but all the Brackett lines show clearly double-peaked emission profiles. This is the only object in our sample where the double-peaked structure is resolved. This source is not in the MIRES catalogue.

\emph{NGC\,6357-F57:} The CO bandheads are clearly present in emission and they show a pronounced blue shoulder. The Br lines seem to be filled in by disk emission. Br$\gamma$ shows a central emission, but the most pronounced part of this feature is very narrow and therefore is likely caused by residuals from the sky subtraction. This source is part of the MIRES catalogue and it is classified as a stage II YSO, dominated by optically thick circumstellar disks.

\emph{M8-32:} The CO bandheads are clearly detected. The Br lines in this object are in emission; Br-11 and Br-10 seem to be contaminated by residuals of the data reduction process and Br-12 and Br$\gamma$ show emission. The Br lines are single-peaked and the CO bandheads lack a pronounced blue shoulder indicative of a high inclination angle of the disk. This source is part of the MIRES catalogue and it is classified as a stage II YSO, dominated by optically thick circumstellar disks.

\emph{NGC\,6357-33:} This object shows weak CO bandheads in emission. The Br lines seem absent, but this could be due to veiling. The mYSO nature of this object is unclear. This source is not in the MIRES catalogue.

%__________________________________________________________________

\section{Spectral classification}
\label{P4:sec:specClass}

The spectral classification of the \TotalSpec\ observed stars was carried out on the basis of their KMOS $H$ and $K$ band spectra, spanning the wavelength range $1.5 - 2.4~\mu$m. By a first visual inspection of the objects with no spectral signatures of them being mYSOs, we checked for the absence or presence of CO absorption bands; in this way all stars were categorised as either early (OBA) or late (FGKM) type, respectively (or both in case of doubt). For a more precise classification two spectral libraries were employed: (1) 56 normalised, O3 to mid-B type, $H$ and $K$ band spectra from \citet{1996ApJS..107..281H}, \citet{2005A&A...440..121B}, and \citet{2005ApJS..161..154H},  for the early types; and (2) 190 flux-calibrated, F0 to M9 type, $H$ and $K$ band spectra from \citet{2009ApJS..185..289R} \footnote{These template spectra can be found in \url{http://irtfweb.ifa.hawaii.edu/~spex/IRTF_Spectral_Library/index.html}} for the late types. Both libraries span all luminosity classes. We note that for our three giant \Hii\ regions, given their young age, the late type stars probably correspond to low- and intermediate-mass PMS stars.

After initial separation in early and late types the spectral classification was done by two independent methods. On the one hand all spectra were classified by visual inspection. On the other hand, a computer code was developed to automatically classify spectra in each category (early or late), making use of the same spectral libraries as for the visual inspection. We now discuss each method separately and subsequently compare their results. 

\subsection{An automatic method to classify stars in the NIR}

In the presentation of their template library \citet{2009ApJS..185..289R} demonstrate that for a few spectral features (\ion{Ca}{ii}, \ion{Na}{i}, \ion{Al}{i} and \ion{Mg}{i}; their Figure~35) a relation exists between equivalent width (EW) and spectral type. They suggest that these relations can be used for spectral classification. Based on this idea and in order to have a reproducible and efficient method at hand, we wrote a Python code to do an automatic spectral classification using $H$ and $K$ band spectra.\footnote{The Python scripts are available upon request.} To the set of features from \citet{2009ApJS..185..289R} we add features from literature, adjusting the definitions where necessary and including a few of our own. Though the spectral features are often centred on a line, they can also involve multiple lines, a molecular bandhead or only a part of a line, depending on the wavelength range and spectral type considered. The features and continuum regions that were used are listed in Table~\ref{P4:tab:EW_late} for the late-type spectra and in Table~\ref{P4:tab:EW_early} for the early-type spectra. 

The automatic method was tested on the spectra of the template libraries themselves. In each case the spectrum under examination was excluded from the fits mentioned in step two below. Taking the root-mean-square of the errors on the results we determined the mean error of the method for the template spectra to be two spectral subtypes for the late, and one spectral subtype for the early types.

\begin{figure*}
\centering
{
\includegraphics[width=0.85\hsize]{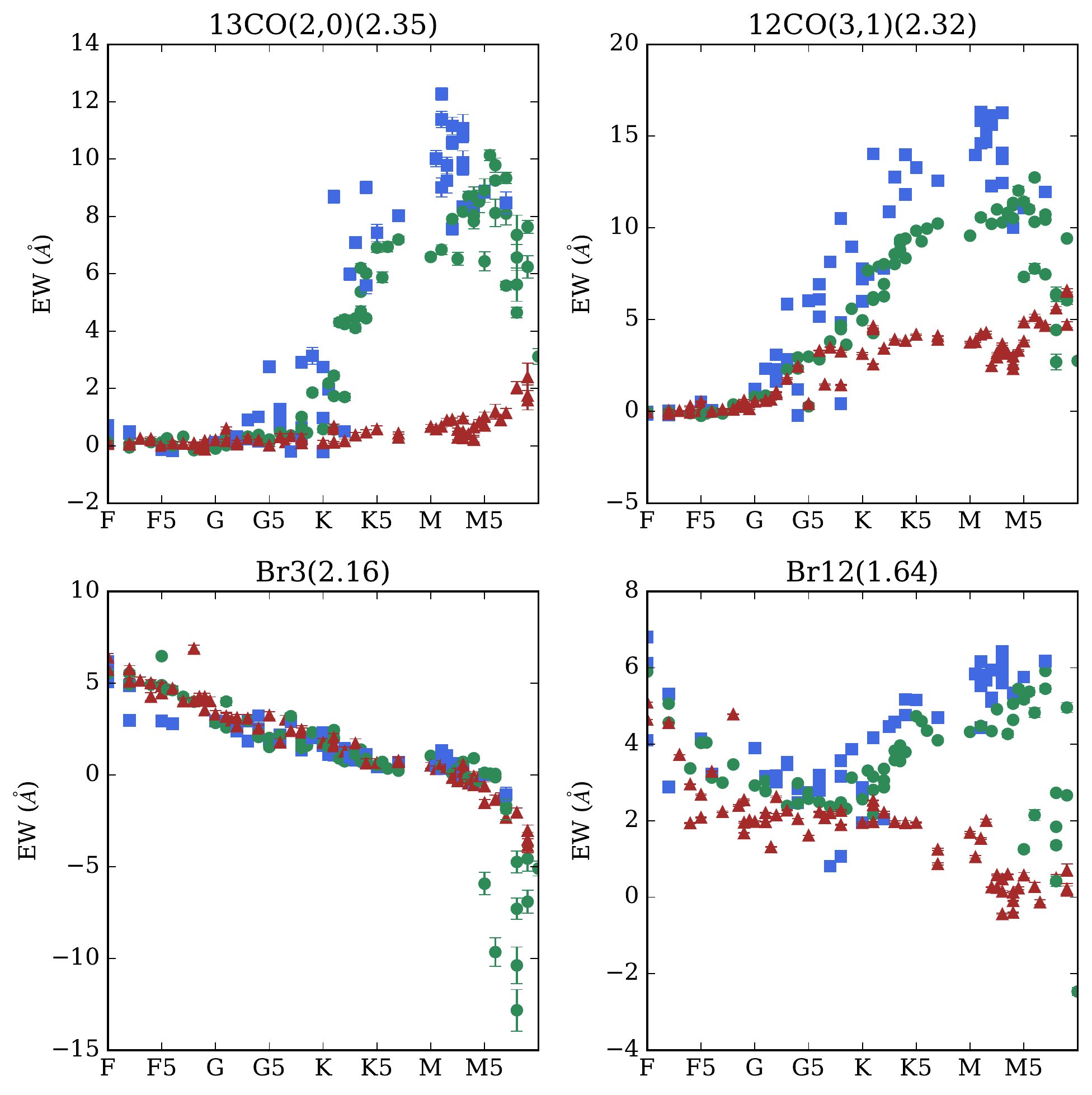}
}
{
\includegraphics[width=0.88\hsize]{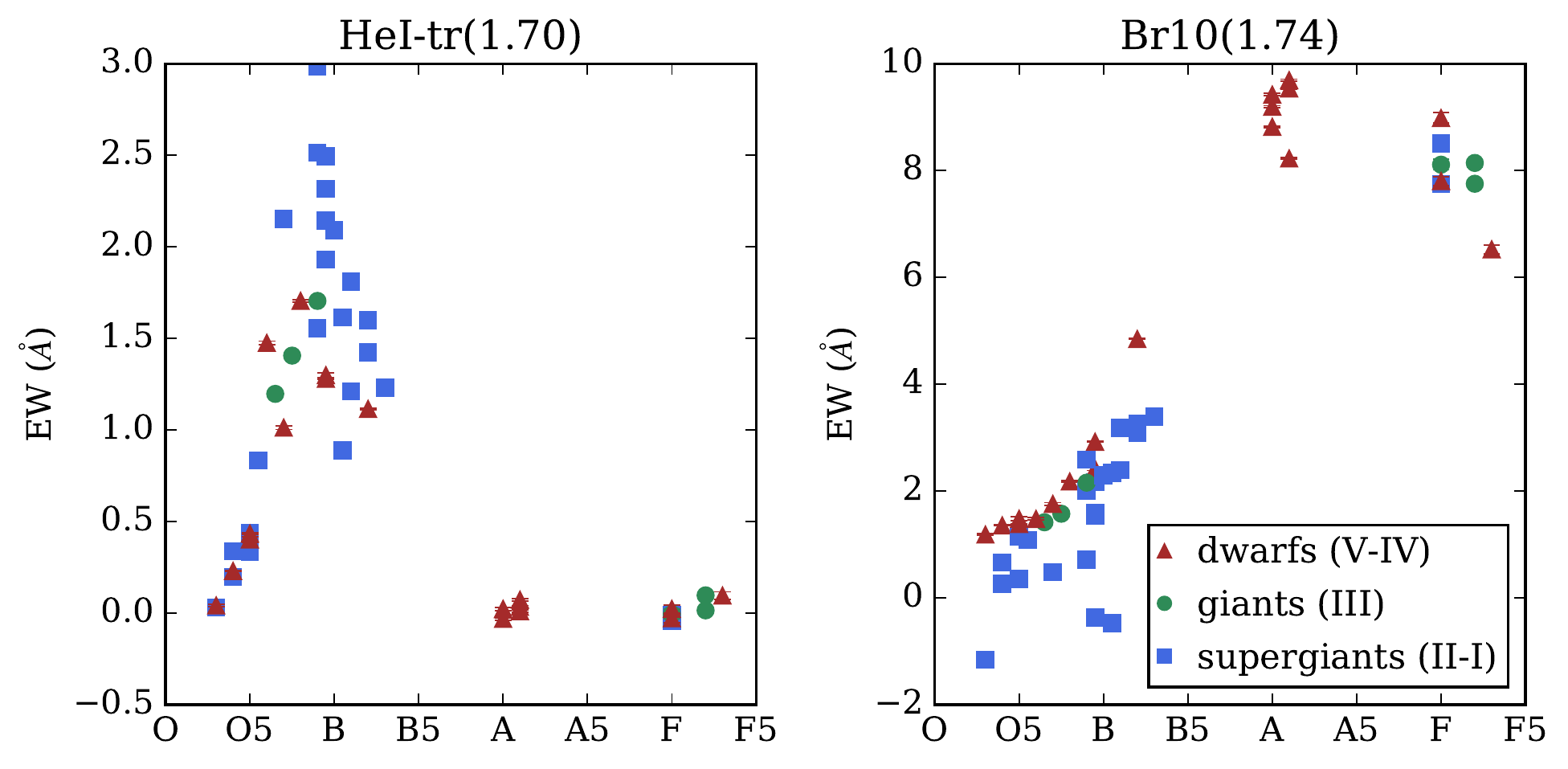}
}
\caption{Relation between EW-value and spectral type on the basis of late type (four top panels) and early type (bottom two panels) templates for four features. The symbols and colours represent  different luminosity classes: red triangles for dwarfs (V-IV), green circles for giants (III) and blue squares for supergiants (II-I).}
\label{P4:fig:EW_SpT}
\end{figure*}

The method treats the early- and late-type stars separately, but the basic steps of the procedure are the same for both cases, as follows.

\vspace {2mm}

    First, the EWs, together with their variance $\sigma^2_{\rm EW}$, of the defined spectral features are calculated for all template spectra. Following \citet{2005ApJ...623.1115C} they are given by: 

    \begin{align}
        \rm EW &= \sum \left[1-\frac{f(\lambda)}{f_{\rm c}(\lambda)}\right] \Delta \lambda \\
        \sigma^2_{\rm EW} &= \sum \Delta \lambda^2 \left[\frac{\sigma^2(\lambda)}{f_{\rm c}^2(\lambda)} + \frac{f^2(\lambda)}{f_{\rm c}^4(\lambda)}\sigma_{\rm c}^2(\lambda) \right]
        ,
    \end{align}

    where $f$ and $f_{\rm c}$ are the observed feature and estimated continuum flux, $\sigma$ and $\sigma_{\rm c}$ are the errors on $f$ and $f_{\rm c}$, respectively, and the summations are over the number of wavelength intervals $\Delta \lambda$ across the feature. The $\Delta \lambda$ intervals were chosen such that the apparent Doppler shifts in the observed spectra do not affect the EW measurements. We follow the procedure as described in the appendix of \citet[][eqs. A1-A20]{1992ApJS...83..147S}. Wavelengths are converted to velocities with the centre of the feature at 0~${\rm km\,s^{-1}}$, and further normalised to the highest velocity so that all (velocity) values fall within the interval [-1,1]. The latter in order to simplify the error calculation \citep[for more details see][]{1992ApJS...83..147S}; conversion back to \AA\ is done at the end. We estimate $f_{\rm c}$ by fitting a first degree polynomial through the defined continuum regions and $\sigma$ (the error on the flux $f$) is determined as the standard deviation of the observed values around these continuum fits, in the fitted regions. $\sigma_{\rm c}$ is calculated from the covariance matrix produced by the fit, as detailed in \citet{1992ApJS...83..147S}.

\vspace {2mm}

    Second, a relation between EW value and spectral type is established per luminosity class. As an example, in Figure~\ref{P4:fig:EW_SpT} we visualise this relation by plotting the EW values of all templates in the library for four different features for the late types and two features for the early types. The relation is quantified by fitting a 5th (for the late types) or 3rd (for the early types) degree polynomial $p_{\rm fit}(\rm SpT)$ through the EWs for each given luminosity class. Thus, for a certain EW value of a particular feature there is one or a few likely spectral types per luminosity class. 

\vspace {2mm}

Third, the EWs are determined for a science spectrum. Taking into account the errors, each EW is mapped to a likelihood for spectral types, using the polynomial fits obtained in the previous step. In order to quantify how far the determined EW lies from the polynomial, the fitted polynomial $p_{\rm fit}(\rm SpT)$ is mapped to a Gaussian $G(x;\mu, \sigma)$ with $\mu= \rm EW$ and $\sigma=\sigma_{\rm EW}$:

     \begin{equation}\label{P4:eq:Lspt}
        L(\rm SpT)_{EW} = G\left(p_{\rm fit}(\rm SpT);\rm EW,\sigma_{\rm EW}\right)
        .
    \end{equation}
 
The resulting likelihood distribution $L(\rm SpT)_{EW}$ (one for each feature) peaks at the most likely spectral type(s) according to the EW for that feature. To avoid very sharp peaks a minimum $\sigma$ of $0.2$\,\AA\ is applied. To combine the information contained in each distribution and obtain one final likelihood distribution for each luminosity class we tried several approaches among which: (1) taking the logarithm of the $L(\mathrm{ SpT})_{EW}$ before adding all distributions and (2) adding one to $L(\mathrm{SpT})_{EW}$ to avoid numerical zero's before multiplying all distributions. In both cases all EWs were weighted equally. Testing the results on the template spectra as described before, the lowest errors were obtained when applying method (2):

\begin{equation}\label{P4:eq:Lspt_plus1}
	L(\mathrm{SpT})_{\mathrm{tot}} = \prod_{\mathrm{EW}} \left[ L(\rm SpT)_{EW} + 1 \right ]
	.
\end{equation}

\begin{figure}
\centering
{
\includegraphics[width=0.98\hsize]{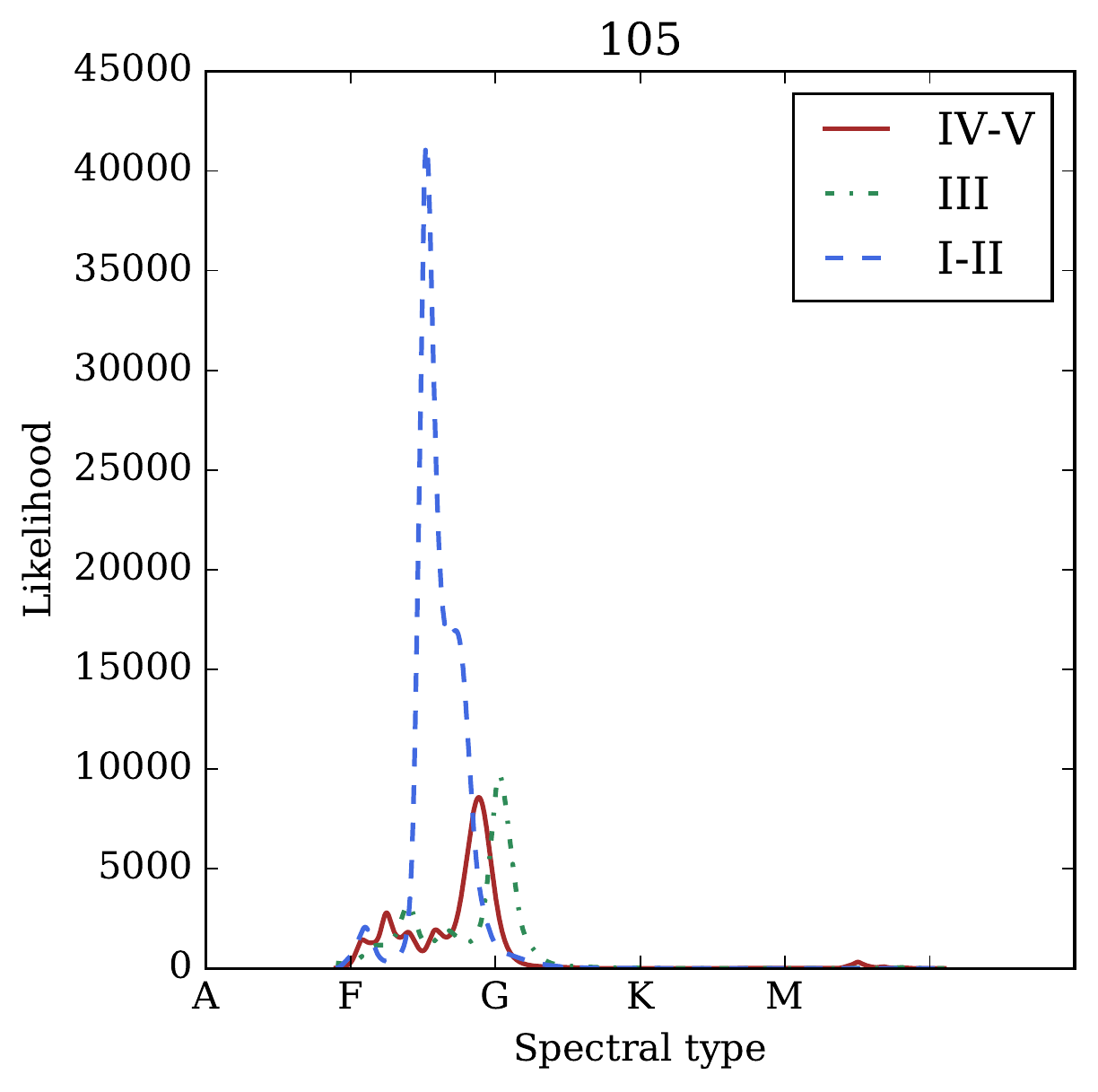}
}
\caption{Example of a final likelihood distribution resulting from the automatic classification method for star 105 in \Hii\ region \ngc6357. The different coloured lines are for different luminosity classes: red for dwarfs (IV,V), green for giants (III) and blue supergiants (I,II). The method returns the spectral type corresponding to the highest peak, in this case F5 I-II. We assign an uncertainty of two spectral sub types to this value for all the classified stars.}
\label{P4:fig:totals_105}
\end{figure}

A detailed explanation is provided in Appendix~\ref{P4:ap:plusOne}. An example of a final distribution is shown in Figure~\ref{P4:fig:totals_105}.

\vspace {2mm}

Finally, the method returns the value of the highest peak of the final distributions as the spectral type of the object and we assign an uncertainty of two spectral sub-types to the obtained value. The luminosity class is determined by the distribution with the highest peak. 

\vspace {2mm}

For the late types, both EWs and ratios of EWs are taken into account. Steps two and three of the procedure are applied in the same way to the EW ratios. The resulting likelihood distributions are simply multiplied together with those from the EWs, so that one final distribution is obtained for each object: one that includes the information both from the EWs and the ratios. Ratios of which at least one of the EWs has a negative value are excluded in order to avoid negative ratios. Since no real new information is added, one could view the inclusion of ratios as a way of giving extra weight to certain lines or features, specifically to those that do not have negative EWs. We list the ratios used for the classification in Table~\ref{P4:tab:lineRatios}.

\subsection{Spectral classification by eye}
\label{P4:sec:class_byeye}

As an additional test on the method all science spectra were classified by eye. Each spectrum was individually overplotted with all available templates of the early or late type library. For each star the best matching type or (in the majority of cases) a range of types was selected, within the limitations of the quality of the KMOS spectrum and the completeness of the template spectra. Some of our stars were not classified in this way due to poor data quality, which was a result either of the ripple patterns caused by the data reduction procedure (see Section~\ref{P4:sec:datred}) or of poor  S/N  due to extinction, or both. The  S/N  in the $H$ band is about 20 on average for the regions G333 and M8; for \ngc6357\ this is close to 30. In the $K$ band a  S/N  of around 30 was reached in all regions, being somewhat lower for G333 and somewhat higher for \ngc6357\. The spectra that were not classified by eye due to poor  S/N  typically had values less than ten in the $H$ band.

For the early types, hydrogen (e.g. \ion{H}{i} $2.16~\mu$m) and helium (e.g. \ion{He}{i} $2.11~\mu$m and \ion{He}{ii} $2.19~\mu$m) lines were the only distinguishing features. For the late types both hydrogen and atomic metallic lines (the most prominent being \ion{Mg}{i}, \ion{Si}{i}, \ion{Ca}{i}, \ion{Al}{i}, and \ion{Fe}{i}) were used, as well as CO ro-vibrational bandheads for the lower temperature types (mostly detected in the $K$ band). 

For F-type stars the spectrum is dominated by the hydrogen lines but some metallic lines start to appear. The hydrogen lines significantly weaken at early G spectral types so that by G0 the spectra are dominated by metals. The ratio of \ion{Mg}{i}/Br$\gamma$ is a good temperature indicator for F-type stars. In order to distinguish the stars in luminosity class, the width of the hydrogen lines was considered: giants and supergiants have narrower lines than dwarf stars.

In the case of G-K stars the $K$ band is best suited to do the spectral classification. The ratios between Br$\gamma$ and \ion{Na}{i} at 2.26~$\rm \mu m$, and Br$\gamma$ and \ion{Ca}{i} around 2.20~$\rm \mu m$ can be used, as these metallic lines are more sensitive to the decrease in temperature than others \citep{2009ssc..book.....G}. For later types, the CO bandheads both in the $H$ and in the $K$-band were mostly used as temperature and luminosity class indicators.

\subsection{Comparison of the automatic method with the classification by eye}

Before we use the spectral classification to determine the age of the studied regions, we first compare the results of the two methods. Stars for which the two spectral types did not match within error bars we carefully reinspected and a final decision was made (see columns three and four of Tables~\ref{P4:tab:catalogues}). For stars that could not be classified by eye due to poor data quality, we kept the results of the automated classification, unless we judged any result to be untrustworthy based on the spectral quality. We label our final decision as `corrected classification'.

\begin{figure}
{
\includegraphics[width=1.0\hsize]{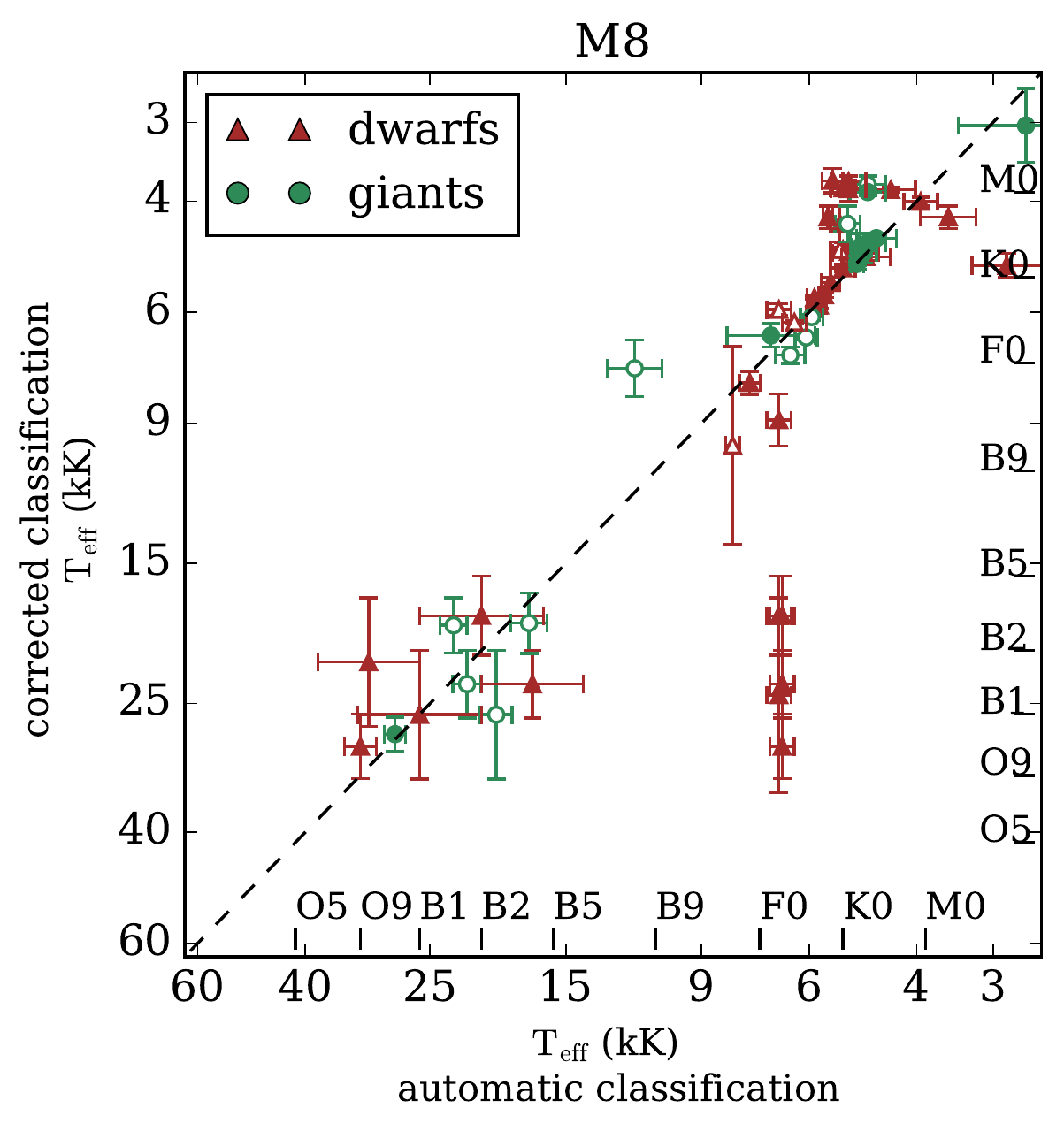}
}
\caption{Comparison of the automatic classification and the one after re-inspection of the spectra (corrected classification) for all stars classified by both methods for M8. The corrected classification corresponds to either the automatic classification or the classification by eye depending on each specific case (see column four in Table~\ref{P4:tab:catalogues}). The red triangles represent dwarfs (IV-V), and the green circles (super)giants (I-III) according to the automatic classification. Full symbols signify agreement in luminosity class with the corrected classification. The outliers in the bottom-right of the plot correspond to the early B-type stars that are misclassified as F-type by the automatic method.}
\label{P4:fig:M8_compare}
\end{figure}

\begin{figure}[ht!]
{
\includegraphics[width=1.0\hsize]{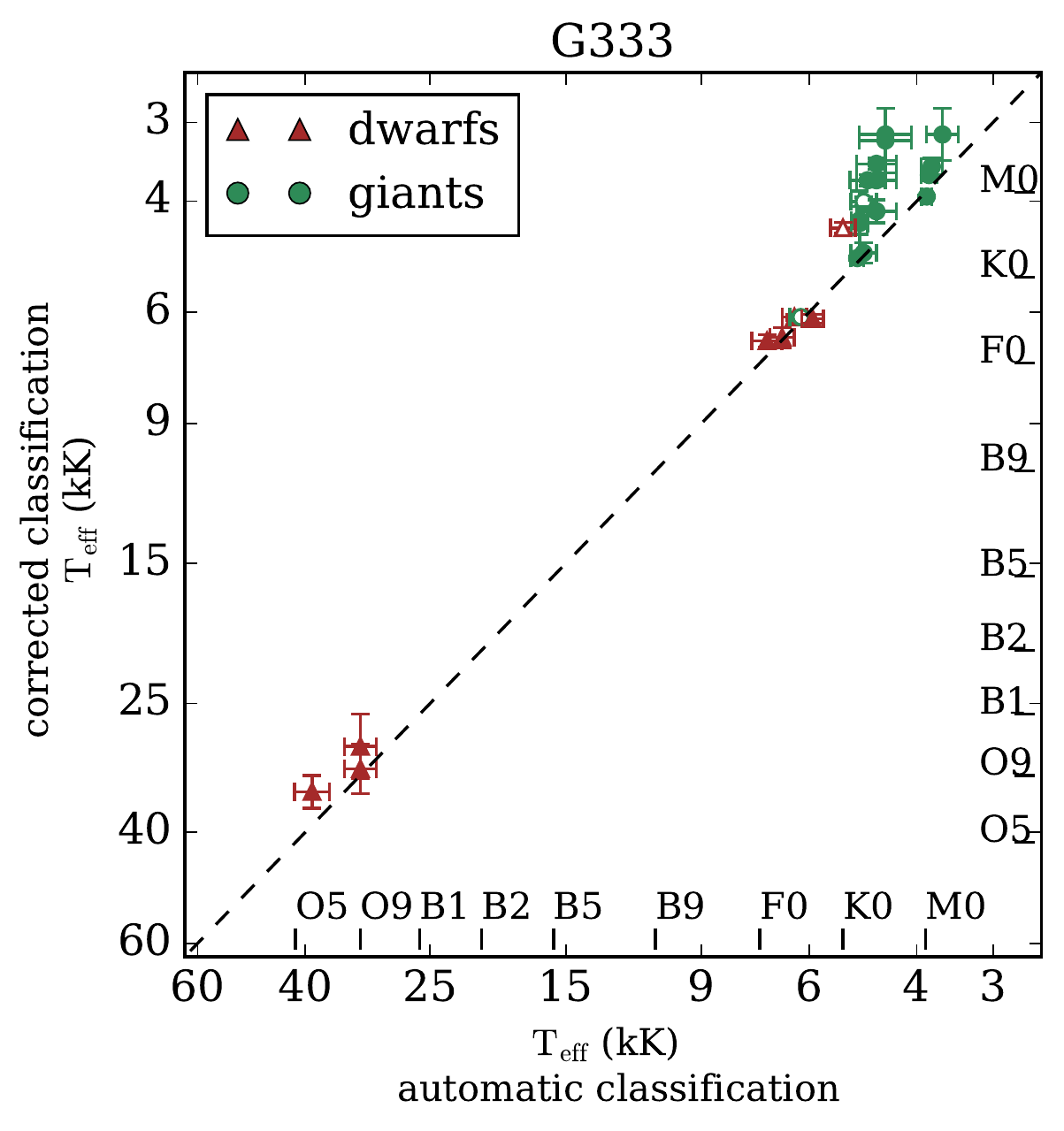}
}
\caption{Same as Figure~\ref{P4:fig:M8_compare}, but for \g333.}
\label{P4:fig:G333_compare}
\end{figure}

\begin{figure}[ht!]
\centering
{
\includegraphics[width=1.0\hsize]{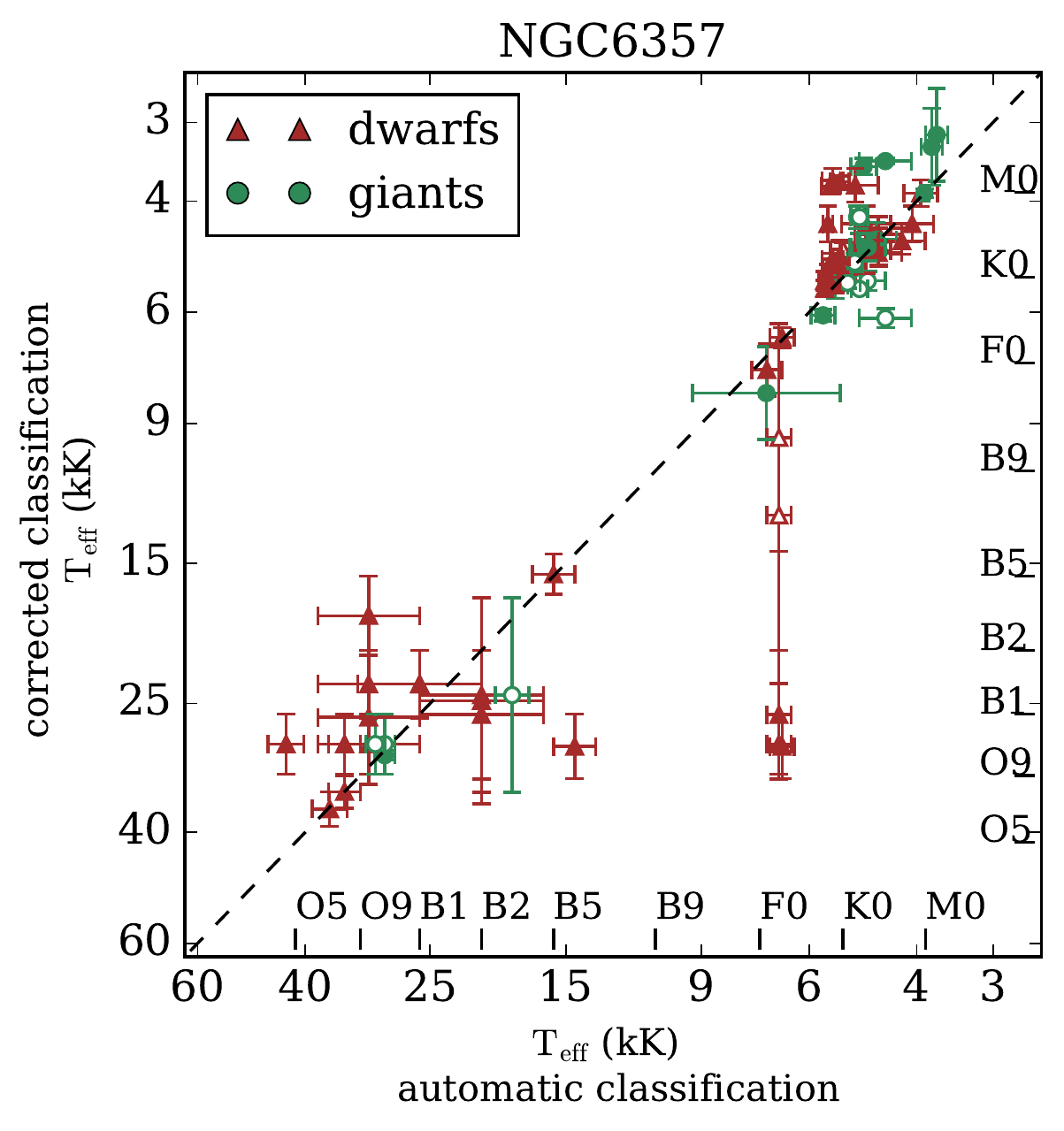}
}
\caption{Same as Figure~\ref{P4:fig:M8_compare}, but for \ngc6357.}
\label{P4:fig:NGC6357_compare}
\end{figure}

Figures \ref{P4:fig:M8_compare}, \ref{P4:fig:G333_compare} and \ref{P4:fig:NGC6357_compare} present the comparison of the corrected classification with the automatic method for each region. The comparison is shown in effective temperature rather than spectral type (see Section~\ref{P4:sec:HRD}), to better reflect effects on the final results. There is an overall good agreement except for some B stars (that are misclassified as early F types by the automatic method) and some late M stars (that are misclassified as K types by the automatic method). These are outliers for different reasons: in case of the early types the miss-classification is due to a lack of late-B and A-type templates, requiring an interpolation from B3 to F0-F3 with only a few A0-A1 templates in between.

This interpolation helps to classify the late-A and early-F type stars correctly, but causes some B-types to be mistaken for F-types. In case of the late types it is the (relative) importance of the $^{13}$CO feature that causes the confusion for M types: detecting this feature by eye assures one that the star must be M type, however, no special weight is assigned to this feature by the automatic method, which takes into account 23 features for the late types.  

To quantify the accuracy of the automatic method further, we compare the number of stars it misclassifies with the number of stars misclassified by initial visual inspection (Section \ref{P4:sec:class_byeye}) taking the final corrected classification as reference. The results are shown in Table~\ref{P4:tab:method_comparison}. We see that over the sample that we test, 74\% is correctly classified by the automatic method and 73\% by initial visual inspection.

\begin{figure*}
{
\includegraphics[width=\hsize]{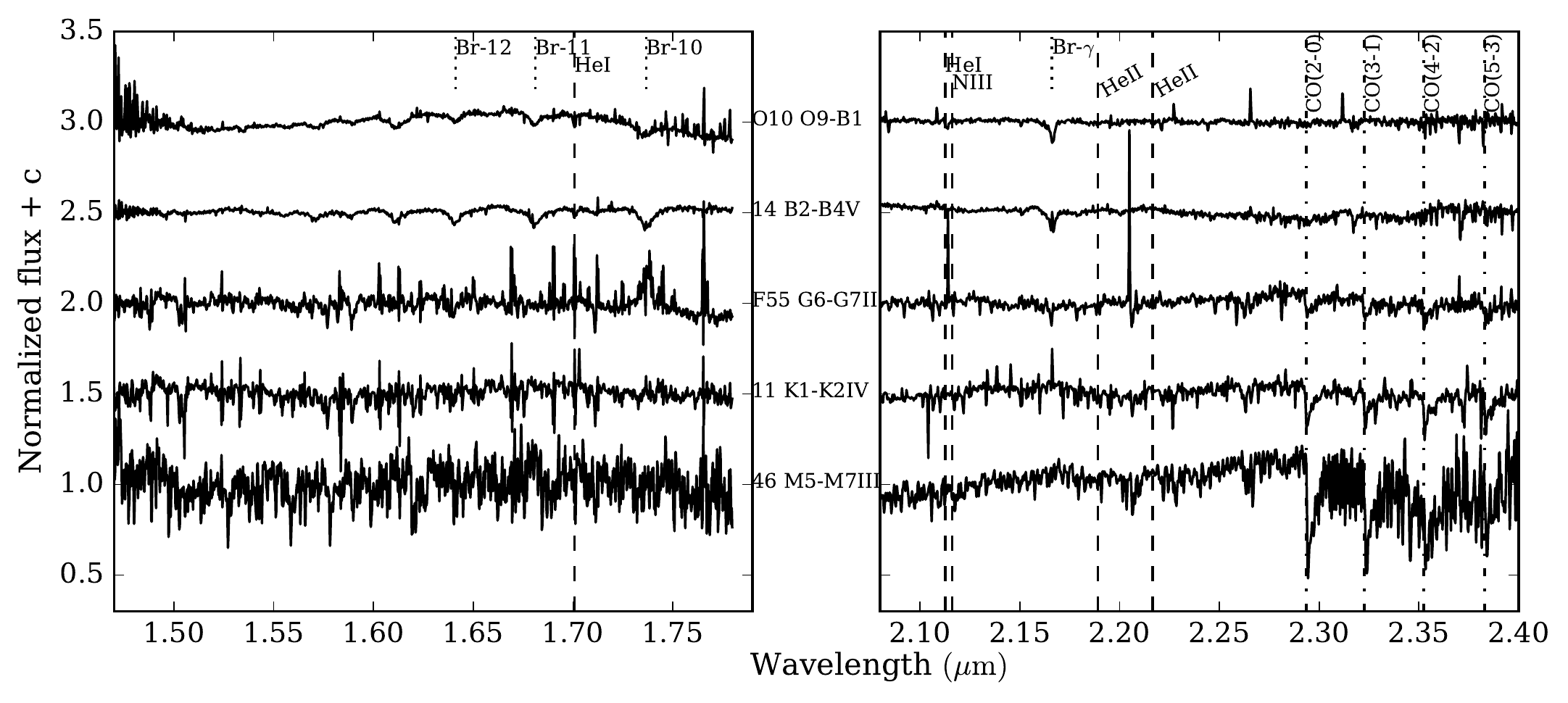}
}
\caption{Examples of the $H-$ and $K-$ band KMOS spectra across the spectral range. The figures containing the full spectra are available in the online only version of this paper.}
\label{P4:fig:M8_SpecExamples}
\end{figure*}

We present the summary of the spectral classification for all the studied sources in Table~\ref{P4:tab:SpClass}. The tabulated spectral types can be found in the online Tables~\ref{P4:tab:catalogues} and the classified spectra are presented in the online only version of this paper. Some examples of spectra are shown in Figure~\ref{P4:fig:M8_SpecExamples}.

\begin{table}
\centering
\caption{Comparison of the performance of the automatic classification (method one) and classification by eye (method two).}
\begin{minipage}{\hsize}
\centering
\renewcommand{\arraystretch}{1.4}
\setlength{\tabcolsep}{2.5pt}
\begin{tabular}{lllll}
\hline
\hline
Region & \# stars & classified & misclassified & misclassified \\
 & & by both & by method one & by method two \\
\hline                       
M8 & \Mtot & 60 & 16 (27 \% ) & 17 (28 \%) \\
\g333\ & \Gtot & 27 &  7 (26 \%) & 5 (18 \%) \\
\ngc6357\ & \NGCtot & 80 & 20 (25 \%) & 23 (29 \%) \\
total & \TotalSpec & 167 & 43 (26 \%) & 45 (27 \%)\\
\vspace{-15pt}	
\end{tabular}
\renewcommand{\footnoterule}{}
\end{minipage}
\label{P4:tab:method_comparison}
\end{table}	

\begin{table}
\caption{Spectral classification.}
\label{P4:tab:SpClass}
\centering
\renewcommand{\arraystretch}{1.4}
\setlength{\tabcolsep}{4pt}
\begin{tabular}{c c c c c c}
\hline\hline
\Hii\ region & OBA & YSOs  & Late type& Unclassified & Total\\
\hline
   M8         & \Moba\ & \Myso\ & \Mlt\ & \Muncl\ & \Mtot\ \\
   \g333\ & \Goba\ & \Gyso\ & \Glt\ & \Guncl\ & \Gtot\ \\ 
   \ngc6357\   & \NGCoba\ & \NGCyso\ & \NGClt\ & \NGCuncl\ & \NGCtot\ \\
\hline
\end{tabular}
\end{table}

\begin{table}
\centering
\caption{Properties of the KMOS targets. This table is available in its entirety in the online journal.}
\begin{minipage}{\hsize}
\centering
\renewcommand{\arraystretch}{1.9}
\setlength{\tabcolsep}{2.5pt}
\begin{tabular}{lllll}
\hline
\hline
 Col. & Label & Description \\
\hline                        % inserts single \vspace{-15pt}								
(1) & Name & Source name in KMOS catalogue \\
(2) & X-ray ID & Source X-ray ID (CXOU J) \\
(3) & kSpT & $K$-band spectral type \\
(4) & SpT$\_$source & \specialcell[t]{source of the spectral classification \\ (\textit{b} if both
methods agree, \textit{e} for the \\ classification by eye, and \textit{a} for the\\automatic classification)}\\
(5) & $\alpha$(J20000) & Right ascension (J2000, hh:mm:ss)\\
(6) & $\delta$(J2000) & Declination (J2000, dd\degr:mm\arcmin:ss\arcsec) \\
(7) & J & $J$-band magnitude (mag) \\
(7) & J$\_$err & $J$-band magnitude error (mag) \\
(8) & H & $H$-band magnitude (mag) \\
(8) & H$\_$err & $H$-band magnitude  error (mag) \\
(9) & K & $K$-band magnitude (mag) \\
(9) & K$\_$err & $K$-band magnitude error (mag) \\
(10) & [3.6] & 3.6 $\mu$m magnitude (mag) \\
(11) & [3.6]$\_$err & 3.6 $\mu$m magnitude error (mag) \\
(12) & [4.5] & 4.5 $\mu$m magnitude (mag) \\
(13) & [4.5]$\_$err & 4.5 $\mu$m magnitude error (mag) \\
(14) & [5.8] & 5.8 $\mu$m magnitude (mag) \\
(15) & [5.8]$\_$err & 5.8 $\mu$m magnitude error (mag) \\
(16) & [8.0] & 8.0 $\mu$m magnitude (mag) \\
(17) & [8.0]$\_$err & 8.0 $\mu$m magnitude error (mag) \\
\hline                        % inserts single 						
\end{tabular}
\renewcommand{\footnoterule}{}
\end{minipage}
\label{P4:tab:catalogues}
\end{table}

\subsection{Spectral classification in M8}

We observed four fields in M8 which resulted in spectra for \Mtot\ stars. From the $K$-band spectral classification we confirm that \Moba\ are OBA stars (\Mob\ OB and \Ma\ A). We have \Myso\ candidate YSO (see Section~\ref{P4:sec:mYSOs} for details about the mYSO classification) and \Mlt\ late-type stars, identified by the presence of CO band-heads in absorption. We were not able to classify \Muncl\ stars due to severe ripples in their spectra or to insufficient signal to noise. The earliest type star in our sample is the source O10, which we classified as a O9-B1 type star.
The $H$ and $K$-band spectra are displayed in Figure~\ref{P4:fig:M8_SpecExamples} of the online only version of this paper. At the left side of the plots the name of the star is shown, at the right side we list the NIR-based spectral classification.

In the top section of Table~\ref{P4:tab:M8physicalProp} we present the spectral classification from this work and compare it with the literature classification (columns four and 5); our $K$-band classification is consistent with the previous classification by \citet{2000MNRAS.319..771W} and \citet{2009yCat....1.2023S} for the overlapping sources. We also list the two candidate OB stars presented by \citet{2017ApJ...838...61P} and observed by us (second panel); we confirm that source 24 is effectively an OB star, while source 48 presents weak CO bandheads in absorption and no IR excess, and therefore it could be either a foreground low-mass PMS star or a background field giant. 
In the third section of Table~\ref{P4:tab:M8physicalProp} we list four new OB and four A stars classified by us that are not listed as massive stars in the literature. The last three columns of this table list the $K$-band extinction ($A_K$), effective temperature (${\rm T_{\rm eff}}$), and luminosity ($\log{{\rm L/L_{\odot}}})$); see Section~\ref{P4:sec:HRD}.

\begin{table*}
\centering
\caption{Spectral types and physical properties of the O, B, and A stars in the M8 KMOS sample.}           % title of Table
\begin{minipage}{\hsize}
\centering
\renewcommand{\arraystretch}{1.1}
\setlength{\tabcolsep}{3pt}
\begin{tabular}{cccccccc}
\hline
\hline
ID & X-ray ID (CXOU J) & Other ID & previous ST & ST this work & $A_K$ (mag) & ${\rm T_{eff}}$ (K) & ${\rm \log{L/L_{\odot}}}$ \\
\hline
\multicolumn{8}{c}{Previously classified OB stars in M8 included in our KMOS sample} \\ 
\hline
14  &  180345.10-242205.2  &  CD-24 13810  & B...\footnote{\label{P4:foot:notMSIMBAD}\citet{2000A&AS..143....9W}} &  B2-B4V & $0.23\pm0.01$ & $18650\pm1950$ & $3.54\pm0.14$ \\ 
33  &  180411.16-242145.1  &  NGC 6530 32  & B3Ve\footnote{\label{P4:foot:notMskiff}\citet{2009yCat....1.2023S}} &  B1-B3 & $0.16\pm0.03$ & $21500\pm4500$ & $3.35\pm0.3$ \\ 
36  &  180415.03-242327.8  &  HD 315032  & B1V\footref{P4:foot:notMskiff} &  B1-B2Vz & $0.16\pm0.02$ & $23300\pm2700$ & $3.84\pm0.17$ \\ 
O5  &  180423.99-242126.7  &  NGC 6530 60  & B1Ve\footref{P4:foot:notMskiff} &  O9-B1-SB2 & $0.17\pm0.01$ & $29250\pm3250$ & $4.03\pm0.13$ \\ 
O6  &  180424.29-242059.5  &  NGC 6530 61  & B2Ve\footref{P4:foot:notMskiff} &  B1-B2V & $0.1\pm0.02$ & $23300\pm2700$ & $3.47\pm0.17$ \\ 
O7  &  180425.55-242044.9a  &  NGC 6530 66  & B2Vep\footref{P4:foot:notMskiff} &  B0-B2V & $0.13\pm0.03$ & $26050\pm5450$ & $3.63\pm0.28$ \\ 
O8  &  180427.19-242249.6  &  NGC 6530 70  & B2Ve\footref{P4:foot:notMskiff} &  B1-B2V & $0.19\pm0.02$ & $23300\pm2700$ & $3.5\pm0.17$ \\ 
60  &  180428.98-242526.2  &  NGC 6530 74  & B2.5Ve\footref{P4:foot:notMskiff} &  B2-B5 & $0.1\pm0.02$ & $18150\pm2450$ & $3.0\pm0.19$ \\ 
O10  &  180432.98-242312.5a  &  NGC 6530 80  & B1V\footref{P4:foot:notMskiff} &  O9-B1 & $0.17\pm0.01$ & $29250\pm3250$ & $4.12\pm0.13$ \\ 
73  &  180434.20-242200.6  &  NGC 6530 86  & B2.5V\footref{P4:foot:notMskiff} &  B0-B3V & $0.3\pm0.04$ & $24250\pm7250$ & $3.78\pm0.4$ \\ 
O11  &  180439.92-242244.4  &  NGC 6530 99  & B2.5Vne\footref{P4:foot:notMskiff} &  B2-B5V & $0.15\pm0.02$ & $18150\pm2450$ & $2.97\pm0.19$ \\ 
79  &  180441.68-242058.5  &  HD 164947A  & B1/2Vne\footref{P4:foot:notMskiff} &  B2-B5V & $0.11\pm0.02$ & $18150\pm2450$ & $3.38\pm0.19$ \\ 
\hline
\multicolumn{8}{c}{Candidate OB stars in M8 included in our KMOS sample} \\ 
\hline
24
% \footnote{Non-cluster member based on its astrometric properties.}  
&  180400.80-242334.6  & $-$ & $-$ & B0-B2V & $0.84\pm0.03$ & $26050\pm5450$ & $4.03\pm0.28$ \\ 
48  &  180423.55-242116.8  & $-$ & $-$ & G8-K0IV-V & $0.21\pm0.01$ & $5385\pm105$ & $1.73\pm0.03$ \\ 
\hline
\multicolumn{8}{c}{New OB and A stars in M8} \\ 
\hline
16  &  180346.16-242149.4  & $-$ & $-$ &  B2-B5V & $0.19\pm0.02$ & $18150\pm2450$ & $2.84\pm0.19$ \\ 
% 24  &  180400.80-242334.6  & $-$ & $-$ &  B0-B2V & $0.84\pm0.03$ & $26050\pm5450$ & $4.03\pm0.28$ \\ 
51  &  180424.29-242059.6  & $-$ & $-$ &  B2-B3V & $0.07\pm0.01$ & $18800\pm1800$ & $3.18\pm0.12$ \\ 
55  &  180425.72-241949.6  & $-$ & $-$ &  A5-F2III-IV & $0.17\pm0.01$ & $9752\pm2948$ & $2.29\pm0.21$ \\ 
69  &  180433.03-242252.9  & $-$ & $-$ &  A0-A5V & $0.01\pm0.01$ & $8890\pm810$ & $1.74\pm0.12$ \\ 
F19  &  180404.31-242130.4  & $-$ & $-$ &  A5-F4V & $0.02\pm0.01$ & $7940\pm140$ & $1.42\pm0.03$ \\ 
F36  &  180420.48-242303.9  & $-$ & $-$ &  A8-F2III & $1.38\pm0.01$ & $8055\pm1251$ & $2.4\pm0.12$ \\ 
F44  &  180422.10-242546.6  & $-$ & $-$ &  B0-B3 & $0.63\pm0.04$ & $24250\pm7250$ & $3.19\pm0.4$ \\ 
F47  &  180426.73-242351.7  & $-$ & $-$ &  A5-A9V & $0.2\pm0.01$ & $7760\pm320$ & $1.53\pm0.05$ \\ 
\hline
\multicolumn{8}{c}{Previously classified OB stars in M8 not included in our KMOS sample} \\ 
\hline
$-$ & 180340.32-242242.8  &  HD 164740  & O7.5V+O9V+B0.5V\footref{P4:foot:notMskiff} & $-$ & $-$ & $36000$ & $-$ \\ 
$-$ & 180352.45-242138.5  &  HD 164794  & O4V((f))\footref{P4:foot:notMskiff} & $-$ & $-$ & $42000$ & $5.82$ \\ 
$-$ & 180356.85-241845.0  &  HD 164816  & O9.5V+B0V\footref{P4:foot:notMskiff} & $-$ & $-$ & $31500$ & $4.94$ \\ 
$-$ & 180414.50-241437.0  &  HD 315026  & B1.5V\footref{P4:foot:notMskiff} & $-$ & $-$ & $24000$ & $-$ \\ 
$-$ & 180420.55-241354.7  &  NGC 6530 55  & B2.5Ve\footref{P4:foot:notMskiff} & $-$ & $-$ & $19000$ & $3.4$ \\ 
$-$ & 180421.27-242118.2  &  NGC 6530 56  & B0.5V\footref{P4:foot:notMskiff} & $-$ & $-$ & $28000$ & $4.25$ \\ 
$-$ & 180422.74-242209.8  &  NGC 6530 58  & B2Ve\footref{P4:foot:notMskiff} & $-$ & $-$ & $21000$ & $-$ \\ 
$-$ & 180423.22-242616.7  &  HD 315033  & B2Vp\footref{P4:foot:notMskiff} & $-$ & $-$ & $21000$ & $3.99$ \\ 
$-$ & 180425.83-242308.4  &  HD 164906  & B0Ve\footref{P4:foot:notMskiff} & $-$ & $-$ & $29500$ & $-$ \\ 
$-$ & 180428.02-242142.9  &  HD 315031  & B1V\footref{P4:foot:notMskiff} & $-$ & $-$ & $26000$ & $4.42$ \\ 
$-$ & 180432.91-241844.6  &  NGC 6530 83  & B2.5Vne\footref{P4:foot:notMskiff} & $-$ & $-$ & $19000$ & $3.41$ \\ 
$-$ & 180435.99-241952.2  &  HD 315021  & B1V\footref{P4:foot:notMskiff} & $-$ & $-$ & $26000$ & $-$ \\ 
$-$ & 180441.65-242055.1  &  HD 164947B  & B2.5V\footref{P4:foot:notMskiff} & $-$ & $-$ & $19000$ & $3.42$ \\ 
$-$ & 180448.62-242635.5  &   NGC 6530 SJ 75  & B2.5\footref{P4:foot:notMSIMBAD} & $-$ & $-$ & $19000$ & $3.5$ \\ 
$-$ & 180456.26-242403.4  &  HD 315034  & B3\footref{P4:foot:notMSIMBAD} & $-$ & $-$ & $17500$ & $2.8$ \\ 
$-$ & 180459.03-242723.1  &  HD 315035  & B3V\footref{P4:foot:notMskiff} & $-$ & $-$ & $17500$ & $-$ \\ 

\hline
\vspace{-15pt}																	
\end{tabular}
\renewcommand{\footnoterule}{}
\end{minipage}
\label{P4:tab:M8physicalProp}
\end{table*}	

\subsection{Spectral classification in \g333 }

We observed two fields in \g333\ and obtained spectra of \Gtot\ stars. From the $K$-band spectral classification we confirm that \Goba\ are OBA stars (\Gob\ OB and \Ga\ A), \Gyso\ candidate YSO and \Glt\ late-type stars, identified by the presence of CO band-heads in absorption. We were not able to classify \Guncl\ spectra. The most massive star in our sample is 6829462 which is an O6 star. The candidate YSO (source 720) does not present CO band-head emission, Br$\gamma$ seems to be in emission with a central nebular component, and the other lines of the Brackett series show a photospheric absorption line with a central emission component. The $H$ and $K$-band spectra are available in Figure~\ref{P4:fig:M8_SpecExamples} of the online only version of this paper. We present four new OB stars and six A stars in our observed sample. In Table~\ref{P4:tab:G333physicalProp} we list these stars together with their derived $K$-band extinction, effective temperature, and luminosity (see Section~\ref{P4:sec:HRD}).

\begin{table*}
\centering
\caption{Spectral types and physical properties of the newly classified O, B, and A stars in the \g333\ KMOS sample.}
\begin{minipage}{\hsize}
\centering
\renewcommand{\arraystretch}{1.1}
\setlength{\tabcolsep}{5pt}
\begin{tabular}{cccccc}
\hline
\hline
ID & X-ray ID (CXOU J) & ST & $A_K$ (mag) & ${\rm T_{eff}}$ (K) & ${\rm \log{L/L_{\odot}}}$ \\
\hline
6829201  & 162231.65-500815.6 & O7-O9 & $1.37\pm0.01$ & $34500\pm2000$ & $4.85\pm0.04$ \\ 
6829462  & 162222.28-500736.7 & O6-O7 & $1.72\pm0.01$ & $37750\pm1250$ & $5.33\pm0.03$ \\ 
7034854  & 162221.87-500335.6 & A8-F2V & $0.75\pm0.01$ & $7155\pm345$ & $2.88\pm0.06$ \\ 
7117881  & 162159.55-500743.5 & O8-B0.5Vz & $1.25\pm0.01$ & $31750\pm2750$ & $4.71\pm0.08$ \\ 
7348675  & 162205.99-500421.5 & A8-F2V & $1.04\pm0.01$ & $7155\pm345$ & $2.64\pm0.06$ \\ 
378  & - & O9-B1 & $-$ & $29250\pm3250$ & $-$ \\ 
767  & - & A8-F2V & $-$ & $7155\pm345$ & $-$ \\ 
789  & 162218.25-500514.8 & A9-F3III & $1.92\pm0.02$ & $7398\pm777$ & $3.44\pm0.11$ \\ 

\hline
\vspace{-15pt}																	
\end{tabular}
\renewcommand{\footnoterule}{}
\end{minipage}
\label{P4:tab:G333physicalProp}
\end{table*}	

\subsubsection{X-ray counterparts for sources in \g333}
\label{P4:sec:xraysG333}

Recently, \citet{2018ApJS..235...43T} presented the second instalment of the Massive Star-forming Regions Omnibus X-ray Catalogue (MOXC2) where they list X-ray point sources in several star forming regions including \g333. By cross-matching their catalogue with the coordinates of our sources we find that 22 out of our initial 39 sources have X-ray counterparts. Those sources are included with Chandra IDs in the second column of Table~\ref{P4:tab:catalogues}. 
Source 378 is a late-O, early-B type star, but it does not have an X-ray counterpart. We assume that this massive star is part of \g333.
Given that we are not certain of the cluster membership of the other (low-mass) stars with no X-ray counterparts, in the remainder of this paper we treat those sources, except 378, as non members of \g333\ and do not take them into account when estimating the age of the region (Section~\ref{P4:sec:HRD}). 

\subsection{Spectral classification in \ngc6357 }

We observed six fields in \ngc6357\ resulting in \NGCtot\ spectra. From the IR spectral classification we confirm that \NGCoba\ are OBA stars (\NGCob\ OB and \NGCa\ A), we have \NGCyso\ candidate YSOs and \NGClt\ late-type stars, identified by the presence of CO band-heads in absorption. We were not able to classify \NGCuncl\ stars from the obtained spectra. The $H$ and $K$-band spectra are available in the online only version of this paper (Figure~\ref{P4:fig:M8_SpecExamples}), in the left side of the plots, the name of the stars is shown, and in the right part we show the NIR-based spectral classification.

In Table~\ref{P4:tab:NGC6357physicalProp} we show the previously classified OB stars in \ngc6357\ that overlap with our sample. The spectral classification obtained for these sources agrees with the previous studies. Out of the 26 new candidate OB stars presented by \citet{2017ApJ...838...61P} we observed 17 with KMOS. We confirm that most of these stars are OB stars with the exception of B3 which presents CO bandheads in absorption and is therefore classified as a K4-K5V type star, and B14 where the spectrum is more consistent with it being an F star. We were not able to classify B12 due to the very strong ripples affecting its spectrum nor B16 because the signal-to-noise is too low to recognise any spectral features. Finally, five newly classified OB stars and nine A stars are also listed in Table~\ref{P4:tab:NGC6357physicalProp}. As in Tables~\ref{P4:tab:M8physicalProp} and~\ref{P4:tab:G333physicalProp}, the last three columns show the physical properties derived in Section~\ref{P4:sec:HRD}.

\begin{table*}
\centering
\caption{Spectral types and physical properties of the O, B, and A stars in the \ngc6357\ KMOS sample.}           % title of Table
\begin{minipage}{\hsize}
\centering
\renewcommand{\arraystretch}{1.19}
\setlength{\tabcolsep}{2pt}
\begin{tabular}{cccccccc}
\hline
\hline
ID & X-ray ID (CXOU J) & Other ID & previous ST & ST this work & $A_K$ (mag) & ${\rm T_{eff}}$ (K) & ${\rm \log{L/L_{\odot}}}$ \\
\hline
\multicolumn{8}{c}{Previously classified OB stars in NGC6357 included in our KMOS sample} \\ 
\hline
16  &  172436.03-341400.4  &  Pis 24 10  & O9V\footnote{\label{P4:foot:notNskiff}\citet{2009yCat....1.2023S}} &  O9-B0 & $0.62\pm0.01$ & $32000\pm500$ & $4.73\pm0.02$ \\ 
O0  &  172442.26-341141.1  &  Pis 24 12  & B1V\footref{P4:foot:notNskiff} &  B1-B2 & $0.46\pm0.02$ & $23300\pm2700$ & $3.87\pm0.17$ \\ 
O1  &  172443.29-341142.1  &  Pis 24 18  & B0.5V\footref{P4:foot:notNskiff} &  O9-B1 & $0.74\pm0.01$ & $29250\pm3250$ & $4.59\pm0.13$ \\ 
O2  &  172443.68-341140.7  &  Pis 24 19  & B1V\footref{P4:foot:notNskiff} &  B1-B2 & $0.93\pm0.02$ & $23300\pm2700$ & $4.12\pm0.17$ \\ 
\hline
\multicolumn{8}{c}{Candidate OB stars in NGC6357 included in our KMOS sample} \\ 
\hline
B0  &  172434.78-341318.0  & $-$ & $-$ & O9.5-B1V & $0.66\pm0.01$ & $29000\pm3000$ & $4.22\pm0.11$ \\ 
B1  &  172440.49-341206.3  & $-$ & $-$ & O9.5-B2V & $0.66\pm0.03$ & $26300\pm5700$ & $4.31\pm0.29$ \\ 
B2  &  172442.87-340911.6  & $-$ & $-$ & O9.5-B1V & $1.03\pm0.01$ & $29000\pm3000$ & $4.59\pm0.11$ \\ 
B3  &  172447.56-341048.6  & $-$ & $-$ & K1-K5V & $1.07\pm0.03$ & $4790\pm380$ & $1.93\pm0.1$ \\ 
B4  &  172455.08-341111.5  & $-$ & $-$ & O9.5-B1V & $0.88\pm0.01$ & $29000\pm3000$ & $4.67\pm0.11$ \\ 
B6  &  172528.51-342652.5  & $-$ & $-$ & O9.5-B1V & $0.55\pm0.01$ & $29000\pm3000$ & $4.23\pm0.11$ \\ 
B7  &  172547.33-342154.1  & $-$ & $-$ & B2-B5V & $0.55\pm0.02$ & $18150\pm2450$ & $3.52\pm0.19$ \\ 
B8  &  172554.16-342156.4  & $-$ & $-$ & B0-B3V & $0.68\pm0.04$ & $24250\pm7250$ & $3.88\pm0.4$ \\ 
B9  &  172557.29-341850.8  & $-$ & $-$ & O9-B3V & $0.85\pm0.05$ & $24750\pm7750$ & $4.3\pm0.42$ \\ 
B10  &  172558.08-341606.1  & $-$ & $-$ & O7-O9V & $0.77\pm0.01$ & $34500\pm2000$ & $4.4\pm0.04$ \\ 
B11  &  172558.15-342119.2  & $-$ & $-$ & O9-B1 & $0.92\pm0.01$ & $29250\pm3250$ & $4.61\pm0.13$ \\ 
B12  &  172602.07-341757.6  & $-$ & $-$ & $-$ & $-$ & $-$ & $-$ \\ 
B13  &  172604.04-341520.0  & $-$ & $-$ & O6-O8V & $1.0\pm0.01$ & $36750\pm2250$ & $4.99\pm0.06$ \\ 
B16  &  172604.25-341557.0  & $-$ & $-$ & $-$ & $-$ & $-$ & $-$ \\ 
B14  &  172604.71-341526.3  & $-$ & $-$ & F2-F6V & $1.43\pm0.01$ & $6575\pm235$ & $2.66\pm0.05$ \\ 
B15  &  172604.72-341611.2  & $-$ & $-$ & O9.5-B1V & $1.03\pm0.01$ & $29000\pm3000$ & $4.22\pm0.11$ \\ 
118  &  172601.88-341631.1  & $-$ & $-$ & B0-B3 & $0.75\pm0.04$ & $24250\pm7250$ & $4.26\pm0.4$ \\ 
\hline
\multicolumn{8}{c}{New OB and A stars in NGC6357} \\ 
\hline
14  &  172434.67-341425.2  & $-$ & $-$ &  A5-F3IV-V & $0.25\pm0.01$ & $7400\pm680$ & $1.65\pm0.12$ \\ 
22  &  172439.41-341526.8  & $-$ & $-$ &  B1-B2 & $0.61\pm0.02$ & $23300\pm2700$ & $3.84\pm0.17$ \\ 
34  &  172444.62-341100.6  & $-$ & $-$ &  A9-F3V & $0.53\pm0.01$ & $7080\pm360$ & $1.98\pm0.06$ \\ 
45  &  172451.14-341201.0  & $-$ & $-$ &  A9-F3V & $0.51\pm0.01$ & $7080\pm360$ & $1.8\pm0.06$ \\ 
65  &  172522.78-342404.6  & $-$ & $-$ &  A5-F5III-IV & $0.39\pm0.01$ & $9477\pm3222$ & $2.34\pm0.29$ \\ 
70  &  172530.30-342455.1  & $-$ & $-$ &  A8-F2III & $0.44\pm0.01$ & $8055\pm1251$ & $2.76\pm0.12$ \\ 
71  &  172532.29-342346.5  & $-$ & $-$ &  A0-F2III-IV & $0.57\pm0.05$ & $12579\pm5775$ & $2.68\pm0.37$ \\ 
73  &  172533.82-342708.5  & $-$ & $-$ &  B0-B2V & $1.02\pm0.03$ & $26050\pm5450$ & $4.1\pm0.28$ \\ 
88  &  172545.29-342535.2  & $-$ & $-$ &  O9.5-B1V & $0.59\pm0.01$ & $29000\pm3000$ & $4.34\pm0.11$ \\ 
107  &  172556.17-341510.4  & $-$ & $-$ &  B4-B6V & $0.4\pm0.01$ & $15600\pm1100$ & $2.92\pm0.1$ \\ 
108  &  172556.81-341727.0  & $-$ & $-$ &  A8-F2V & $0.73\pm0.01$ & $7155\pm345$ & $2.11\pm0.06$ \\ 
112  &  172559.18-341919.4  & $-$ & $-$ &  B0-B2V & $0.76\pm0.03$ & $26050\pm5450$ & $4.03\pm0.28$ \\ 
\hline
\multicolumn{8}{c}{Previously classified OB stars in NGC6357 not included in our KMOS sample} \\ 
\hline
$-$ & 172428.95-341450.6  &  Pis 24 15  & O7.5Vz\footnote{\label{P4:foot:notNJMA16}\citet{2016ApJS..224....4M}} & $-$ & $-$ & $35000$ & $4.93$ \\ 
$-$ & 172442.30-341321.4  &  Pis 24 3  & O8V\footref{P4:foot:notNskiff} & $-$ & $-$ & $35000$ & $4.96$ \\ 
$-$ & 172443.28-341243.9  &  Pis 24 2  & O5.5V((f))\footref{P4:foot:notNskiff} & $-$ & $-$ & $40000$ & $5.37$ \\ 
$-$ & 172443.49-341156.9  &  Pis 24 1 SW+NEab  & O4III(f+)+O3.5If*\footnote{\label{P4:foot:notNJMA07}\citet{2007ApJ...660.1480M}} & $-$ & $-$ & $42000$ & $6.22$ \\ 
$-$ & 172444.48-341158.8  &  Pis 24 16  & O7.5V\footref{P4:foot:notNskiff} & $-$ & $-$ & $36000$ & $-$ \\ 
$-$ & 172444.72-341202.6  &  Pis 24 17  & O3.5III(f*)\footref{P4:foot:notNskiff} & $-$ & $-$ & $44500$ & $5.92$ \\ 
$-$ & 172445.77-340939.8  &  Pis 24 13  & O6.5V((f))\footref{P4:foot:notNskiff} & $-$ & $-$ & $38000$ & $5.25$ \\ 
$-$ & 172508.85-341112.4  &  HD 157504  & WC6+O7/9\footref{P4:foot:notNskiff} & $-$ & $-$ & $37000$ & $-$ \\ 
$-$ & 172529.16-342515.6  &  [N78] 51  & O6Vn((f))\footref{P4:foot:notNJMA16} & $-$ & $-$ & $41000$ & $5.35$ \\ 
$-$ & 172534.23-342311.7  &  [N78] 49  & O5.5IV(f)\footref{P4:foot:notNJMA16} & $-$ & $-$ & $39000$ & $5.87$ \\ 
$-$ & 172538.60-340843.5  &  LS 4151  & O6/7III\footref{P4:foot:notNskiff} & $-$ & $-$ & $38000$ & $5.41$ \\ 
$-$ & 172626.45-341700.9  &  HD 319788  & B2III:\footref{P4:foot:notNskiff} & $-$ & $-$ & $21000$ & $4.22$ \\ 

\hline
\vspace{-15pt}																	
\end{tabular}
\renewcommand{\footnoterule}{}
\end{minipage}
\label{P4:tab:NGC6357physicalProp}
\end{table*}	

%__________________________________________________________________

\subsection{Relation between photometry, astrometry, and spectral classification}
\label{P4:sec:PhotVsSpt}

Here we discuss some individual objects whose position in the CMDs and CCDs (Figures~\ref{P4:fig:CMD_CCD_M8}, \ref{P4:fig:CMD_CCD_G333}, and~\ref{P4:fig:CMD_CCD_NGC6357}) or their astrometry from Gaia DR2 data (Section~\ref{P4:sec:GaiaDR2}) suggest that they have peculiar properties.

Source 46 in M8 seems to be too bright for its spectral type, one possibility is that the X-ray an IR sources were incorrectly matched. If the sources are correctly matched this source could be a cool giant foreground star where the X-ray emission is originated by binary interaction. Sources 24 and 48 seem to be highly reddened in comparison to the rest of the population in the NIR CMD, but a strong excess is not apparent in the NIR nor MIR CMDs. Source 24 is also amongst the candidate non-members according to its Gaia DR2 astrometric properties, but the quality of the astrometry is insufficient to firmly exclude membership of M8. 
From the photometric properties of sources F13 and F24 it does not appear obvious that these stars are not members of M8, but we discard them as cluster members based on our results from the Gaia DR2 data. 
We consider these four sources not to belong to M8 and exclude them from the extinction calculation and age determination.

In the NIR CCD of \ngc6357\ source 33 is below the reddening line for $R_V=3.1$, and shows evidence for the presence of a circumstellar disk which indicates that it might be a massive YSO with a NIR excess (see Section~\ref{P4:sec:mYSOs}). 
Sources 91 and 8 seem to be too bright for their spectral types, which is consistent with the astrometric solution asserting these stars to be non cluster members. We exclude these stars from the extinction and age determination. 

In \g333\, the stars classified as M sources correspond to some of those which do not have X-ray counterparts and were selected only on the basis of their NIR colours. From the NIR diagrams we conclude that these stars are cool giants and are not part of the cluster. This shows the importance of the \mys\ project in order to identify cluster members. Stars 515756828027 and 515757115402 have X-ray counterparts but their position in the CMD and CCD seems inconsistent with their spectral type, this in agreement with the astrometric solution showing that they are not cluster members.

Our spectral classification agrees with the SED classification by \citet{2013ApJS..209...31P} in the MIR CCDs: the objects classified as OB type stars are located where the authors find the visible photospheres, and the sources that we identify as pre-main-sequence stars are likely YSOs. Objects 22, 88, and 112 in \ngc6357\ are an exception because they are in the region dominated by YSOs, but do not show any gaseous disk signatures in the NIR diagram, nor in the $H$- and $K$-band spectra. This could be explained by a relatively large inner hole.

From the photometry, the spectral classification, and the X-ray counterparts (see also Section~\ref{P4:sec:phot}), it becomes clear that there are some stars that are not cluster members. These have been marked with grey symbols in the HRDs and have not been included in the extinction histograms nor taken into account for the age determination. Unfortunately, the quality of the astrometric fit for these stars is poor, and therefore we are not able to evaluate their membership with Gaia DR2 data.  
We do not discard the stars classified as giants because the fact that they have low surface gravities could mean that they are PMS swollen stars contracting towards the Zero age main-sequence (ZAMS).

%__________________________________________________________________

\section{Age of the giant \Hii\ regions}
\label{P4:sec:HRD}

After the spectroscopic classification, we can place our KMOS sources in the Hertzsprung-Russell diagram (HRD) and match them with MESA isochrones and evolutionary tracks obtained from the MIST project \citep{2011ApJS..192....3P, 2013ApJS..208....4P, 2015ApJS..220...15P, 2016ApJS..222....8D, 2016ApJ...823..102C}. The isochrones and tracks used have solar metallicity and initial rotational velocity $v_{\rm ini}=0.4 v_{\rm crit}$, where $v_{\rm crit}$ is the critical velocity, reached when the gravitational acceleration is exactly compensated by the centrifugal force.

The effective temperature, intrinsic colours $(V-Ks)_0$, $(J-H)_0$, and $(H-K)_0$, and bolometric correction of the luminosity class V stars are taken from \citet{2013ApJS..208....9P}; for the luminosity class III stars we used the calibrations from \citet{1999A&AS..140..261A} for spectral types F0-F9, and for G4-M5 stars we used the calibration of \citet{2000asqu.book.....C}. With the effective temperature and bolometric correction in the $V$-band, $BC_V$, we derive the absolute magnitude in the $V$-band. In order to calculate the luminosity of the observed targets we used the absolute $K$-band magnitude assuming the \citet{2005ApJ...619..931I} extinction law to derive $A_K$,

\begin{equation}
    A_K = \frac{(H-K)-(H-K)_0}{0.55} \, ,
\end{equation}

\noindent and calculated the absolute $K$-band magnitude by scaling it to the region's distance and correcting it by the extinction $A_K$. We obtained the bolometric correction in the $K$-band, $BC_K$, from $BC_V$ and the absolute $V$ and $K$-band magnitudes ($BC_K = BC_V + M_V - M_K$).

The $K$-band extinction, temperature, and luminosity derived for the OB and A stars in M8, \g333, and \ngc6357\ are listed in the last three columns of Tables~\ref{P4:tab:M8physicalProp}, \ref{P4:tab:G333physicalProp}, and \ref{P4:tab:NGC6357physicalProp}.

\subsection{Extinction $A_K$}

\begin{figure*}
\setlength{\tabcolsep}{-1pt}
     \centering
        \subfigure[M8]{%
           \includegraphics[width=0.33\hsize]{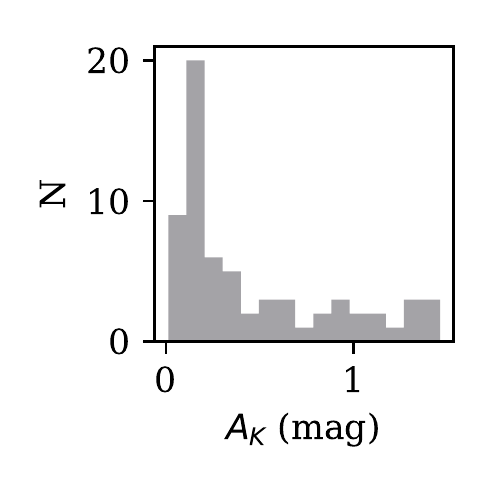}
        }%
          \subfigure[\g333]{%
            \includegraphics[width=0.33\hsize]{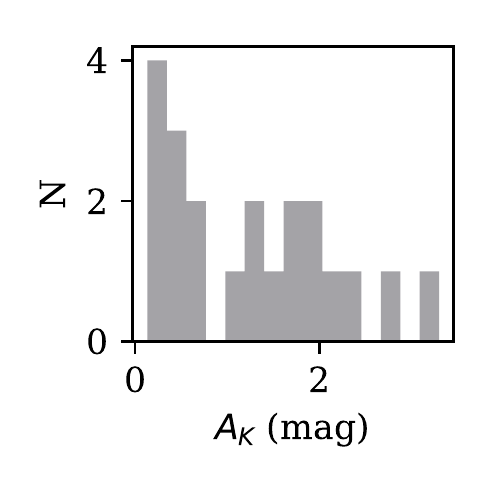}
        }%
        \subfigure[\ngc6357]{%
           \includegraphics[width=0.33\hsize]{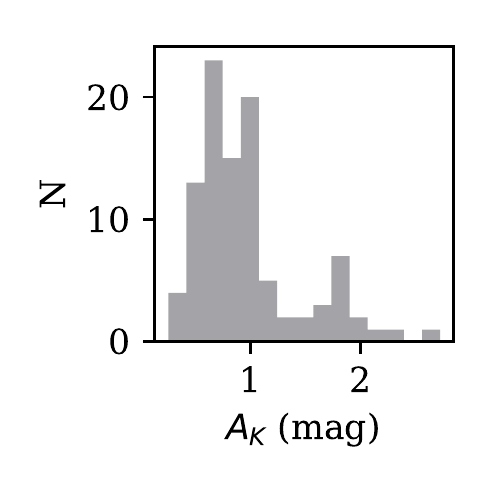}
        }\\%
    \caption[]{Distribution of the extinction $A_K$ for the three regions studied using the \citet{2005ApJ...619..931I} extinction law.
    }
\label{P4:fig:Ak_histograms}
\end{figure*}

To asses the sensitivity of our results to the choice of the extinction law, we also used the \citet{2009ApJ...696.1407N} extinction law ($A_H/A_K = 1.62$) to derive the luminosity as described in the previous section. With this extinction law the mean $A_K$ values are slightly lower than the values derived with the \citet{2005ApJ...619..931I} extinction law; the peak of the $A_K$ distribution shifts at most 0.1~mag (Figure~\ref{P4:fig:Ak_histograms}). Given this small difference, adopting the \citet{2005ApJ...619..931I} extinction law seems secure.

The extinction in M8 peaks at around $A_K = 0.1$~mag and has a tail extending to values of $A_K=1.5$~mag. The object with the highest extinction is F5; its position in the CMD is consistent with having such a high extinction. With an extinction $A_K = 2.5$~mag the star F9 may be a non-member, thus we have excluded it from the histogram and the age classification. Although the trend is not very strong, the extinction seems to be higher in the western part of this region, close to the location of the massive multiple system Herschel~36.

In the case of \g333\ the $A_K$ values have an almost flat distribution ranging from 0.2 to 3.5; this is consistent with what is observed in the CMD and CCD. Most of the OB stars have an extinction ranging from 1 to 2~mag. The objects with the highest extinction are found at the centre of the cluster while the lowest extinction is measured for the four objects to the north-east of the cluster centre. 

The mean value of $A_K$ for \ngc6357\ is higher than for M8, which is also evident from the CMDs. In this region we can distinguish two peaks, one around $A_K=0.7$ and a smaller one at about $A_K=1.7$. The highest extinction corresponds to F67 which is also one of the reddest objects in the CMD and CCDs. The stars in the north-west region in \ngc6357\ are on average less extincted in the $K$-band than those in the south-east region, but other than that there is no obvious trend of extinction with location.

\subsection{Age determination}
\label{P4:sec:age}

\begin{figure*}
\centering
        {\includegraphics[width=0.95\hsize]{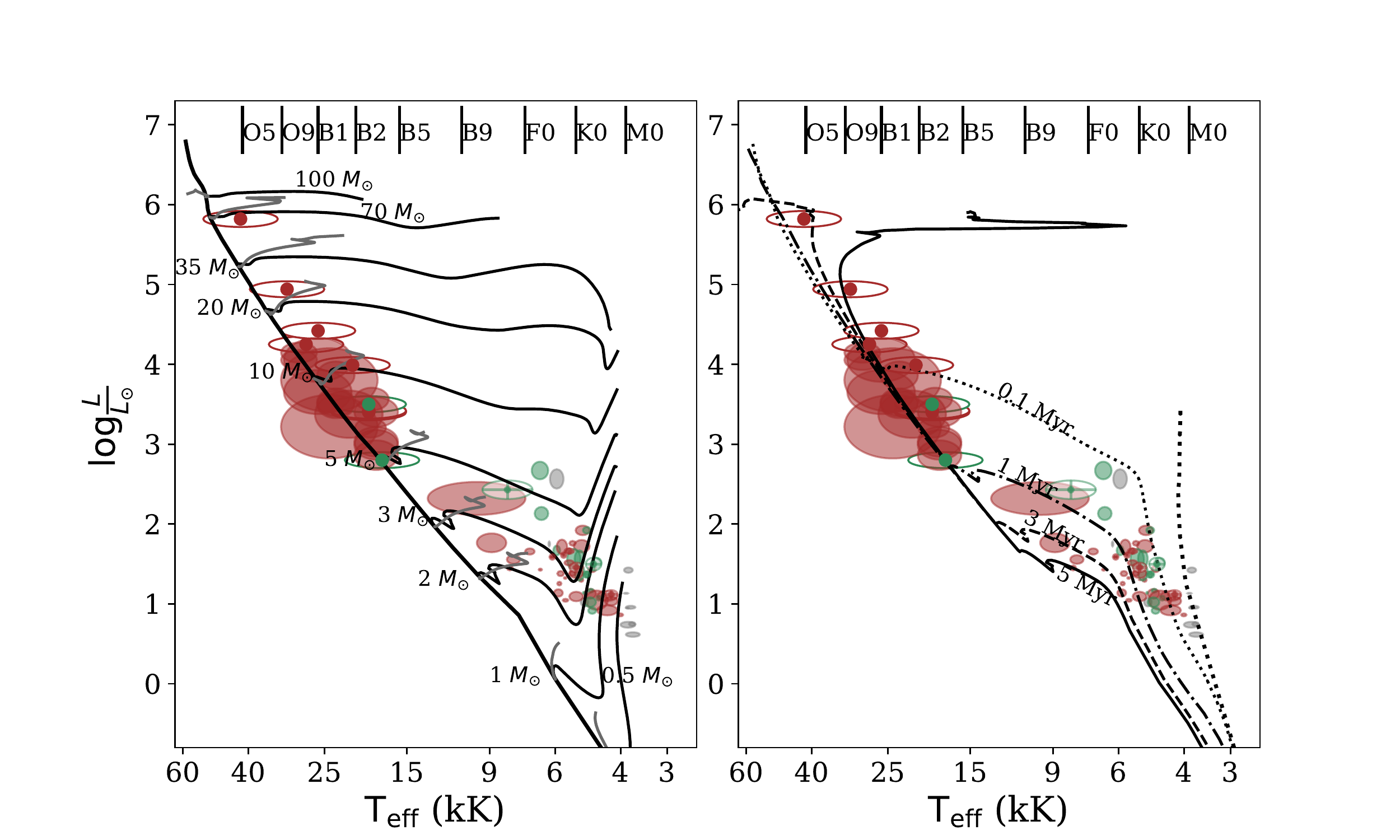}
        }
  \caption[]{Hertzsprung-Russell diagram (HRD) for the stars in M8. In the left panel we show the pre-main-sequence (black) and post-main-sequence (grey) tracks for stars of masses of 1, 2, 3, 5, 10, 20, 35, 70, and 100~\msun, respectively. In the right panel we plot the 0.1, 1, 3, and 5~Myr isochrones. The ellipses show the position of the observed stars. The dots show the position of previously classified OB stars reported in with 0.2~dex error ellipses representing the scatter about the ZAMS \citep{2017ApJ...838...61P}.
  The luminosity class IV and V stars are shown in red, the stars with luminosity class III are shown in green. In grey we show the late-K and M-type stars. The empty ellipses with crosses show the objects located below the reddening line in the NIR CCD, which have been excluded from the age determination. The age of M8 is estimated to be between \minagem\ and \maxagem~Myr.

}
     \label{P4:fig:M8_HRD}
\end{figure*}

\begin{figure*}
\centering
        {\includegraphics[width=0.95\hsize]{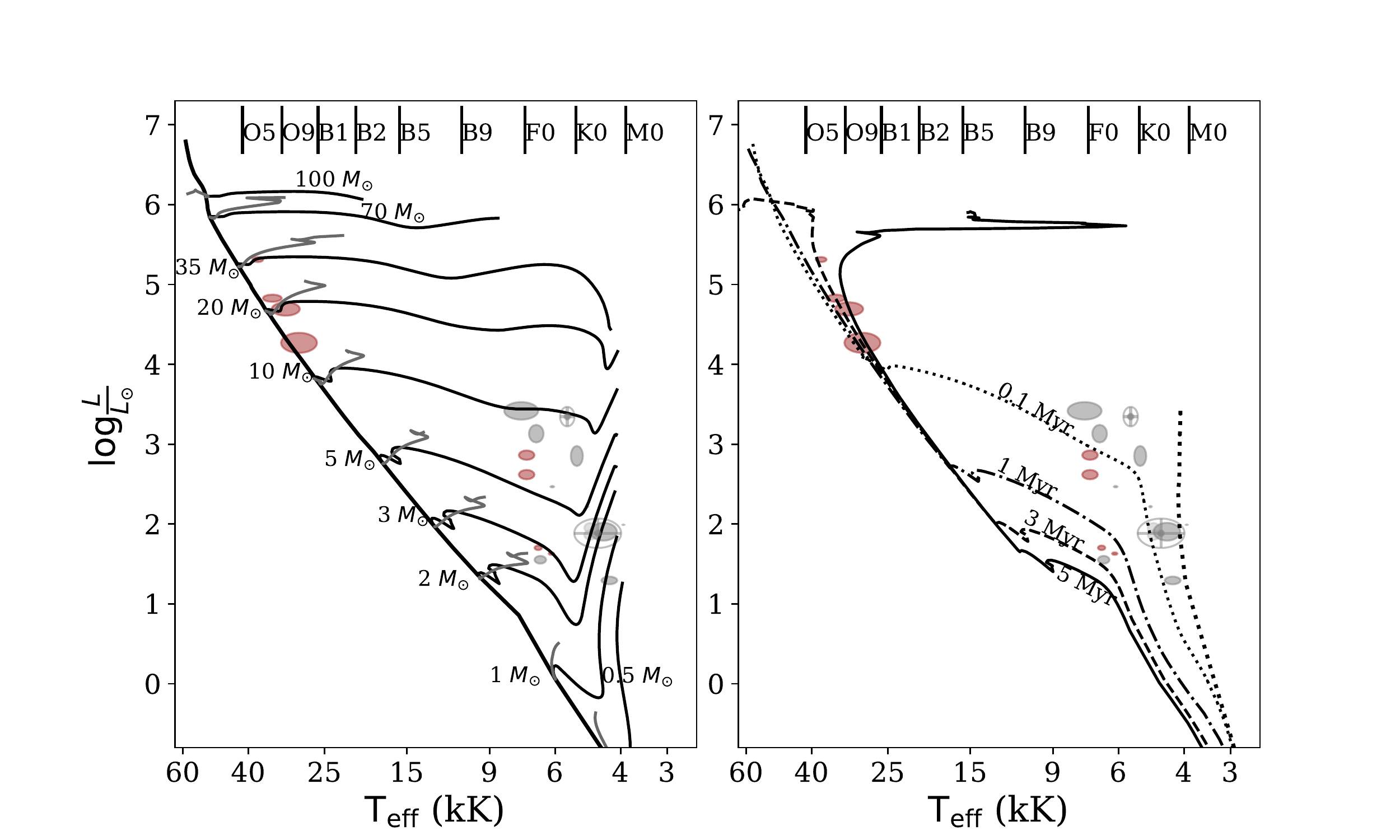}
        }
  \caption{Same as Figure~\ref{P4:fig:M8_HRD} but for the classified stars in \g333. We estimate an age $< \maxageg$~Myr.}
     \label{P4:fig:G333_HRD}
\end{figure*}

\begin{figure*}
\centering
        {\includegraphics[width=0.95\hsize]{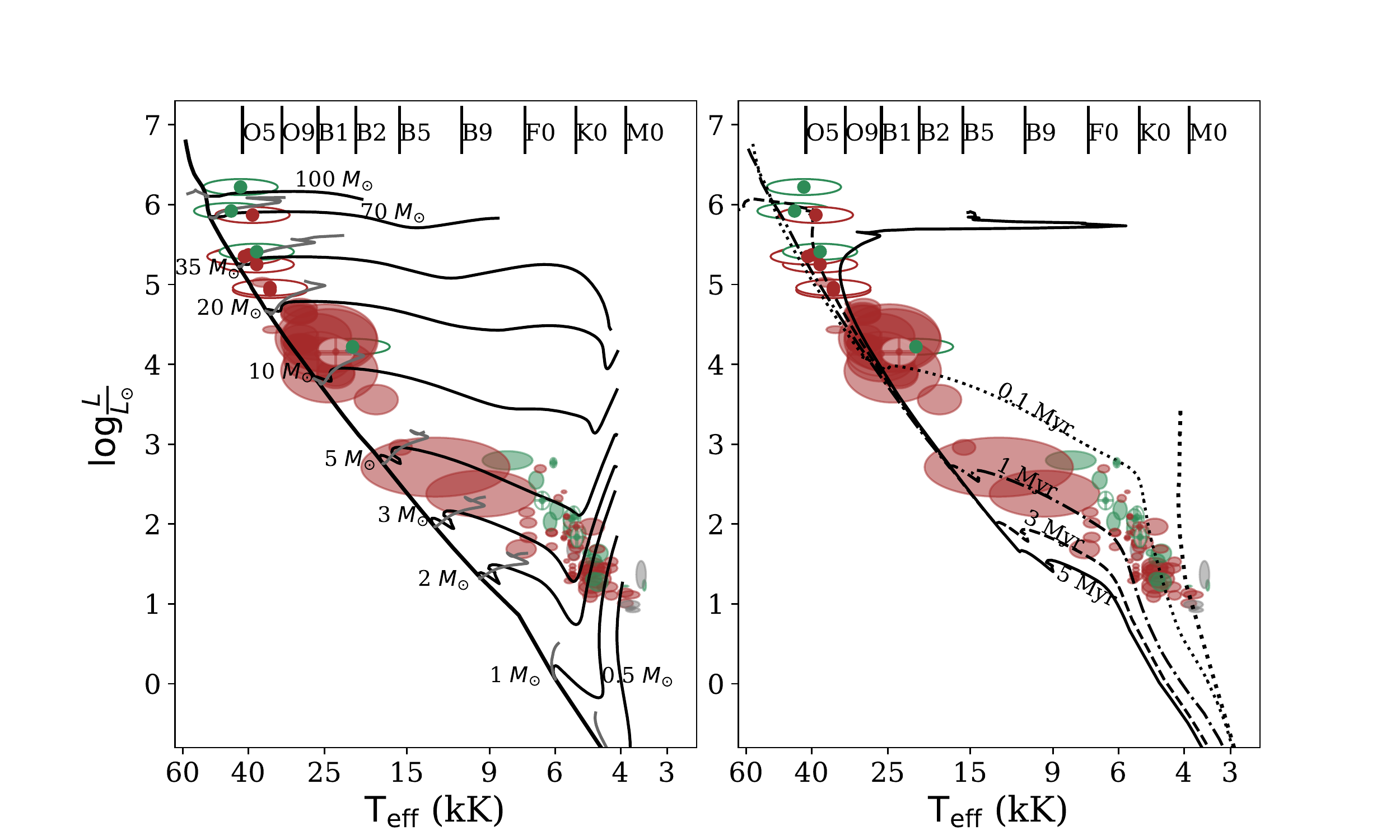}
        }
  \caption{Same as Figure~\ref{P4:fig:M8_HRD} but for the classified stars in \ngc6357. The age is estimated to be $\minagengc-\maxagengc$~Myr.}
     \label{P4:fig:NGC6357_HRD}
\end{figure*}

Figures~\ref{P4:fig:M8_HRD}, \ref{P4:fig:G333_HRD}, and \ref{P4:fig:NGC6357_HRD} show the HRDs for the three regions. We show the position of the observed targets as ellipses, where the height and width of the ellipses represent the errors in the temperature and luminosity resulting from the uncertainty in the spectral classification. The left panels show the MESA pre-main-sequence (black) and post-main-sequence (dark grey) tracks for stars of 1, 2, 3, 5, 10, 20, 35, 70, and 100~\msun, respectively; the thick black line shows the position of the zero-age main sequence. In the right panels we show the isochrones for 0.1, 1, 3, and 5~Myr with dotted, dash-dotted, dashed, and solid lines, respectively. 
The thick dotted line shows the birth line for stars until 8~\msun. 
The accuracy of the locations of the stars in the HRD is dominated by the uncertainty in the spectral classification.

To get a more complete picture of the studied regions, we add the previously known OB stars from the MPCM catalogue listed in \citet{2017ApJ...838...61P}. The derived properties of these objects are listed in the last section of Tables~\ref{P4:tab:M8physicalProp} and \ref{P4:tab:NGC6357physicalProp} and shown as solid green and red dots corresponding to the giant and dwarf stars, respectively. We also plot 0.2~dex error circles to represent the scatter reported by \citet{2017ApJ...838...61P} around the ZAMS. The infrared spectral classification of the most massive stars (spectral types O3-O4) is normally degenerate \citep[see e.g.][]{2014A&A...568L..13W,2016A&A...589A..16W, 2019arXiv190205460B} so it could affect the lower age limit. To avoid this, we determined the age based on the whole population (including the low-mass PMS-stars) instead of focusing only on the massive stars. 

In addition to the sources mentioned in Section~\ref{P4:sec:PhotVsSpt} that were discarded as members of the regions, we marked the $K$-band excess sources (below the reddening line in the NIR CCDs) with empty ellipses filled with crosses. Their NIR excess emission could cause an overestimation of $L_{\rm bol}$ and affect the age estimation, therefore we did not take them into account when assessing the age of the regions. 

The main sequence stars in M8 have masses between $\sim$5 and $\sim$70~\msun. The PMS stars will become ZAMS stars of $\sim$0.5-3~\msun. From comparing the observed population with MIST isochrones we conclude that the age of M8 is between \minagem\ and \maxagem~Myr. 

For \g333\ the main sequence population ranges in mass between $\sim$5 and $\sim$35~\msun. As in this region the membership of the stars is more uncertain, we had to focus on the massive stars only to estimate the age. For this reason, we only provide an upper limit. By comparing the few confirmed OB stars with X-ray counterparts with MIST isochrones we estimate an age \mbox{$<\maxageg$~Myr} for this region.     

The massive stars in \ngc6357\ have main sequence masses between $\sim$10 and $\sim$100~\msun. From comparing the position of the stars with MESA isochrones, we conclude that the mean age of the PMS and main-sequence populations of \ngc6357\ is probably somewhat younger than that of M8. We determine the age of \ngc6357\ to be $\minagengc-\maxagengc$~Myr.

We also tested the effect of the different extinction laws on the age determination. We conclude that the \citet{2009ApJ...696.1407N} and the \citet{2005ApJ...619..931I} extinction laws result in similar age estimates for the studied regions.

\section{Discussion}
\label{P4:sec:Discussion}

In this section we discuss the relation between the results obtained from photometry and spectroscopy. We consider some of the uncertainties that could affect the age determination and the impact that would have on our results. 

\subsection{The effectiveness of the \mys\ selection}

When selecting cluster members based on their NIR colours, typically half of the sources are discarded after spectral classification because they are background or foreground stars. For example, \citet{2016A&A...589A..16W} selected 44 candidate OB stars belonging to the cluster W49 to perform a spectroscopic follow up. They found that 22 of the observed spectra (50\%) are dominated by CO in absorption, which led them to conclude that they are foreground or background stars. \citet{2012A&A...542A..74G} studied the CN15/16/17 molecular complex; they aimed at observing the massive stars in the complex and by classifying them spectroscopically, they constrained the distance and extinction towards the stars. The authors selected the spectroscopic sample based on the NIR colours (brightest stars in $K$ and $H-K_s > 0.3$). Their sample consists of 14 stars out of which five (35\%) are field giants polluting the sample of massive stars. Similarly, \citet{2011A&A...531A..27M} studied the massive star population of IRAS~06084-0611. They selected 39 stars to follow up spectroscopically and after looking for highly reddened sources, IR excess in the IRAC bands, evidence of jets or outflows, and the presence of a radio counterpart, they identified 28 objects as cluster members (so 28\% as foreground and background stars).

The importance of the \mys\ selection becomes very clear in this paper where two regions are included in the MPCM list (M8 and \ngc6357) and one is not (\g333). The \mys\ selection is especially important in the case of low-mass PMS stars. Given that they are more abundant than massive stars, when these stars are selected on the basis of their NIR colours only, the probability that these are foreground or background sources is high. Nevertheless, adding the X-ray information points towards these sources being PMS stars. 
In the case of our two \mys\ regions -- M8 and \ngc6357 -- the fraction of stars whose membership is not confirmed based on their spectral type and astrometric properties is very low ($<10\%$), while in \g333\ the fraction of stars with uncertain membership is around 30\%.

In all HRDs a group of stars (grey symbols) is located at the right side of the 0.1~Myr isochrone near the birth line. These objects correspond to late-K and M-type stars and the most likely explanation for them to be at that location is that they are not members of the cluster, and so contaminate the \mys\ MPCM list.

\subsection{Accuracy of the age determination}

We constrained the age of each region in section~\ref{P4:sec:age}. Here we discuss some of the issues that might affect the age determination and compare our age estimates with those published in the literature. These issues are:

\noindent {\em i)} Binarity, which could affect the luminosity determination by making stars look brighter than they actually are. No information about the multiplicity properties of our sources is however available. {\em ii)} The use of different evolutionary tracks, which could lead to different age estimates, e.g, \citet{2013A&A...560A..16M} show that the main-sequence age for different models could vary by $\sim0.7~{\rm Myr}$. In our case, this uncertainty is likely smaller than that caused by the uncertainty on the spectral type. {\em iii)} The adopted extinction law, which may impact the age determination. We tested the effect of two different extinction laws on the age determination. By using the \citet{2009ApJ...696.1407N} extinction law, we obtain lower $A_K$ values, and therefore slightly fainter objects. However, the implied difference in luminosity does not significantly affect the age determination. 

\citet{2005A&A...430..941P, 2019A&A...623A.159P} estimated the age of the pre-main sequence stars in M8 between $\sim 0.8 - 2.5 ~{\rm Myr}$, in agreement with our estimate -- including the massive stars -- of $\minagem-\maxagem~{\rm Myr}$. \citet{2015A&A...573A..95M} determined the age of \ngc6357\ at $1-3~{\rm Myr}$ based on the visual and NIR CMD and CCDs. By comparing the whole population of stars with MIST isochrones, we estimate an age of \minagengc\ to \maxagengc~Myr for this cluster. For \g333\ we could not find a previous age determination.

The age obtained for the studied regions by considering only the massive stars is consistent with the age obtained only from the low-mass PMS stars. If the late-K and M-type stars near the Hayashi track were cluster members, we would obtain a younger age ($\sim0.1-1~{\rm Myr}$ younger) for the PMS stars compared to that of the massive stars only.

\subsection{Ionisation rates}

\citet{2018ApJ...864..136B} did a multi-wavelength study of Galactic massive star-forming regions where they studied their energy budgets and star formation rates. They obtained the total infrared luminosity processed by dust from the radio free-free continuum and the global infrared SEDs ($3.6 -500~\mu$m). In the cases of M8 and \ngc6357\ they used a similar sample of ionising stars as we do in this paper, but in the case of G333 they had to rely on an incomplete stellar population. They concluded that in the first two cases there are more than enough stars to explain both the observed ionisation and the bolometric luminosity. For the entire G333 complex (of which \g333\ is the brightest component), they calculated that around a dozen O5V stars are necessary to explain the total luminosity of the complex, and therefore concluded that the principal ionising stars in the complex remain undiscovered. 
Due to the ambiguity in the stellar population of G333 \citet{2018ApJ...864..136B} assumed the presence of five O5V stars in \g333\ based on a photometric study by \citet{2005MNRAS.356..801F}. We only classified three new OB stars in \g333, all of which have spectral types later than O5V (hence produce less ionising photons). Therefore, we conclude that there are probably more OB stars to be discovered in this region.

%__________________________________________________________________

\section{Conclusions}
\label{P4:sec:Conclusion}

We have presented a photometric and spectroscopic study of massive stars in three \ion{H}{ii} regions and characterised their stellar populations in terms of age and mass ranges.

\begin{itemize}

    \item M8, \g333\ and \ngc6357\ contain young stellar populations: we detect O and B stars, several mYSOs as well as a low-mass PMS stellar population. 
    
    \smallskip

    \item We develop an automatic spectral classification method in the $H$- and $K$-band, that is accurate to within two spectral subtypes. We conclude that this method is at least as accurate as a classification by eye, while having the obvious merits of being fast and reproducible.

    \smallskip

    \item The most massive stars in our KMOS sample are objects 73 (O9-B1V) in M8, 6829462 (O6V) in \g333, and 16 (O9-B0V) in \ngc6357. The most massive stars published in the literature and not included in our sample for these clusters are HD~164794 (O4V((f))) in M8, and Pismis 24 17  (O3.5III(f*)) in \ngc6357.
    
    \smallskip
    
    \item We cross matched our catalogues with the Gaia DR2 database in order to further constrain the cluster membership of our sources based on their astrometric properties. For each of the clusters we found two sources that are not cluster members. F13 and F24 in M8;  515756828027 and 515757115402 in \g333; 91 and 8 in \ngc6357.

    \smallskip
    
    \item The sample selection for \g333\ was made based on the NIR-colours only. By classifying the stars spectroscopically and by using the information provided by Gaia (when available), we conclude that $\sim$30\% of the stars observed in \g333\ are not cluster members, but luminous background stars. For the other regions we used the \mys\ selection \citep{2013ApJS..209...26F} in order to select cluster members; in these case the fraction of non-members is less than 10\%. The latter demonstrates the importance and effectiveness of the \mys\ analysis when identifying cluster members.
    
    \smallskip
    
    \item Based on the presence of CO bandheads in emission and/or (double-peaked) emission lines in the spectra, we find one mYSO candidate in \g333 (720), one in M8 (32), and four in \ngc6357 (33, F57, 114, and F68).

    \smallskip
    
    \item We determined the $K$-band extinction of the observed stars by comparing their intrinsic near-infrared colours with the observed ones. In all regions $A_K$ varies from $\sim$0 to $\sim$3~mag. The mean extinction of \ngc6357 is higher than that of M8. We do not find strong evidence for a correlation between the extinction value and the location of the stars in the clusters suggesting that the extinction varies on small spatial scales, that is, is patchy. 

    \smallskip
    
    \item The observed stellar population of M8 has an age of $\sim$\minagem-\maxagem~Myr, and that of \ngc6357\ is $\minagengc-\maxagengc$~Myr old. From the OB stars in \g333\ we estimate an age $<\maxageg$~Myr. Given that this region did not profit from the \mys\ analysis, we only use the main sequence stars to determine the age. A more accurate determination is at present not possible given the uncertainty of the spectral classification.

\end{itemize}

\noindent
With present-day 8-10\,m class ground-based telescopes we may resolve stellar populations in the Magellanic Clouds. With future extremely large telescopes, like the ELT and TMT, we are going to resolve young stellar populations in more distant galaxies. With the techniques used and introduced here we may characterise these regions in terms of content, mass, age, and multiplicity properties even in highly obscured regions. This will allow us to probe massive star formation in galaxies spanning a wide range of types, masses, and metallicities. 

%-------------------------------------------------------------------

\begin{acknowledgements}
Based on observations collected at the European Organisation for Astronomical Research in the Southern Hemisphere under ESO program 095.C-0048. 
The authors thank Leisa Townsley, Eric Feigelson, Matthew Povich and Patrick Broos for valuable discussions on the \mys\ catalogue and assistance with the construction of the final target lists for the KMOS observations. 
The authors thank Richard Davies, Ray Sharples, Nadine Neumayer and Michael Hilker for discussions on KMOS data reduction. \\
JP and LK acknowledge support from their NWO/PAPESP grant for Advanced instrumentation (MOSAIC). 
HB acknowledges support from the European Research Council under the European Community's Horizon 2020 frame-work program (2014-2020) via the ERC Consolidator grant ‘From Cloud to Star Formation (CSF)' (project number 648505). AP acknowledges  support from the Sonderforschungsbereich SFB 881 `The Milky Way System' (subproject B5) of the German Research Foundation (DFG). 
O.H.R.A. acknowledges funding from the European Union's Horizon 2020
research and innovation programme under the Marie Sklodowska-Curie
grant agreement No 665593 awarded to the Science and Technology Facilities
Council.\\
This research made use of Astropy, a community-developed core Python package for Astronomy \citep{2018arXiv180102634T},
APLpy, an open-source plotting package for Python \citep{2012ascl.soft08017R},
NASA's Astrophysics Data System Bibliographic Services (ADS), the SIMBAD database, operated at CDS, Strasbourg, France \citep{2000A&AS..143....9W}, and the cross-match service provided by CDS, Strasbourg. 
The VISTA Data Flow System pipeline processing and science archive are described in \citet{2004SPIE.5493..411I}, \citet{2008MNRAS.384..637H}, and \citet{2012A&A...548A.119C}. We have used data from the 2nd data release of the VVV survey \citep{2012A&A...537A.107S}. 
This work has made use of data from the European Space Agency (ESA) mission
{\it Gaia} (\url{https://www.cosmos.esa.int/gaia}), processed by the {\it Gaia}
Data Processing and Analysis Consortium (DPAC,
\url{https://www.cosmos.esa.int/web/gaia/dpac/consortium}). Funding for the DPAC
has been provided by national institutions, in particular the institutions
participating in the {\it Gaia} Multilateral Agreement.

\end{acknowledgements}

%-------------------------------------------------------------------

\bibliography{references}

%-------------------------------------------------------------------

\begin{appendix}
\section{`Plus one' method to combine the likelihood distributions}
\label{P4:ap:plusOne}

To combine the likelihoods of spectral type prediction by each feature's EW and obtain the final likelihood distribution for each luminosity class, one needs to multiply all the likelihood functions for all possible EW available:
\begin{equation}
	L(\mathrm{SpT})_{\mathrm{tot}} = \prod_{\mathrm{EW}} L(\mathrm{SpT})_{\mathrm{EW}}
\end{equation}
(To avoid numerical zeros, the multiplication was in fact carried out as a sum of logarithmic values $\sum_{\mathrm{EW}}\log(L_{\mathrm{EW}})$ and then converted back to linear values.)
The resulting distribution function displays very sharp peaks, as the multiplication amplifies the overlapping peaks of different $L(\mathrm{SpT})_{\mathrm{EW}}$ by orders of magnitude while oppressing other peaks to nearly zero. Therefore, the final distribution obtained by this method does not provide a good uncertainty estimate for the spectral type predictions. 

\bigskip

Alternatively, we combine the ``plus one'' quantity $(L(\mathrm{SpT})_{\mathrm{EW}} + 1)$ by multiplication (Equation~\ref{P4:eq:Lspt_plus1}). The outcome of this operation gives a better likelihood distribution overall, with peaks and reasonable spreads around them to be used as the uncertainty estimation for the spectral type prediction.

This is not only a numerical trick to broaden the widths of the distribution functions and avoid numerical zeros, but also has physical meanings. To shorten the notation, the formula above is written as $L = \prod_i (G_i + 1)$ where $i$ is the index of a list of $n$ EWs. 

Expand the product as a binomial expansion we can obtain
\begin{equation}
	\prod_{i=1}^{n} (G_i + 1) = 1 + \sum_{i=1}^{n} G_i + \sum_{i,j} G_i G_j + \sum_{i,j,k} G_i G_j G_k + ... + \prod_{i=1}^{n} G_i
\end{equation}
where each collected term is the collection of terms with the same number of $G$-functions multiplied together, runs from zero to $n$. For example, $\sum_{i,j} G_i G_j$ means the sum of all terms with two $G$-functions like $G_i G_j$ taking all possible choices of indices ($i, j$) for $i \neq j \in \{1, 2, ..., n\}$ and the different sequence of the indices with the same numbers are not counted as extra terms. 

We can immediately find out the last term $\prod_{i=1}^{n} G_i$ is exactly the same as the result of direct multiplication: the combined likelihood function considering the predictions of all available EWs. Then, the terms with just one $G$-function missing represent all the possible cases where we have one of the EWs information missing, and use the rest $n-1$ EWs to determine the combined likelihood. After that, we come to the terms with all the cases where two $G$-functions replaced by 1 and multiply the rest $n-2$ $G$s together... all the way down to $\sum_{i=1}^{n} G_i$, in which cases only one EW is available and the combined distributions are the same as the $G$-functions. Summing up all the terms, we obtain a likelihood function which not only considers the combination of all available EWs, but all the possible cases where one or more of the EW information are missing. This method introduces robustness when there are a few unknown ``bad'' measurements mixed in the EW but the majority of the data are properly produced, as the $G$-function of each EW appears in only half of the terms in the binomial expansion, the contribution of a low quality EW measurement is reduced to half, at the cost of having a larger uncertainty range than the direct combination method.

\clearpage
\onecolumn
\section{Spectral features}
\label{P4:ap:features}
\bigskip

\begin{table}[h]
\centering
\caption{Spectral features used for equivalent width (EW) calculations for the F-M type stars. All wavelengths are in $\mu$m.}             % title of Table
\begin{minipage}{\hsize}
\centering
\renewcommand{\arraystretch}{1.4}
\begin{tabular}{ccccc}
\hline
\hline
Line ($\lambda_c$ in $\mu$m) & Feature & Continuum 1 & Continuum 2 & Continuum 3 \\ % table heading 
\hline                        % inserts single horizontal line
MgI(1.485) \footnote{\label{P4:foot:Rayner}\citet{2009ApJS..185..289R}}   & 1.475 - 1.497   & 1.477 - 1.485   & 1.491 - 1.497                      \\     % & 1.485
Mg I (1.50) \footnote{\label{P4:foot:Ivanov}\citet{2004yCat..21510387I}}  & 1.5020 - 1.5060 & 1.4990 - 1.5020 & 1.5060 - 1.5090                    \\     % & 1.5040 
Fe I (1.58) \footref{P4:foot:Ivanov}                                     & 1.5810 - 1.5850 & 1.5790 - 1.5810 & 1.5850 - 1.5865                    \\     % & 1.5830 
Si I (1.59) \footref{P4:foot:Ivanov}                                     & 1.5870 - 1.5910 & 1.5830 - 1.5870 & 1.5910 - 1.5950                    \\     % & 1.5890 
CO (1.62)   \footref{P4:foot:Ivanov}                                     & 1.6175 - 1.6220 & 1.6145 - 1.6175 & 1.6255 - 1.6285                    \\     % & 1.6198 
Br12 (1.64)            & 1.6385 - 1.6439 & 1.6281 - 1.6294 & 1.6503 - 1.6510                    \\     % & 1.6412 
Br11 (1.68)            & 1.6781 - 1.6849 & 1.6700 - 1.6704 & 1.702  - 1.7053                    \\     % & 1.6811 
Mg I a (1.71)  \footref{P4:foot:Ivanov}                                   & 1.7100 - 1.7130 & 1.7085 - 1.7100 & 1.7130 - 1.7160                    \\     % & 1.7115 
Mg I b (1.711) \footref{P4:foot:Rayner}                                   & 1.695  - 1.726  & 1.702 - 1.708   & 1.715 - 1.720                      \\     % & 1.711
Na I (2.206)   \footref{P4:foot:Rayner}                                   & 2.185 - 2.230   & 2.192 - 2.198   & 2.213 - 2.220                      \\     % & 2.206
Mg I (2.11)       \footref{P4:foot:Ivanov}                              & 2.1040 - 2.1110 & 2.1000 - 2.1040 & 2.1110 - 2.1150                    \\     % & 2.1075 
Br$\gamma$ (2.16) \footref{P4:foot:Ivanov}                              & 2.1639 - 2.1686 & 2.0320 - 2.034  & 2.0907 - 2.0951 & 2.2873 - 2.2925  \\     % & 2.1662 
Na I a (2.21)     \footref{P4:foot:Ivanov}                              & 2.2040 - 2.2110 & 2.2140 - 2.2200 & 2.2330 - 2.2370                    \\     % & 2.2075 
Na I b (2.21)     \footref{P4:foot:Ivanov}                              & 2.2053 - 2.2101 & 2.0907 - 2.0951 & 2.2873 - 2.2925                    \\     % & 2.2077 
Ca I a (2.26)     \footref{P4:foot:Ivanov}                              & 2.2580 - 2.2690 & 2.2490 - 2.2530 & 2.2700 - 2.2740                    \\     % & 2.2635 
Ca I b (2.26)     \footref{P4:foot:Ivanov}                              & 2.2611 - 2.2662 & 2.0907 - 2.0951 & 2.2873 - 2.2925                    \\     % & 2.2637 
Mg I a (2.28)     \footref{P4:foot:Ivanov}                              & 2.2780 - 2.2832 & 2.2720 - 2.2780 & 2.2832 - 2.2882                    \\     % & 2.2806 
Mg I b (2.28)     \footref{P4:foot:Ivanov}                              & 2.2790 - 2.2850 & 2.2700 - 2.2740 & 2.2871 - 2.2901                    \\     % & 2.2820 
CO (2.29)         \footref{P4:foot:Ivanov}                              & 2.2924 - 2.2977 & 2.2873 - 2.2919 & 2.305  - 2.32                      \\     % & 2.2950 
$^{12}$CO (2,0) (2.29) \footref{P4:foot:Ivanov}                          & 2.2930 - 2.3030 & 2.2873 - 2.2919 & 2.305  - 2.32                      \\     % & 2.2980 
$^{12}$CO (3,1) (2.32) \footref{P4:foot:Ivanov}                          & 2.3218 - 2.3272 & 2.2873 - 2.2919 & 2.340  - 2.341                     \\     % & 2.3245 
$^{13}$CO (2,0) (2.35) \footref{P4:foot:Ivanov}                          & 2.3436 - 2.3491 & 2.2873 - 2.2925 & 2.399  - 2.3997                    \\     % & 2.3463 
CO (2.35)             \footref{P4:foot:Ivanov}                          & 2.3100 - 2.4000 & 2.2873 - 2.2925 & 2.399  - 2.3997                    \\     % & 2.3550 
\hline
\vspace{-15pt}
\end{tabular}
\renewcommand{\footnoterule}{}
\end{minipage}
\label{P4:tab:EW_late}
\end{table}

\begin{table}
\centering
\caption{Spectral features used for equivalent width (EW) calculations for the O-A type stars, based on normalized spectra. All wavelengths are in $\mu$m. Lines taken from \citet{2005ApJS..161..154H}.}             % title of Table
\begin{minipage}{\hsize}
\centering
\renewcommand{\arraystretch}{1.4}
\begin{tabular}{cccc}
\hline
\hline
Line ($\lambda_c$ in $\mu$m) & Feature & Continuum 1 & Continuum 2  \\ % table heading 
\hline                        % inserts single horizontal line
Br 12 (1.64)         & 1.6371 - 1.6451 & 1.705 - 1.710  \\                       % 1.6411
Br 11 (1.68)         & 1.6772 - 1.6850 & 1.705 - 1.710  \\                       % 1.6811
He II (1.69)        & 1.6895 - 1.6951 & 1.705 - 1.710  \\                       % 1.6923
He I triplet (1.70) & 1.6993 - 1.7021 & 1.705 - 1.710  \\                       % 1.7007
Br 10 (1.74)         & 1.7327 - 1.7407 & 1.705 - 1.710  \\                       % 1.7367
C IV (2.08)          & 2.0773 - 2.0809 & 2.1813 - 2.1839 & 2.1987 - 2.1991 \\    % 2.0796
He I triplet (2.11) & 2.1113 - 2.1132 & 2.1813 - 2.1839 & 2.1987 - 2.1991 \\    % 2.1125
He I (2.11)         & 2.1132 - 2.1146 & 2.1813 - 2.1839 & 2.1987 - 2.1991 \\    % 2.1138
He I (2.16)         & 2.1596 - 2.1639 & 2.1813 - 2.1839 & 2.1987 - 2.1991 \\    % 2.1623
Br 3 (2.16)          & 2.1639 - 2.1686 & 2.1813 - 2.1839 & 2.1987 - 2.1991 \\    % 2.1661
He II (2.19)         & 2.1861 - 2.1921 & 2.1813 - 2.1839 & 2.1987 - 2.1991 \\    % 2.1891
\hline
\vspace{-15pt}
\end{tabular}
\renewcommand{\footnoterule}{}
\end{minipage}
\label{P4:tab:EW_early}
\end{table}

\begin{table}
\centering
\caption{\label{EW_early} Line ratios used for spectral typing of late-type stars.}             % title of Table
\begin{minipage}{\hsize}
\centering
\renewcommand{\arraystretch}{2.5}
\large
\begin{tabular}{cccccccc}
\hline
  $\mathrm{\frac{CO1(2.29)}{Br3(2.16)}}$ &  $\mathrm{\frac{CO2(2.29)}{Br12(1.64)}}$ &          $\mathrm{\frac{CO(2.35)}{MgI(1.50)}}$ &   $\mathrm{\frac{CO(1.62)}{Br3(2.16)}}$ &   $\mathrm{\frac{Br3(2.16)}{NaI(2.206)}}$ &    $\mathrm{\frac{Br3(2.16)}{MgI(2.11)}}$ &        $\mathrm{\frac{Br12(1.64)}{MgI(2.11)}}$ &   $\mathrm{\frac{13CO(2,0)(2.35)}{MgI(2.11)}}$ \\
 $\mathrm{\frac{CO1(2.29)}{Br11(1.68)}}$ &   $\mathrm{\frac{CO2(2.29)}{SiI(1.59)}}$ &         $\mathrm{\frac{CO(2.35)}{NaI(2.206)}}$ &  $\mathrm{\frac{CO(1.62)}{Br11(1.68)}}$ &   $\mathrm{\frac{Br11(1.68)}{SiI(1.59)}}$ &   $\mathrm{\frac{Br3(2.16)}{CaI1(2.26)}}$ &       $\mathrm{\frac{Br12(1.64)}{CaI1(2.26)}}$ &  $\mathrm{\frac{13CO(2,0)(2.35)}{CaI1(2.26)}}$ \\
 $\mathrm{\frac{CO1(2.29)}{Br12(1.64)}}$ &   $\mathrm{\frac{CO2(2.29)}{MgI(1.50)}}$ &   $\mathrm{\frac{13CO(2,0)(2.35)}{Br3(2.16)}}$ &  $\mathrm{\frac{CO(1.62)}{Br12(1.64)}}$ &   $\mathrm{\frac{Br11(1.68)}{MgI(1.50)}}$ &   $\mathrm{\frac{Br3(2.16)}{CaI2(2.26)}}$ &       $\mathrm{\frac{Br12(1.64)}{CaI2(2.26)}}$ &  $\mathrm{\frac{13CO(2,0)(2.35)}{CaI2(2.26)}}$ \\
  $\mathrm{\frac{CO1(2.29)}{SiI(1.59)}}$ &  $\mathrm{\frac{CO2(2.29)}{NaI(2.206)}}$ &  $\mathrm{\frac{13CO(2,0)(2.35)}{Br11(1.68)}}$ &   $\mathrm{\frac{CO(1.62)}{SiI(1.59)}}$ &  $\mathrm{\frac{Br11(1.68)}{NaI(2.206)}}$ &  $\mathrm{\frac{Br11(1.68)}{NaI1(2.21)}}$ &         $\mathrm{\frac{CO(2.35)}{NaI1(2.21)}}$ &         $\mathrm{\frac{CO(1.62)}{NaI1(2.21)}}$ \\
  $\mathrm{\frac{CO1(2.29)}{MgI(1.50)}}$ &    $\mathrm{\frac{CO(2.35)}{Br3(2.16)}}$ &  $\mathrm{\frac{13CO(2,0)(2.35)}{Br12(1.64)}}$ &   $\mathrm{\frac{CO(1.62)}{MgI(1.50)}}$ &   $\mathrm{\frac{Br12(1.64)}{SiI(1.59)}}$ &   $\mathrm{\frac{Br11(1.68)}{MgI(2.11)}}$ &          $\mathrm{\frac{CO(2.35)}{MgI(2.11)}}$ &          $\mathrm{\frac{CO(1.62)}{MgI(2.11)}}$ \\
 $\mathrm{\frac{CO1(2.29)}{NaI(2.206)}}$ &   $\mathrm{\frac{CO(2.35)}{Br11(1.68)}}$ &   $\mathrm{\frac{13CO(2,0)(2.35)}{SiI(1.59)}}$ &  $\mathrm{\frac{CO(1.62)}{NaI(2.206)}}$ &   $\mathrm{\frac{Br12(1.64)}{MgI(1.50)}}$ &  $\mathrm{\frac{Br11(1.68)}{CaI1(2.26)}}$ &         $\mathrm{\frac{CO(2.35)}{CaI1(2.26)}}$ &         $\mathrm{\frac{CO(1.62)}{CaI1(2.26)}}$ \\
  $\mathrm{\frac{CO2(2.29)}{Br3(2.16)}}$ &   $\mathrm{\frac{CO(2.35)}{Br12(1.64)}}$ &   $\mathrm{\frac{13CO(2,0)(2.35)}{MgI(1.50)}}$ &  $\mathrm{\frac{Br3(2.16)}{SiI(1.59)}}$ &  $\mathrm{\frac{Br12(1.64)}{NaI(2.206)}}$ &  $\mathrm{\frac{Br11(1.68)}{CaI2(2.26)}}$ &         $\mathrm{\frac{CO(2.35)}{CaI2(2.26)}}$ &         $\mathrm{\frac{CO(1.62)}{CaI2(2.26)}}$ \\
 $\mathrm{\frac{CO2(2.29)}{Br11(1.68)}}$ &    $\mathrm{\frac{CO(2.35)}{SiI(1.59)}}$ &  $\mathrm{\frac{13CO(2,0)(2.35)}{NaI(2.206)}}$ &  $\mathrm{\frac{Br3(2.16)}{MgI(1.50)}}$ &   $\mathrm{\frac{Br3(2.16)}{NaI1(2.21)}}$ &  $\mathrm{\frac{Br12(1.64)}{NaI1(2.21)}}$ &  $\mathrm{\frac{13CO(2,0)(2.35)}{NaI1(2.21)}}$ &    \\                           
\hline
\end{tabular}
\renewcommand{\footnoterule}{}
\end{minipage}
\label{P4:tab:lineRatios}
\end{table}

\onecolumn

\section{Spectra}
\label{P4:ap:spectra}

  \begin{figure*}[h]
            {\includegraphics[width=0.9\textwidth]{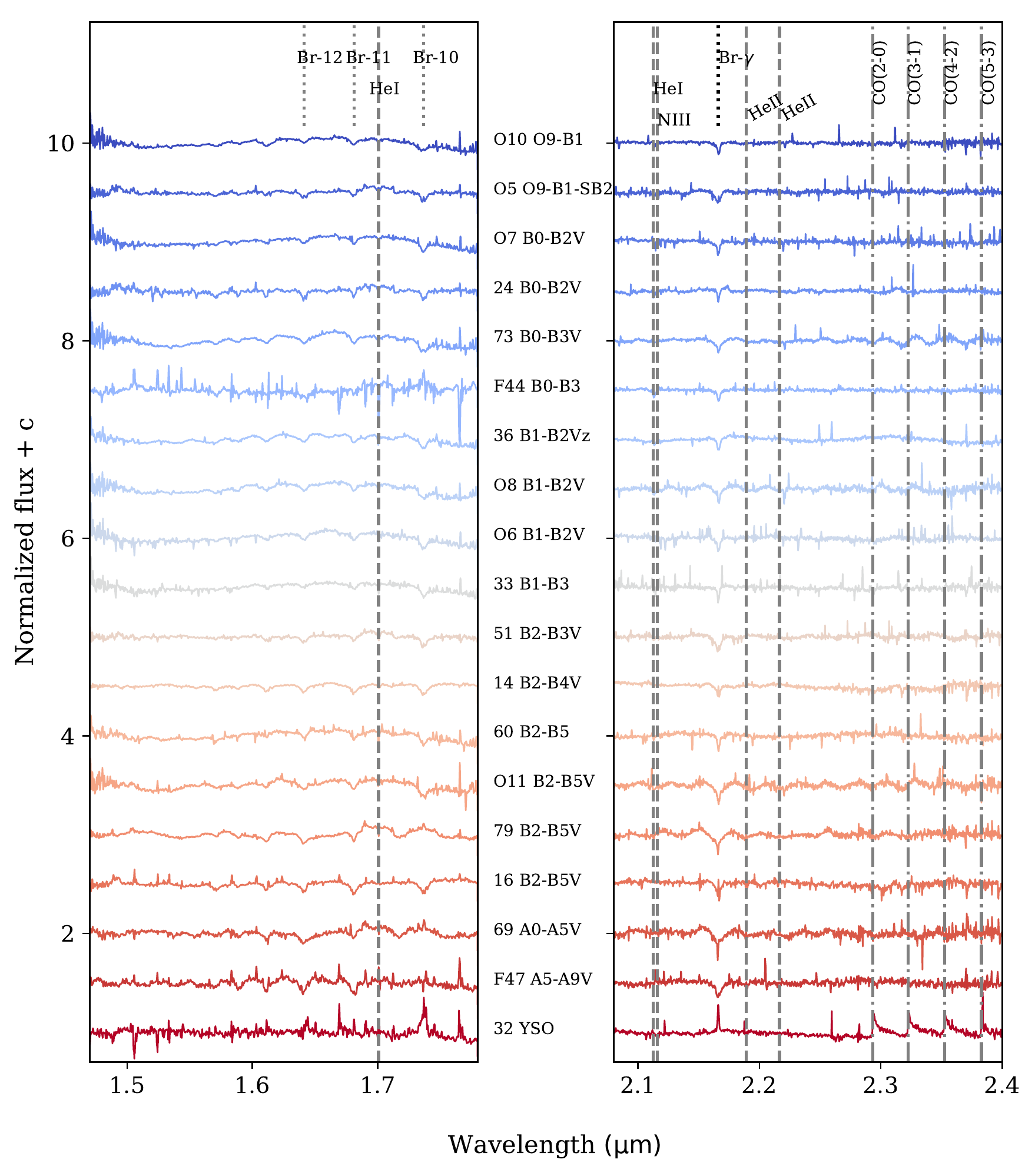}
            }
      \caption{$H$ and $K$-band spectra of the OBA type stars and YSOs in M8.}
         \label{P4:fig:M8_OBA}
  \end{figure*}

  \begin{figure*}
            {\includegraphics[width=0.98\textwidth]{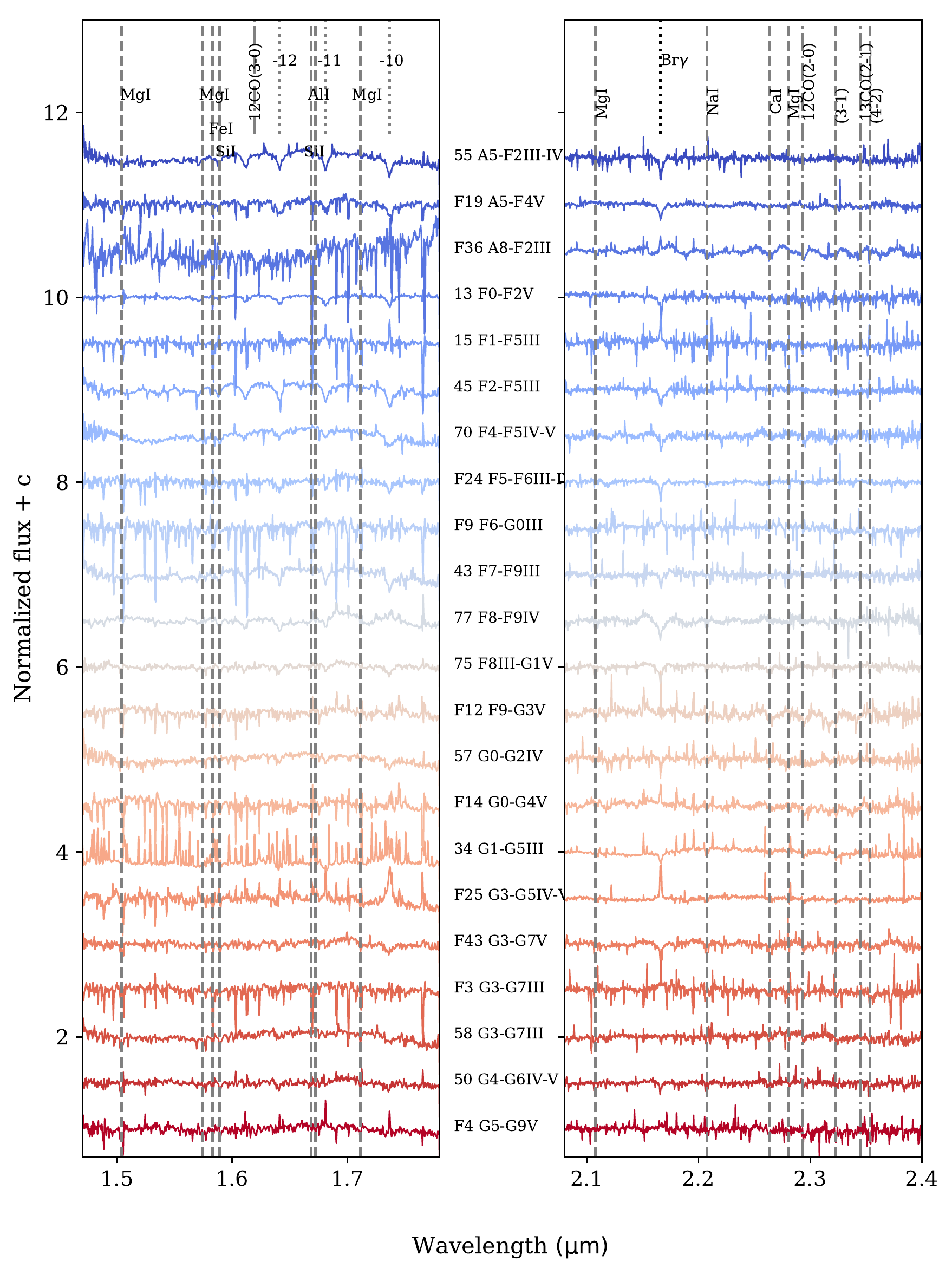}
            }
      \caption{$H$ and $K$-band spectra of the late type stars in M8.}
         \label{P4:fig:M8_Late_type2}
  \end{figure*}

  \begin{figure*}
            {\includegraphics[width=0.98\textwidth]{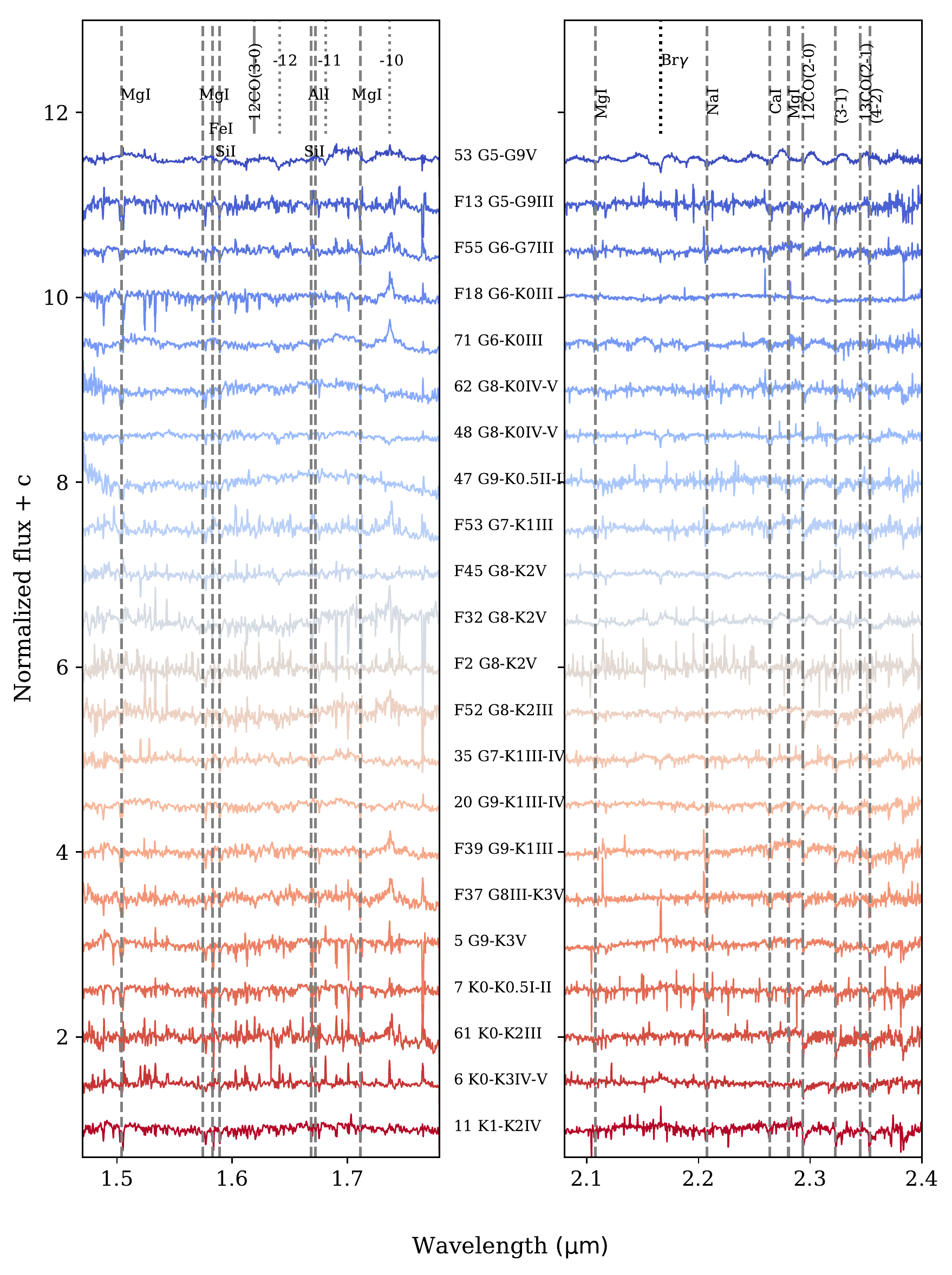}
            }
      \caption{$H$ and $K$-band spectra of the late type stars in M8.}
         \label{P4:fig:M8_Late_type1}
  \end{figure*}

  \begin{figure*}
            {\includegraphics[width=0.98\textwidth]{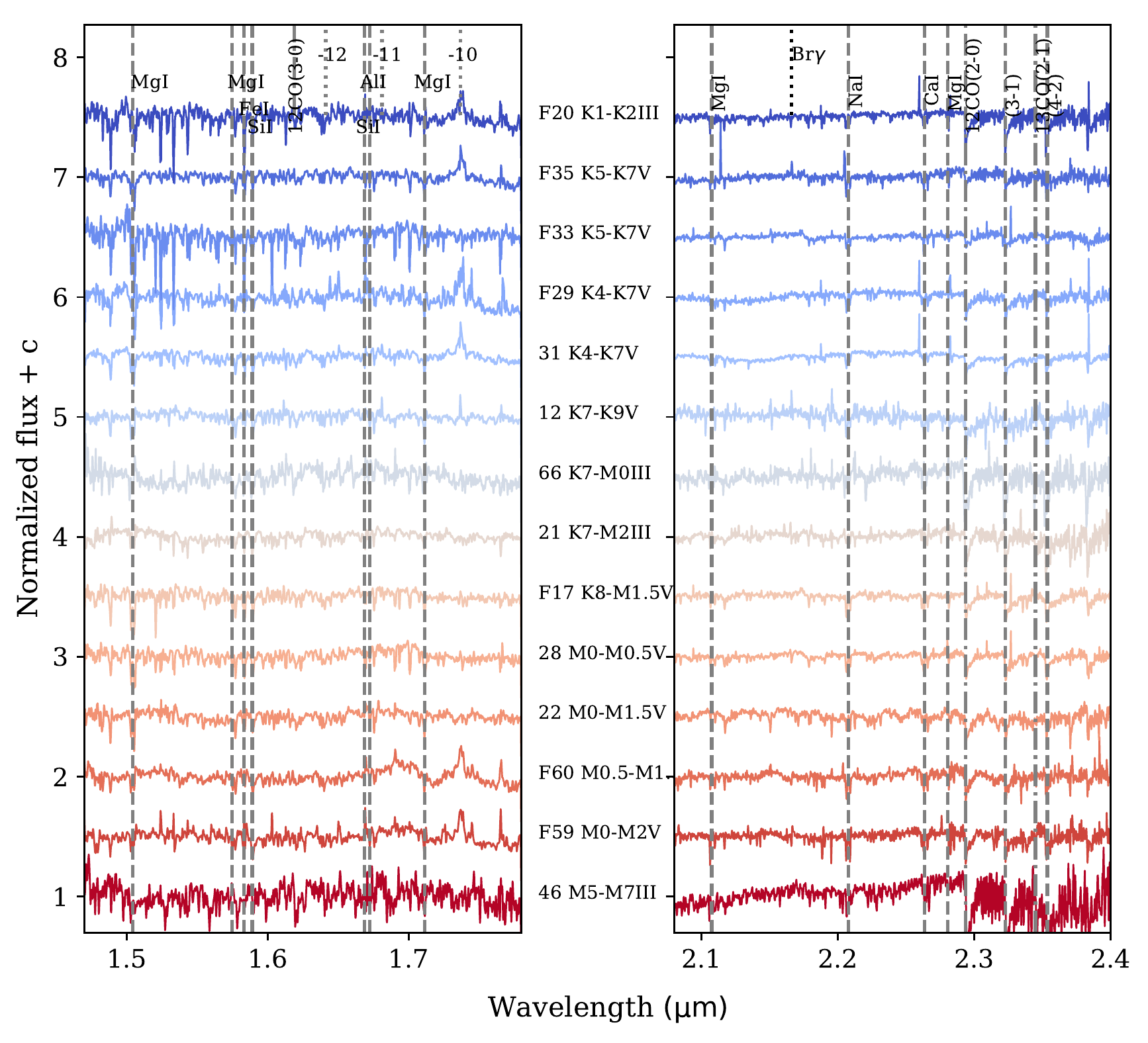}
            }
      \caption{$H$ and $K$-band spectra of the late type stars in M8.}
         \label{P4:fig:M8_Late_type0}
  \end{figure*}

  \begin{figure*}
            {\includegraphics[width=0.98\textwidth]{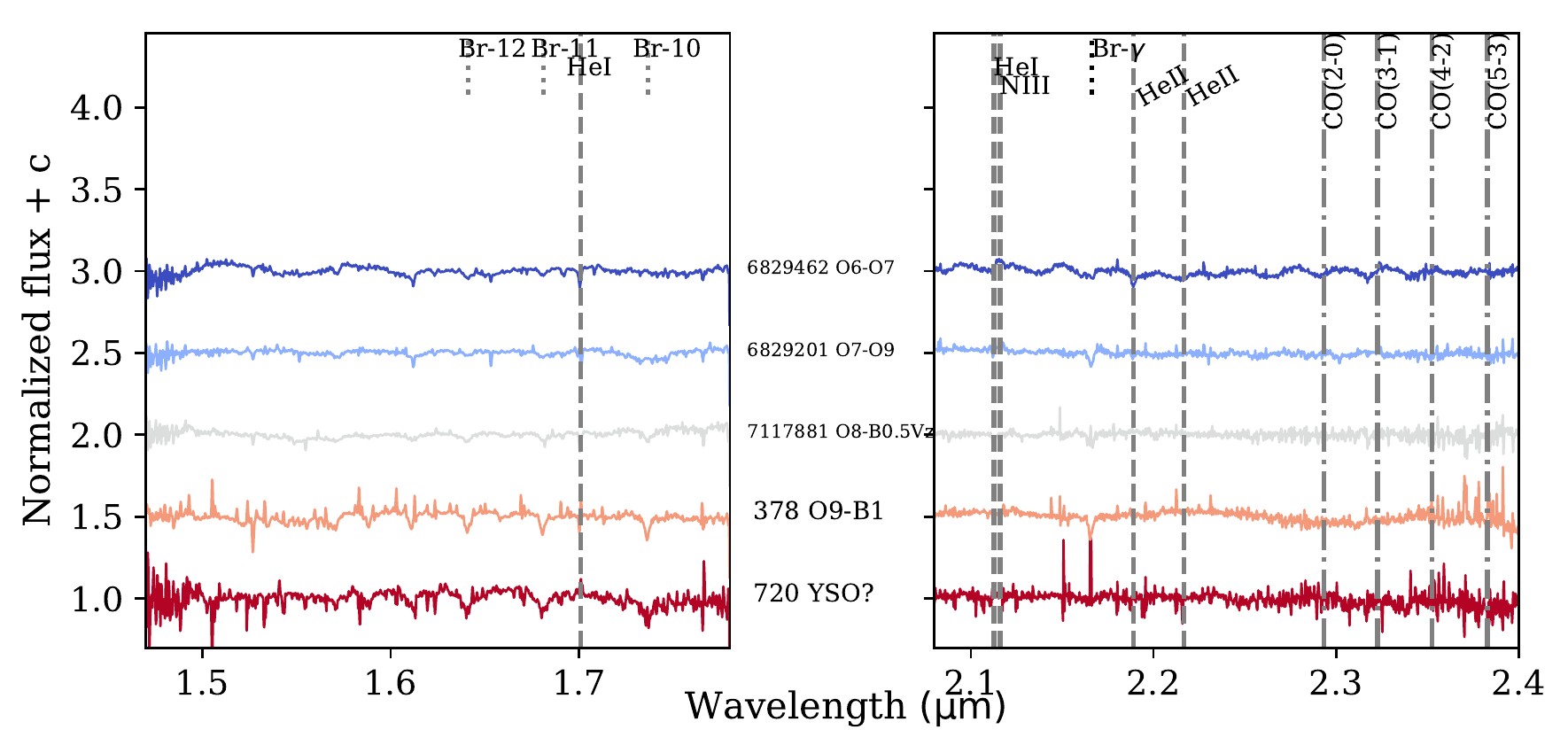}
            }
      \caption{$H$ and $K$-band spectra of the OBA type stars and YSOs in \g333.}
         \label{P4:fig:G333_OBA}
  \end{figure*}

  \begin{figure*}
            {\includegraphics[width=0.98\textwidth]{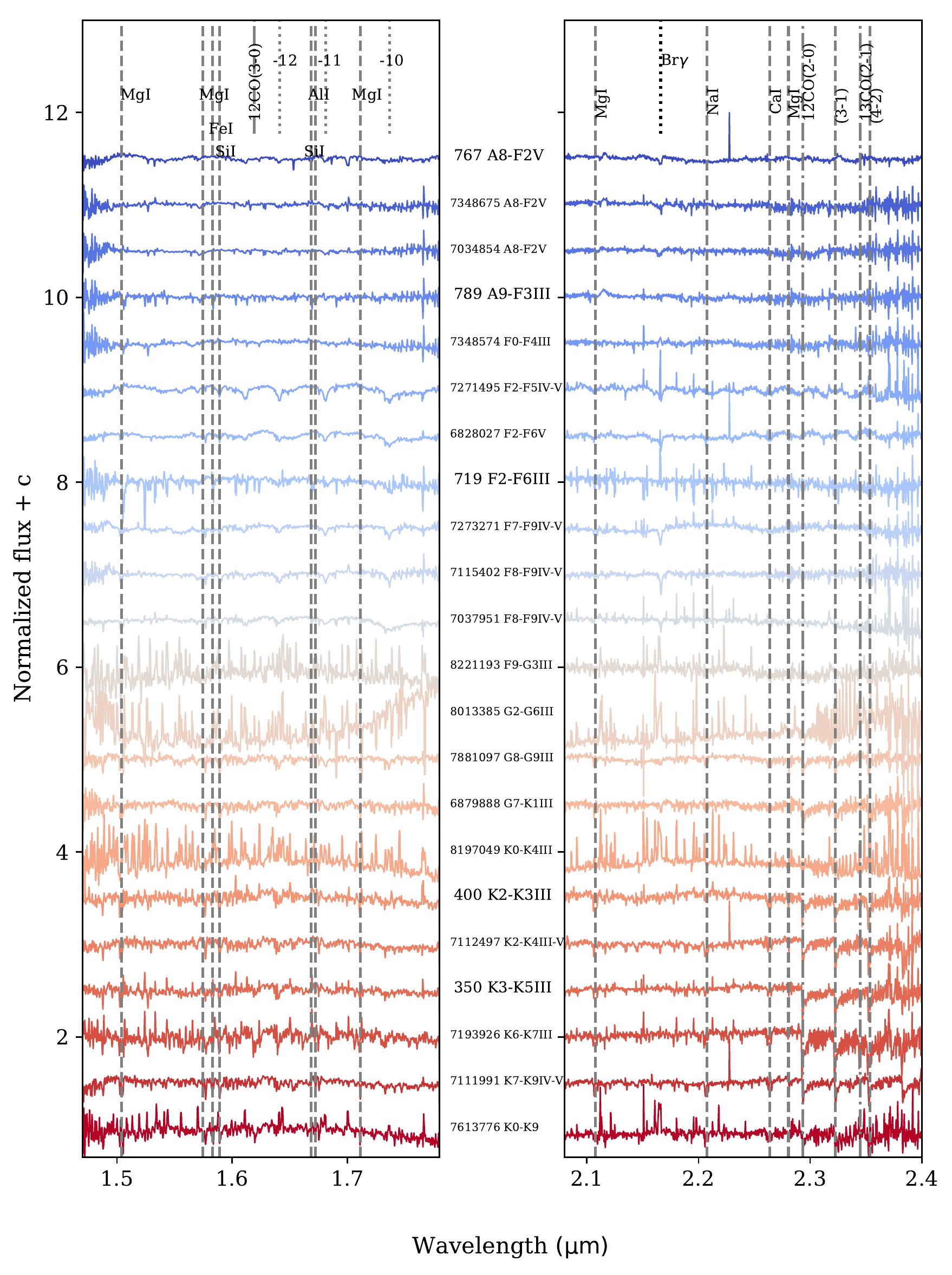}
            }
      \caption{$H$ and $K$-band spectra of the late type stars in \g333.}
         \label{P4:fig:G333_Late_type1}
  \end{figure*}

  \begin{figure*}
            {\includegraphics[width=0.98\textwidth]{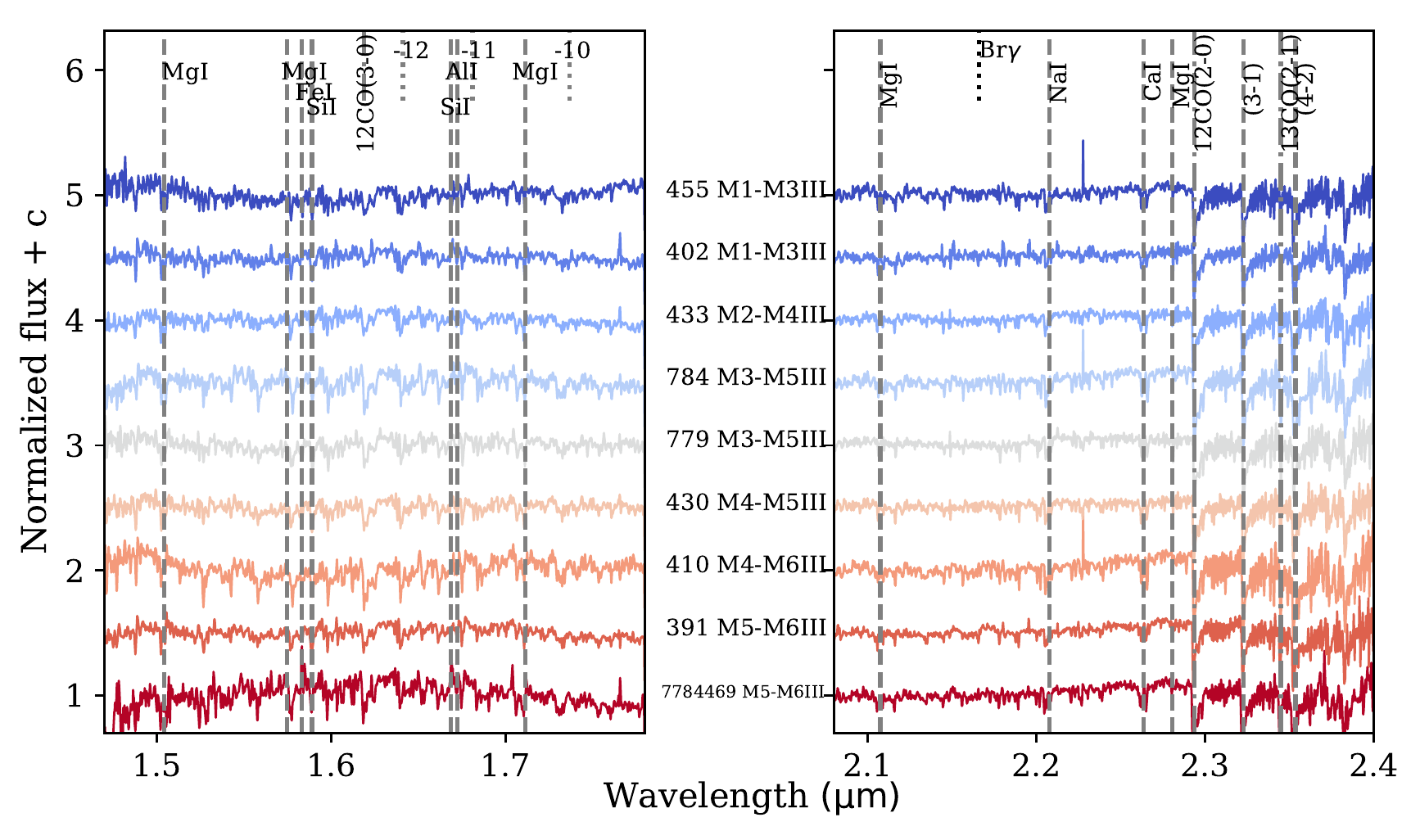}
            }
      \caption{$H$ and $K$-band spectra of the late type stars in \g333.}
         \label{P4:fig:G333_Late_type0}
  \end{figure*}

  \begin{figure*}
            {\includegraphics[width=0.98\textwidth]{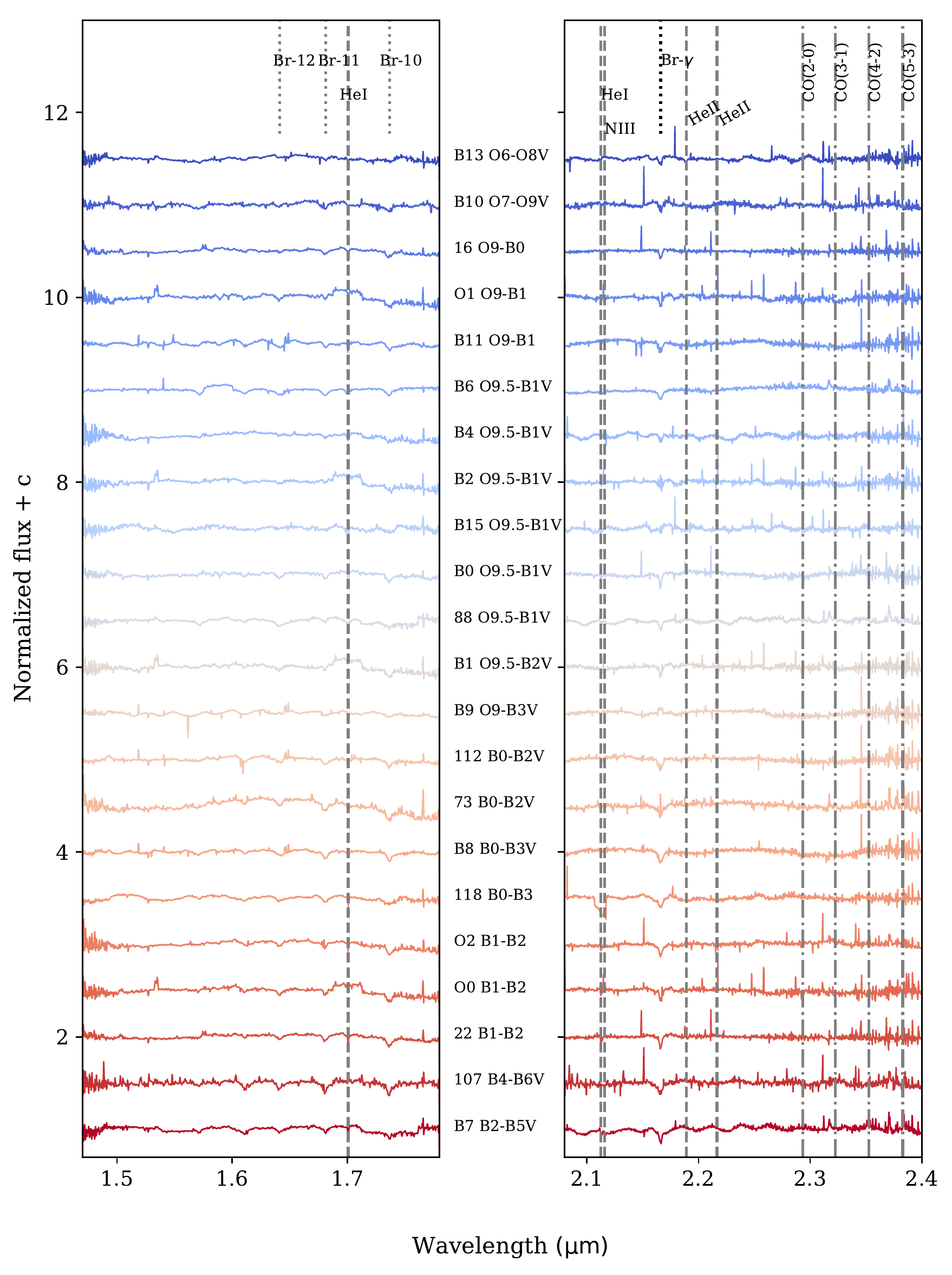}
            }
      \caption{$H$ and $K$-band spectra of the OBA type stars and YSOs in \ngc6357.}
         \label{P4:fig:NGC6357_OBA1}
  \end{figure*}

  \begin{figure*}
            {\includegraphics[width=0.98\textwidth]{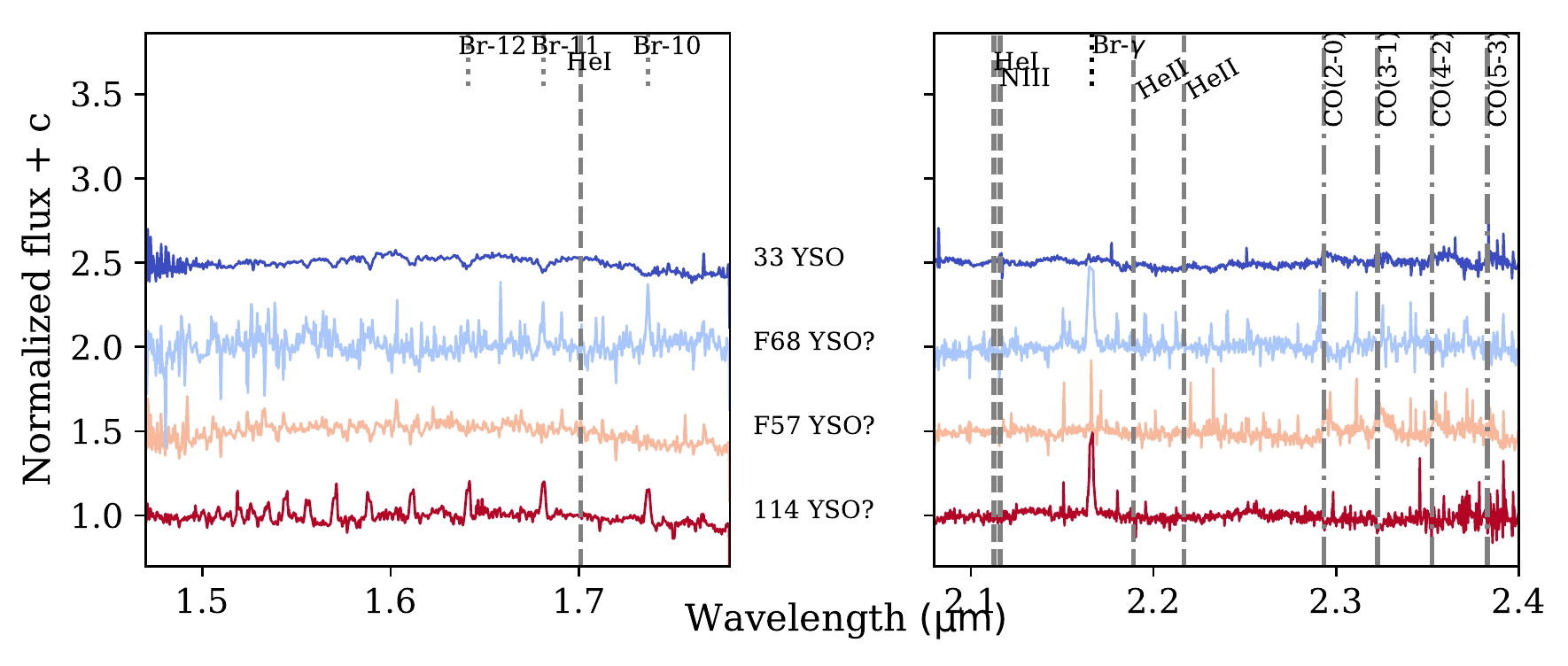}
            }
      \caption{$H$ and $K$-band spectra of the OBA type stars and YSOs in \ngc6357.}
         \label{P4:fig:NGC6357_OBA0}
  \end{figure*}

  \begin{figure*}
            {\includegraphics[width=0.98\textwidth]{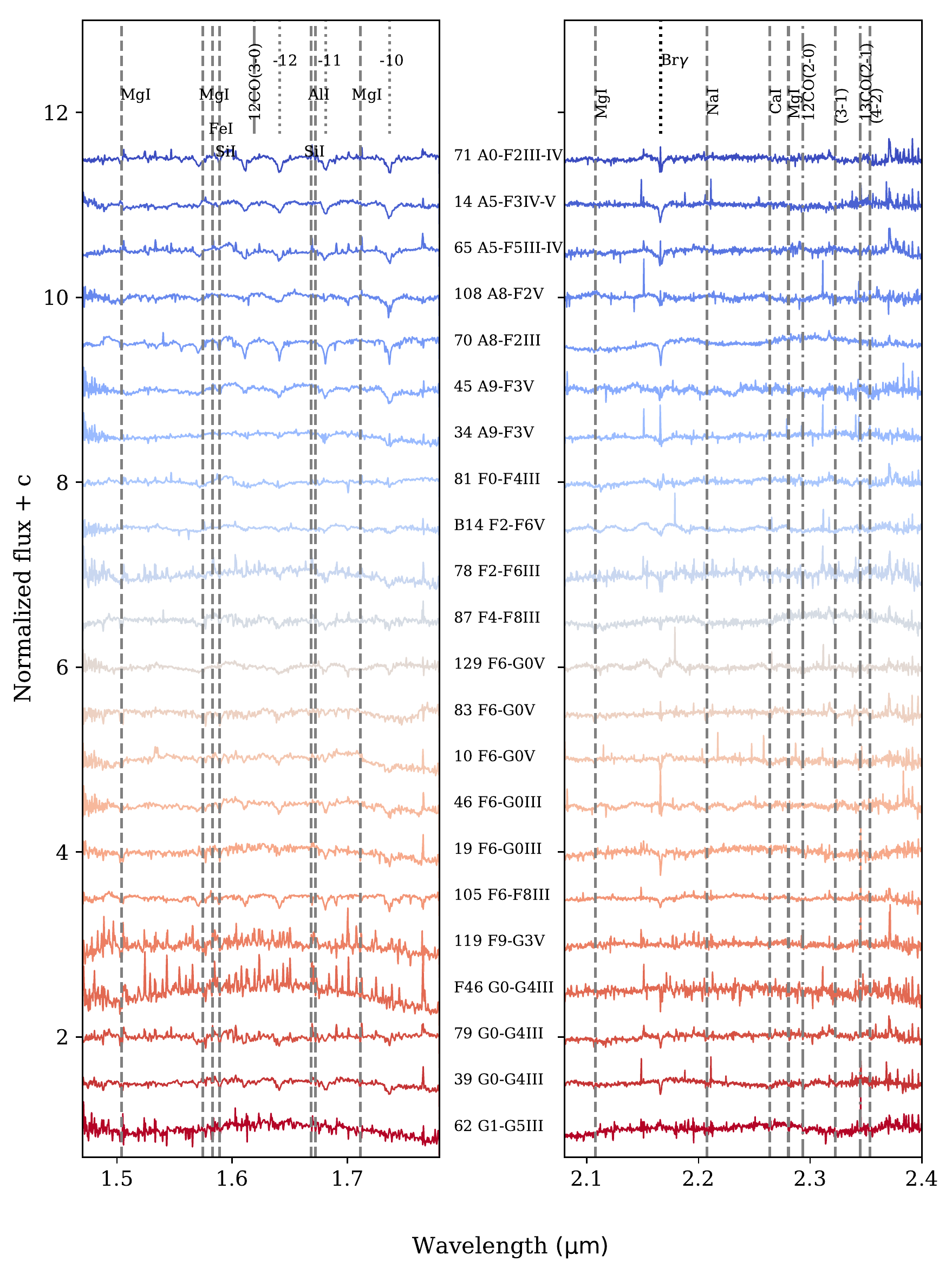}
            }
      \caption{$H$ and $K$-band spectra of the late type stars in \ngc6357.}
         \label{P4:fig:NGC6357_Late_type3}
  \end{figure*}

  \begin{figure*}
            {\includegraphics[width=0.98\textwidth]{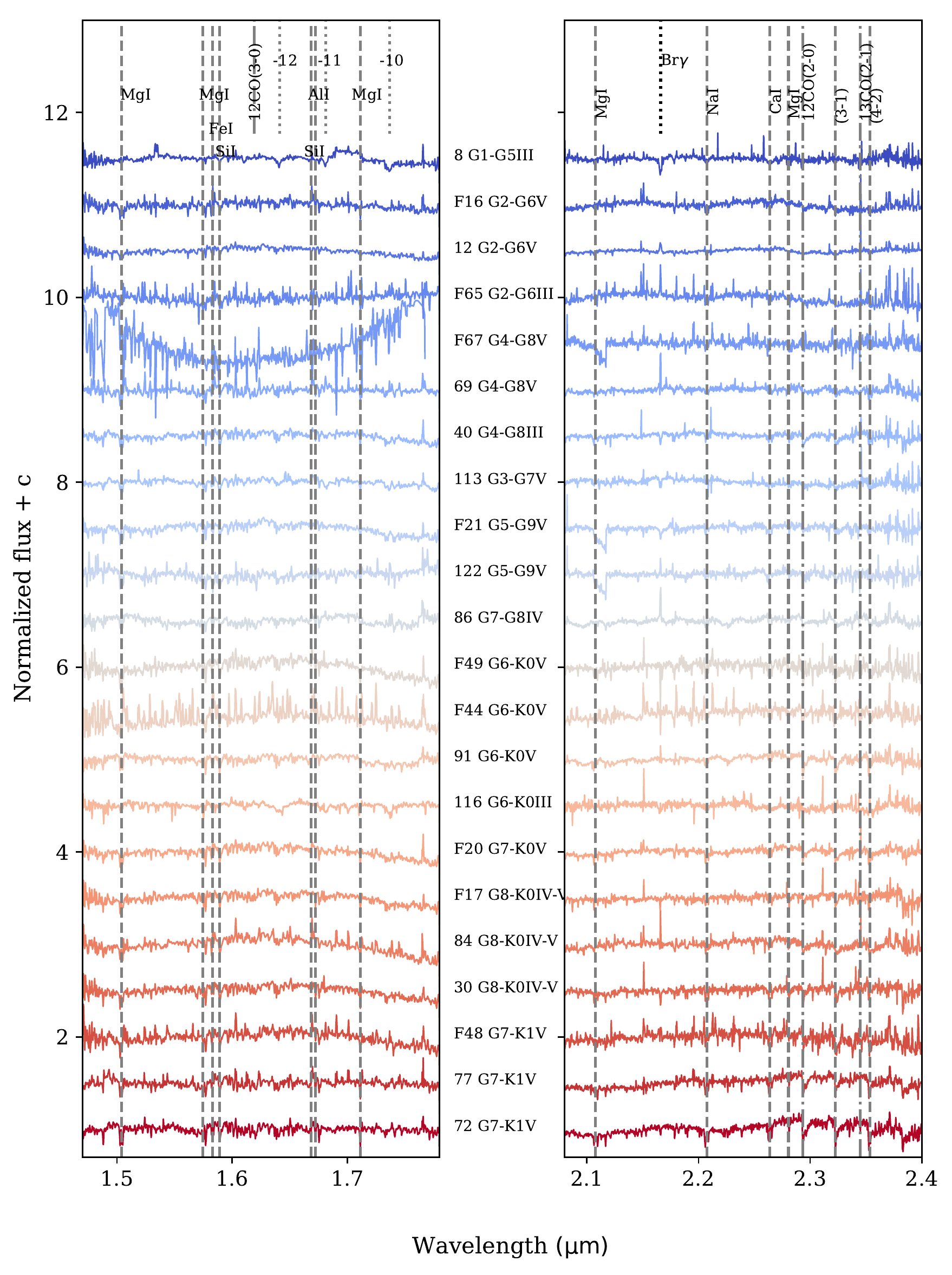}
            }
      \caption{$H$ and $K$-band spectra of the late type stars in \ngc6357.}
         \label{P4:fig:NGC6357_Late_type2}
  \end{figure*}

  \begin{figure*}
            {\includegraphics[width=0.98\textwidth]{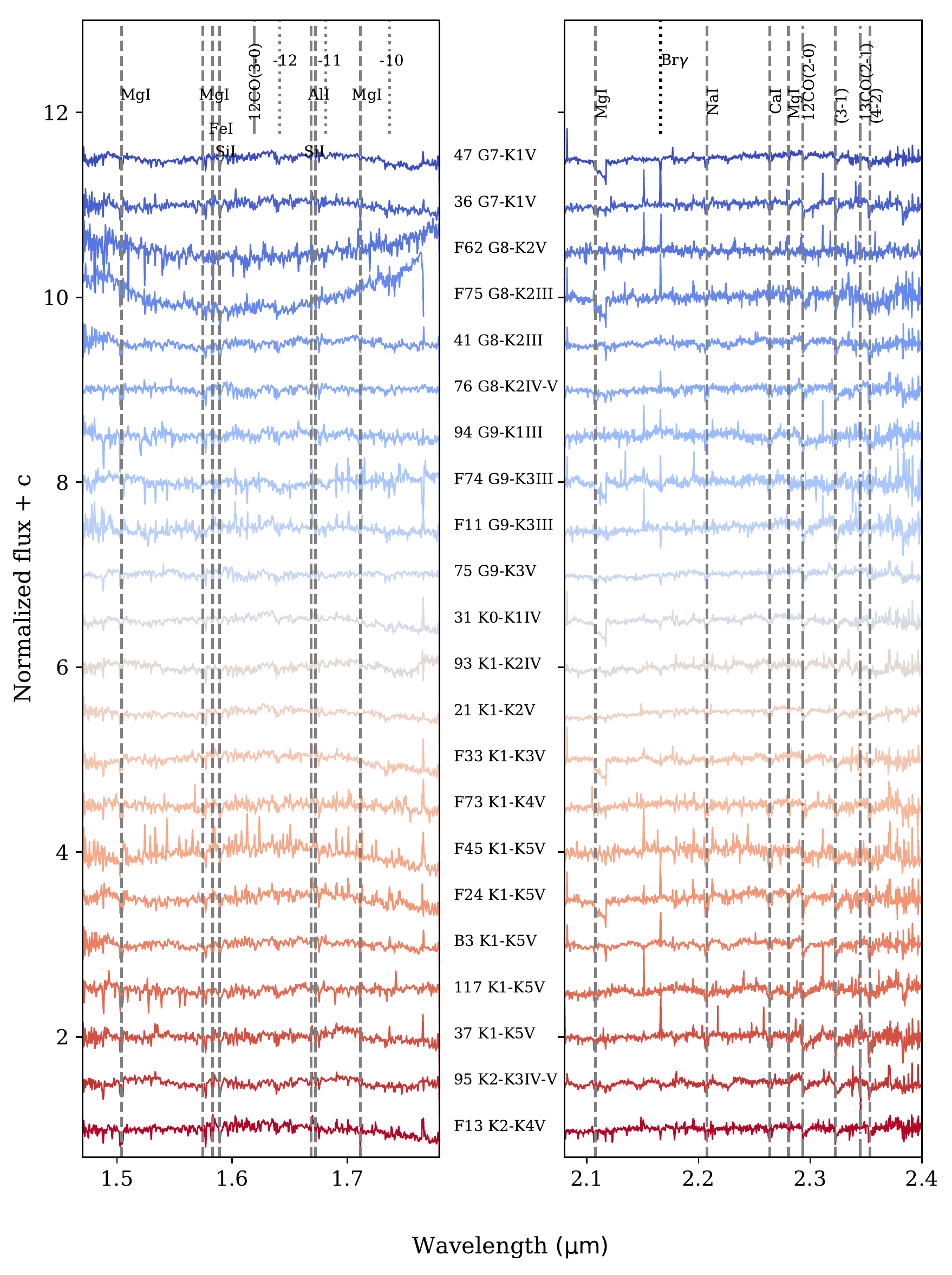}
            }
      \caption{$H$ and $K$-band spectra of the late type stars in \ngc6357.}
         \label{P4:fig:NGC6357_Late_type1}
  \end{figure*}

  \begin{figure*}
            {\includegraphics[width=0.98\textwidth]{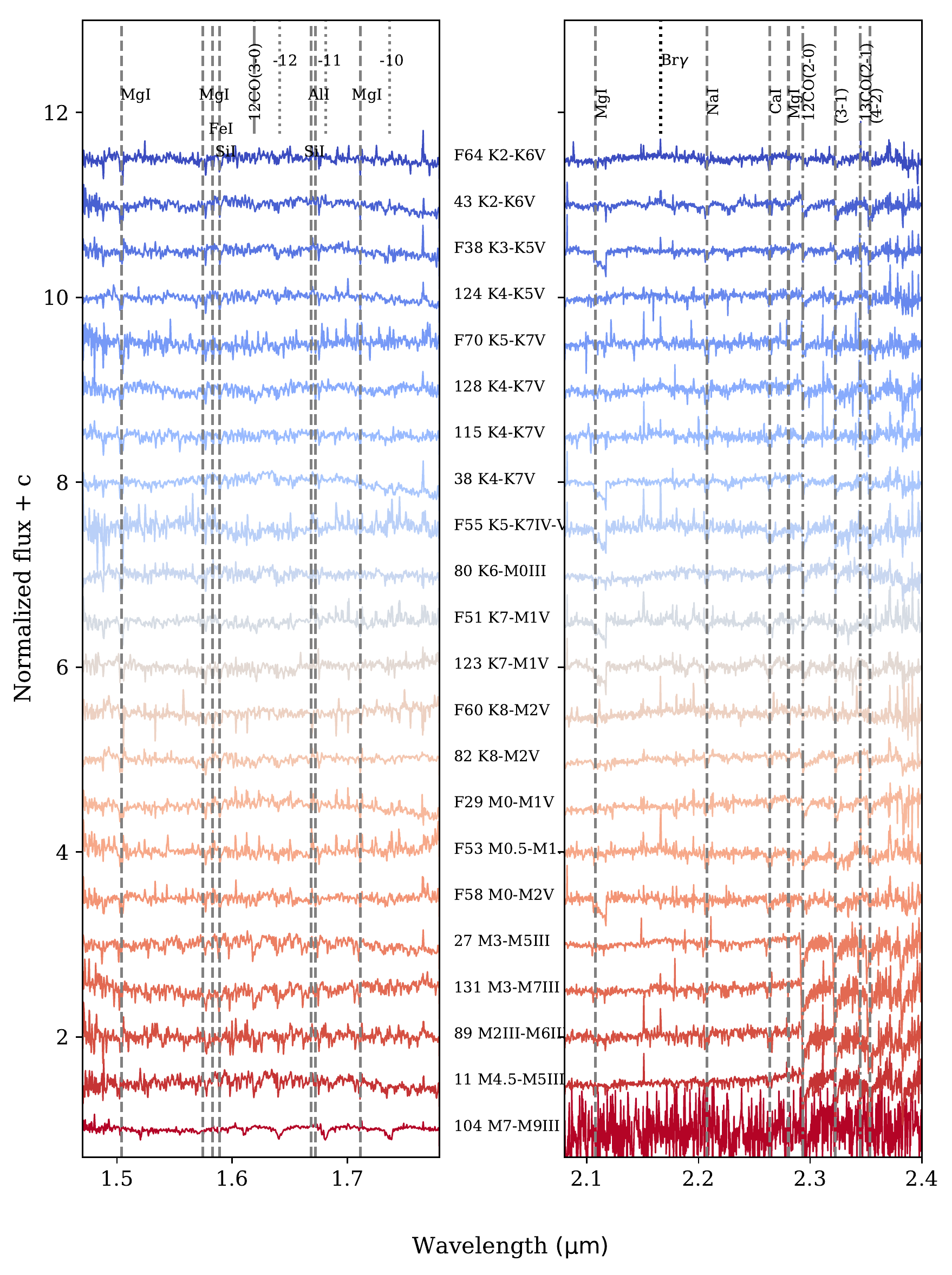}
            }
      \caption{$H$ and $K$-band spectra of the late type stars in \ngc6357.}
         \label{P4:fig:NGC6357_Late_type0}
  \end{figure*}

\clearpage

\twocolumn

\end{appendix}

\end{document}